\documentclass[twocolumn]{aastex7}

\usepackage{longtable}
\usepackage{xargs} 
\usepackage{amsmath}

\newcommand{\rplan}{$R_{\text{p}}$\ }
\newcommand{\rjup}{R_{\text{Jup}}}
\newcommand{\rsol}{$R_{\sun}$\ }

\newcommand{\teff}{T$_{\text{eff}}$\ }
\newcommand{\logg}{{$\log$(g)}\ }

\defcitealias{gilliland2000hstsearch}{G00}
\defcitealias{weldrake2005tuc}{W05}
\defcitealias{fressin2013keplerocc}{F13}
\defcitealias{masuda2017tucreassessment}{MW17}

\received{2024 December 12}
\revised{2025 May 22}
\accepted{2025 May 24}
\submitjournal{AJ}

\begin{document}

\title{The Multiband Imaging Survey for High-Alpha PlanetS (MISHAPS) I: Preliminary Constraints on the Occurrence Rate of Hot Jupiters in 47 Tucanae}
\shorttitle{MISHAPS I: 47 Tuc}

\author[0000-0003-4310-3440]{Alison L. Crisp}
\affiliation{Department of Physics \& Astronomy, Louisiana State University, Baton Rouge, LA 70803}
\affiliation{Department of Astronomy, The Ohio State University, Columbus, OH 43210, USA}
\email{crisp.92@osu.edu}

\author[0000-0002-3469-5133]{Jonas Kl{\"u}ter}
\affiliation{Department of Physics \& Astronomy, Louisiana State University, Baton Rouge, LA 70803}
\email{jonas.klueter@googlemail.com}

\author[0009-0002-1973-5229]{Marz L. Newman}
\affiliation{Department of Physics \& Astronomy, Louisiana State University, Baton Rouge, LA 70803}
\email{mnewm25@lsu.edu}

\author[0000-0001-7506-5640]{Matthew T. Penny}
\affiliation{Department of Physics \& Astronomy, Louisiana State University, Baton Rouge, LA 70803}
\affiliation{Center for Computation \& Technology, Louisiana State University, Baton Rouge, LA 70803}
\email{penny1@lsu.edu}

\author[0000-0002-9539-4203]{Thomas G. Beatty}
\affiliation{Department of Astronomy, University of Wisconsin -- Madison, Madison, WI 53706, USA}
\email{tgbeatty@wisc.edu}

\author[0000-0003-2076-8001]{L. Ilsedore Cleeves}
\affiliation{Department of Astronomy, University of Virginia, Charlottesville, VA 22904, USA}
\affiliation{Department of Chemistry, University of Virginia, Charlottesville, VA 22904, USA}
\email{lic3f@virginia.edu}

\author[0000-0001-6588-9574]{Karen A. Collins}
\affiliation{Harvard-Smithsonian Center for Astrophysics, Cambridge, MA 02138, USA}
\email{karenacollins@outlook.com}

\author[0000-0001-7258-1834]{Jennifer A. Johnson}
\affiliation{Department of Astronomy, The Ohio State University, Columbus, OH 43210, USA}
\affiliation{Center for Cosmology \& Astro Particle Physics, The Ohio State University, Columbus, OH 43210, USA}
\email{johnson.3064@osu.edu}

\author[0000-0002-5099-8185]{Marshall C. Johnson}
\affiliation{Department of Astronomy, The Ohio State University, Columbus, OH 43210, USA}
\email{johnson.7240@osu.edu}

\author[0000-0003-2527-1598]{Michael B. Lund}
\affiliation{NASA Exoplanet Science Institute, California Institute of Technology, Pasadena, CA 91125, USA}
\email{mlund@ipac.caltech.edu}

\author[0000-0002-9144-7726]{Clara E. Mart{\'i}nez-V{\'a}zquez}
\affiliation{Gemini North Observatory, NOIRLab, Hilo, HI 96720}
\email{clara.martinez@noirlab.edu}

\author[0000-0001-5082-6693]{Melissa K. Ness}
\affiliation{Department of Astronomy, Columbia University, New York, NY 10027, USA}
\affiliation{Simons Center for Computational Astrophysics, Flatiron Institute, New York, NY 10004, USA}
\email{mkness@gmail.com}

\author[0000-0001-8812-0565]{Joseph E. Rodriguez}
\affiliation{Center for Data Intensive and Time Domain Astronomy, Department of Physics \& Astronomy, Michigan State University, East Lansing, MI 48824, USA}
\email{rodriguez.jr.joey@gmail.com}

\author[0000-0001-5016-3359]{Robert Siverd}
\affiliation{Institute for Astronomy, University of Hawai'i at M\=anoa, Honolulu, HI 96822, USA}
\email{robert.siverd@gmail.com}

\author[0000-0002-5951-8328]{Daniel J. Stevens}
\affiliation{Department of Physics \& Astronomy, University of Minnesota -- Duluth, Duluth, MN 55812, USA}
\email{djs17@d.umn.edu}

\author[0000-0001-6213-8804]{Steven Villanueva}
\affiliation{Exoplanets \& Stellar Astrophysics Laboratory, NASA Goddard Space Flight Center, Greenbelt, MD 20771, USA}
\email{steven.villanueva@nasa.gov}

\author[0000-0002-0619-7639]{Carl Ziegler}
\affiliation{Department of Physics, Engineering \& Astronomy, Stephen F. Austin State University, Nacogdoches, TX 75962, USA}
\email{carl.ziegler@sfasu.edu}

\begin{abstract}
The first generation of transiting planet searches in globular clusters yielded no detections, and in hindsight, only placed occurrence rate limits slightly higher than the measured occurrence rate in the higher-metallicity Galactic thick disk.
To improve these limits, we present the first results of a new wide field search for transiting hot Jupiters in the globular cluster 47~Tucanae.
We have observed 47~Tuc as part of the Multiband Imaging Survey for High-Alpha Planets (MISHAPS).
Using 24 partial and full nights of observations taken with the Dark Energy Camera on the 4-m Blanco telescope at CTIO, we perform a search on 19,930 stars in the outer regions of the cluster.
Though we find no clear planet detections, by combining our result with the upper limit enabled by \citeauthor{gilliland2000hstsearch}'s~\citeyearpar{gilliland2000hstsearch} \textit{Hubble} search for planets around an independent sample of 34,091 stars in the inner cluster, we place the strongest limit to date on hot Jupiters with periods of $0.8 \leq P \leq 8.3$ days and $0.5~R_{\rm Jup} \leq R_{\rm P} \leq 2.0~R_{\rm Jup}$ of $f_{\rm HJ} < 0.11\%$, a factor of ${\sim}$4 below the occurrence rate in the \textit{Kepler} field.
Our search found 35 transiting planet candidates, though we are ultimately able to rule out each without follow-up observations.
We also found 4 eclipsing binaries, including 3 previously-uncataloged detached eclipsing binary stars.
\end{abstract}

\keywords{Exoplanet astronomy --- Transit photometry --- Hot Jupiters  --- Globular star clusters}

\section{Introduction}\label{sec:intro}
Though giant planet formation rates have been found to correlate with [Fe/H] \citep[e.g., ][]{gonzales1997metallicity, santos2004gpmetal, fischer2005gpfeh, johnson2010gpoccurrence}, the specific elements responsible for the correlation have largely been unresolved. 
Volatiles and silicates are expected to be crucial ingredients for giant planet formation, because the silicate dust grains in protoplanetary disks provide surfaces for the more abundant volatiles to stick to, forming planetesimals \citep{pontoppidan2014volatiles, oberg2021astrochem, nielsen2023MWplanetform}.
Beyond the ice line, these volatiles likely make up the majority of the protoplanetary disk's solid mass.
These compounds largely consist of $\alpha$-elements --- elements formed in $\alpha$-capture reactions --- such as oxygen, silicon, and magnesium, suggesting that the $\alpha$-abundance, [$\alpha$/H] = [Fe/H] + [$\alpha$/Fe], may influence planet occurrence rates more than [Fe/H].
However, our current sample of giant planet hosts prevents us from disentangling a relationship.
Many come from \textit{Kepler} and \textit{TESS}, whose targets primarily lie within the high-[Fe/H], low-[$\alpha$/Fe] thin disk \citep{bensby2003diskI, bensby2005diskII, reddy2006thickdisc, guo2017hjmet, weinberg2019apogee, griffith2021abundance}.
Past observational limits for ground-based radial velocity and transit searches have similarly limited our sample to primarily disk stars.
Some analyses of low-[Fe/H], high-[$\alpha$/Fe] subsamples of stars in the Galactic thick disk such as \citet{adibekyan2012kepharps, adibekyan2012overabundance} have found higher numbers of giant planets than are predicted by a [Fe/H] dependence, which could potentially be explained by the relative $\alpha$-enhancement.
However, the bias of our giant planet sample towards disk stars prevents us from determining a conclusive explanation for this surplus: we lack observations in the full [$\alpha$/Fe]-[Fe/H] parameter space, particularly in high-[Fe/H], high-[$\alpha$/Fe] populations \citep[e.g., ][]{wilson2022chemicalkep}.

Some populations that can potentially be used to separate the influences of different elements are the stars of globular clusters.
Their relatively low metallicities should hinder giant planet formation if it is dependent on only [Fe/H] or other products of Type Ia nucleosynthesis.
But, if any giant planets were to be found, it could indicate that their enhanced $\alpha$-abundances may compensate for their low metallicity.

Additionally, globular clusters allow us to search for planets in dense stellar environments --- a region of parameter space that is largely unexplored. 
This parameter space is complicated by some theoretical uncertainty: we don’t yet know if, e.g., the gravitational interactions produced by these dense fields would inhibit or enhance hot Jupiter formation.
\citet{hamers2017hjgc} posit that only the densest core regions of the clusters would eject all planets, and \citet{winter2020clustering} suggest that the gravitational perturbations in dense fields may initiate planet-planet scattering that actually \textit{enhances} hot Jupiter formation. 
Other studies suggest that conditions in globular clusters beyond gravitational effects may adversely affect the likelihood of planet formation.
\citet{thompson2013hotwater} theorizes that because dense regions such as globular clusters tend to form in hot starbursts, the high temperatures may evaporate the protoplanetary disks more quickly than giant planets can form. 
Further, \citet{lee2020radiation} find that the ambient radiation in star forming regions is extremely high, which could potentially eliminate ice lines in protoplanetary disks entirely. 
Even if the ice line is not entirely eliminated, analyses such as \citet{winter2022photoevap} find that in disks subjected to high far ultraviolet radiation, the growth and migration of giant planets is still suppressed.
Thus, tighter observational constraints on planet occurrence rates in extreme environments are needed to constrain these seemingly contradictory theories.

Globular clusters are also of potential interest for planet searches because of their well-constrained, relatively advanced ages.
For hot Jupiters in particular, studies such as \citet{chen2023hjagedist} suggest that, even if hot Jupiters formed in globular clusters, the systems may be old enough that the planets have already been engulfed by their host stars due to orbital tidal decay.
Improved constraints on hot Jupiter occurrence rates in globular clusters can thus improve our understanding of formation in this parameter space.

To date, planet searches of globular clusters have resulted in one successful detection in M4: a roughly 1.18 $\rjup$  planet orbiting the millisecond pulsar PSR~B1620-26 \citep{richer2003gcplanet}.
Its location towards the core of M4 implies that planet formation in clusters may still be possible, despite their complex stellar dynamics \citep{fregeau2006interactions}.
Other studies suggest it may indicate that --- though globular clusters may have too low a metallicity for typical giant planet formation --- they may be suitable for forming planets from supernova fall-back \citep[e.g., ][]{beer2004m4, sigurdsson2008gcpulsarplanets}.
However, no searches for globular cluster planets orbiting normal main sequence or giant stars have successfully discovered a planet, including a further search of M4 using \textit{K2} data \citep{wallace2020m4}, which was limited to only $\sim$4,000 distinguishable stars due to \textit{Kepler}'s 4"/px resolution.

These unsuccessful searches include surveys of 47~Tucanae and $\omega$~Centauri, some of the most massive and closest clusters to Earth.
47~Tuc in particular has been promoted as a potential planet-hosting cluster, as analyses of pulsar planet dynamics such as \citet{sigurdsson1992gcpulsarplanets} place its density in a dynamical sweet spot where a giant planet could form from supernova ejecta remaining around a pulsar before the material is disrupted by stellar interactions.
Its relatively high abundances -- [Fe/H]$\approx$-0.79~dex and [$\alpha$/Fe]$\approx$-0.3~dex \citep[e.g., ][]{cordero2014tucabundances} -- also make it a compelling case for planet searches of globular clusters.
Three different studies, including two surveys, have therefore assessed 47~Tuc's planet-hosting abilities \citep{gilliland2000hstsearch, weldrake2005tuc, masuda2017tucreassessment}.
$\omega$~Cen was searched by \citet{weldrake2008wcen}, which analyzed $\sim$31,000 main sequence stars in the cluster, finding no planets.
An additional survey by \citet{nascimbeni2012hst} analyzed over 5,000 lightcurves from \textit{HST} observations of another globular cluster, NGC~6397, focusing particularly on a sample of 2,215 M-dwarfs, and again found no planets.

Surveys of 47~Tuc began with an \textit{HST} search by \citet{gilliland2000hstsearch} (hereafter G00).
Using the accepted hot Jupiter occurrence rate in the solar neighborhood for the time --- 0.8-1.0\% --- they expected to find $\sim$17 planets in their observations.
However, despite taking 8.3 days of near-continuous data for $\sim$34,000 stars, they observed no statistically significant transit events.
In a later re-analysis using a simulated stellar sample based on giant planet hosting stars in the \textit{Kepler} sample, \citet{masuda2017tucreassessment} (hereafter MW17) find that, if the occurrence rate of hot Jupiters in 47~Tuc was to match that of \textit{Kepler} field stars, accounting for the host-mass dependence of giant planet occurrence rate \citep{johnson2010gpoccurrence}, \citetalias{gilliland2000hstsearch} could only have expected to find $\sim$2 planets in their sample of 34,000 stars, with a $\sim$15\% chance of finding no planets.
Thus, the null result of the \citetalias{gilliland2000hstsearch} survey is less statistically significant than initially thought.

\citet{weldrake2005tuc} (hereafter W05) followed-up the \citetalias{gilliland2000hstsearch} survey with a 33-night ground-based search using the Australian National University 40 inch Telescope. 
To complement the \citetalias{gilliland2000hstsearch} search, which focused more on the core of the cluster, this survey focused on $\sim$22,000 stars in the outskirts of the cluster.
Using Monte Carlo simulations, they expected to find $\sim$7 planets in their data if the occurrence rate was 0.8\%, and still found none.

With the results of these surveys in mind, we have observed 47~Tuc as part of the Multiband Imaging Survey for High-Alpha PlanetS (MISHAPS) to set more stringent limits on globular cluster planet occurrence rates and test the effects of $\alpha$-abundance on giant planet formation.
In this paper, we present the first results of our search of 19,930 47~Tuc stars using the Dark Energy Camera.
The structure of the paper is as follows.
In Section \ref{sec:survey}, we provide an overview of the MISHAPS survey design and observations. 
In Section \ref{sec:static}, we describe how we extract and calibrate our photometry, and characterize our stellar sample.
In Section \ref{sec:timeseries}, we discuss the timeseries analysis of our data, including de-trending lightcurves.
In Section \ref{sec:vetting}, we describe our transit injections, transit search, efficiency calculations, and vetting process.
In Section \ref{sec:detailed}, we discuss the planet candidates found in the search and  our rejection process.
In Section \ref{sec:occurrence}, we place a new upper limit on our estimate of the hot Jupiter occurrence rate in 47~Tuc.
We discuss our results in Section \ref{sec:Discussion} and give conclusions in Section \ref{sec:conclusions}.

\section{Survey Design \& Observations}\label{sec:survey}
    \subsection{Survey Design}\label{subsec:design}
    Motivated by \citetalias{masuda2017tucreassessment}, our goals are to place stronger constraints on the occurrence rate than \citetalias{gilliland2000hstsearch}'s observations allow, and to conclusively rule out an occurrence rate equal to that in the \textit{Kepler} field (average [M/H]=-0.045$\pm$0.009, e.g., \citet{guo2017hjmet}).
    
    MISHAPS uses the Dark Energy Camera (DECam) on the 4-m V\'ictor M. Blanco telescope at Cerro-Tololo Interamerican Observatory to perform a high-cadence, wide-field survey of 47~Tuc and other $\alpha$-enhanced fields to search for transiting planets.
    DECam's wide field of view and the Blanco's 4-m diameter allow us to pursue an alternative strategy to most transiting planet searches that have relied in part on covering multiple transits and folding to increase the signal-to-noise ratio (S/N) above a detection threshold.
    Instead, the Blanco's aperture enables a clear detection of a Jupiter-radius planet's transit over $\sim$5 magnitudes of the 47~Tuc main sequence.
    We therefore planned a small survey ($\sim$7 nights) to search for single full or partial transits on a larger sample of stars than either \citetalias{gilliland2000hstsearch} or \citetalias{weldrake2005tuc}. 
    Without the necessity of multiple transit detections, the detection probability for short-period planets is therefore $\sim$10's of percent for even 7 nights of data. 
    With the expectation that the number of planet candidates we would find would be small, and that the quality of data would allow many EB false positives to be rejected from survey data alone, we anticipated that the number of targets that required follow-up would be small. 
    For those that did, an estimate of the orbital period from a single transit would limit the amount of observations needed to recover a second transit and measure a period \citep[e.g., ][]{seager2003parameters,yee2008keplptcharacterization,villanueva2019singletransit}.

    We selected DECam for its size, photometric precision, near-infrared $z-$band sensitivity, and low overhead. 
    DECam consists of 62 deep-depletion science CCDs, each of which has a size of 2048x4096 px for a total of 520 Mpx \citep{honscheid2008decam}, though only 60 are currently in use.
    The $\sim$3~deg$^{2}$ FOV allows us to simultaneously monitor the full expanse of 47~Tuc.
    Its photometric precision allows us to resolve stars with magnitudes down to $r \approx24$ and to potentially detect single transits on stars down to $r\approx22$.
    DECam’s extremely high $z-$band sensitivity and low overheads are critical to enabling chromatic timeseries observations, which allow us to use differences in eclipse depths in $r$ and $z$ to rule out false positives caused by secondary eclipses of eclipsing binaries (EBs), diluted primary eclipses of hierarchical triple stars, and blended Small Magellanic Cloud (SMC) EBs.
    The low overheads also allow observations at a high enough cadence to resolve ingresses and egresses of transits, which are important for estimating accurate transit times and host stellar densities \citep[e.g., ][]{seager2003parameters}, as well as increasing the accuracy with which one can estimate periods by adding well-sampled partial eclipses. 

    \subsection{Observations}\label{subsec:obs}
    Preliminary observations of 47~Tuc were taken in short 2-4 hour windows on 17 nights in 2019 and 2021.
    Additionally, eight half-nights of observations of the cluster were conducted in 2022.
    Each night, we took a continuous timeseries of images with 100 s exposure times, alternating between the $r$ and $z$ filters with each exposure.
    This exposure sequence resulted in a $\sim4$-minute cadence in each filter, or $\sim2$ minutes combined.
    These sequences were interrupted roughly once per hour to check and correct the telescope pointing if it drifted by more than 10".
    All images were first processed through the DECam Community Pipeline \citep{valdes2014decamcp}, which applies calibration frames, corrects instrumental signatures, and applies a cosmic ray removal algorithm.
    We performed a preliminary determination of the overall quality of the nights by plotting the background, ellipticity, zeropoint magnitude, and seeing values from the image headers and flagging any nights which consistently had a zeropoint $<$30.0, the magnitude of a star resulting in only 1 photon s$^{-1}$ in the exposure.
    These poor-quality nights were later confirmed via analysis of the nightly median absolute deviation (MAD) of the lightcurve's photometric root-mean-square (RMS) after the images were processed and the photometry was extracted (\S\ref{subsec:dia}).
    Table \ref{tab:obs} shows the log of observing nights and their mean data quality statistics.
    In total, our observations produce 1,288 images in the $z-$band and 1,293 images in the $r-$band.

    \begin{deluxetable*}{cccccccccccccc}[ht!]
        \centering
        \tablecaption{Observations}
        \tablehead{
            \colhead{} & \colhead{} & \colhead{$T_{\rm obs}$} & \colhead{$\quad$ } & \colhead{} & \colhead{$r$} & \colhead{} & \colhead{} & \colhead{$\quad$} & \colhead{} & \colhead{$z$} & \colhead{} & \colhead{} & \colhead{} \\
            \colhead{Date} & \colhead{Night} & \colhead{$(hrs)^{*}$} & \colhead{$\quad$ Ims.} & \colhead{FWHM} & \colhead{Sky} & \colhead{$m_{\rm zp}$} & \colhead{Ellip.} & \colhead{$\quad$ Ims.} & \colhead{FWHM} & \colhead{Sky} & \colhead{$m_{\rm zp}$} & \colhead{Ellip.} & \colhead{Good?} \\
        }
        \startdata
            2019-08-01 & 697 & 2.43 & $\quad$ 35 & 7.49 & 281.80 & 29.76 & 0.45 & $\quad$ 34 & 7.40 & 2040.98 & 29.90 & 0.61 & N \\
2019-08-02 & 698 & 3.42 & $\quad$ 48 & 4.65 & 267.31 & 30.18 & 0.12 & $\quad$ 47 & 3.79 & 1963.30 & 30.25 & 0.10 & Y \\
2019-08-03 & 699 & 2.66 & $\quad$ 37 & 5.38 & 306.57 & 30.11 & 0.08 & $\quad$ 37 & 4.63 & 2450.80 & 30.21 & 0.07 & Y \\
2019-08-04 & 700 & 3.54 & $\quad$ 49 & 4.81 & 304.65 & 30.18 & 0.09 & $\quad$ 49 & 4.05 & 2737.96 & 30.27 & 0.08 & Y \\
2021-07-05 & 1401 & 1.75 & $\quad$ 24 & 7.01 & 387.98 & --- & 0.54 & $\quad$ 24 & 7.15 & 3456.72 & 29.27 & 0.63 & N \\
2021-07-06 & 1402 & 1.91 & $\quad$ 26 & 5.39 & 328.45 & 30.13 & 0.07 & $\quad$ 26 & 4.65 & 2946.13 & 30.20 & 0.07 & Y \\
2021-07-07 & 1403 & 1.91 & $\quad$ 26 & 4.22 & 310.48 & 30.30 & 0.10 & $\quad$ 26 & 3.54 & 2966.94 & 30.33 & 0.07 & Y \\
2021-07-08 & 1404 & 2.03 & $\quad$ 28 & 3.93 & 240.22 & 30.30 & 0.12 & $\quad$ 28 & 3.23 & 1760.58 & 30.33 & 0.09 & Y \\
2021-07-09 & 1405 & 2.13 & $\quad$ 30 & 4.35 & 271.76 & 30.25 & 0.10 & $\quad$ 29 & 3.68 & 2255.65 & 30.31 & 0.08 & Y \\
2021-07-10 & 1406 & 2.48 & $\quad$ 34 & 4.59 & 331.25 & 29.90 & 0.10 & $\quad$ 35 & 3.82 & 2824.49 & 29.87 & 0.09 & N \\
2021-07-11 & 1407 & 2.34 & $\quad$ 32 & 6.47 & 256.85 & 29.74 & 0.13 & $\quad$ 33 & 5.94 & 1930.90 & 29.77 & 0.11 & N \\
2021-07-12 & 1408 & 2.38 & $\quad$ 33 & 6.92 & 230.31 & 29.48 & 0.51 & $\quad$ 33 & 7.32 & 1662.95 & 29.62 & 0.57 & N \\
2021-08-02 & 1429 & 3.73 & $\quad$ 51 & 3.81 & 313.30 & 30.29 & 0.11 & $\quad$ 51 & 3.15 & 2187.50 & 30.33 & 0.09 & Y \\
2021-08-03 & 1430 & 3.88 & $\quad$ 52 & 4.83 & 361.64 & 30.22 & 0.09 & $\quad$ 51 & 4.21 & 2836.70 & 30.24 & 0.09 & Y \\
2021-08-04 & 1431 & 3.87 & $\quad$ 51 & 4.57 & 295.14 & 30.24 & 0.11 & $\quad$ 51 & 3.95 & 2308.42 & 30.22 & 0.09 & Y \\
2021-08-05 & 1432 & 3.99 & $\quad$ 55 & 4.61 & 296.64 & 30.12 & 0.10 & $\quad$ 54 & 3.89 & 2507.44 & 30.19 & 0.10 & Y \\
2021-08-06 & 1433 & 4.07 & $\quad$ 54 & 5.49 & 252.50 & 30.13 & 0.11 & $\quad$ 54 & 4.87 & 2122.65 & 30.19 & 0.10 & Y \\
2022-11-01 & 1885 & 7.13 & $\quad$ 95 & 5.65 & 987.97 & 30.15 & 0.08 & $\quad$ 95 & 4.92 & 3146.69 & 30.20 & 0.08 & Y \\
2022-11-02 & 1886 & 7.02 & $\quad$ 94 & 4.51 & 1163.83 & 30.27 & 0.09 & $\quad$ 94 & 3.86 & 2775.51 & 30.29 & 0.08 & Y \\
2022-11-04 & 1888 & 6.94 & $\quad$ 94 & 4.63 & 1828.89 & 30.26 & 0.09 & $\quad$ 94 & 4.03 & 3251.21 & 30.28 & 0.08 & Y \\
2022-11-05 & 1889 & 6.87 & $\quad$ 93 & 3.99 & 2180.92 & 30.29 & 0.10 & $\quad$ 93 & 3.54 & 3838.18 & 30.30 & 0.09 & Y \\
2022-11-07 & 1891 & 5.45 & $\quad$ 75 & 4.74 & 5213.08 & 29.34 & 0.12 & $\quad$ 73 & 4.33 & 6028.71 & 29.36 & 0.21 & N \\
2022-11-08 & 1892 & 6.49 & $\quad$ 87 & 4.44 & 7452.61 & 29.36 & 0.11 & $\quad$ 87 & 4.12 & 8387.70 & 29.26 & 0.11 & N \\
2022-11-10 & 1894 & 6.66 & $\quad$ 90 & 6.38 & 1579.38 & 29.82 & 0.50 & $\quad$ 90 & 5.90 & 3934.21 & 29.82 & 0.50 & N \\
        \enddata
        \tablecomments{47~Tuc observation log. Observation night numbers are given in terms of JD-2458000.0. The total time observed is rounded to the nearest half-hour. The median FWHM, sky counts subtracted, photometric zeropoints, and ellipticities have been taken over all images in each filter, as measured by the DECam Community Pipeline and recorded in the image headers. Our sky counts increase significantly starting in 2022 because our 47~Tuc-targeted observations can be performed during bright time, while previous MISHAPS observations were during dark time. Survey nights 697, 1401, 1406, 1407, 1408, 1891, 1892, and 1894 all had $m_{\rm zp}<30.0$ for the majority of their observations, which increases the chance of systematics and lower S/N, and were thus deemed unreliable (denoted with an ``N'' in the ``Good?'' column). However, we keep the data in the lightcurves in case those nights can be used to help confirm detections on good nights.}
        \label{tab:obs}
    \end{deluxetable*}

\section{Reference Image Analysis}\label{sec:static}
We produce reference images for each band for use in difference imaging, requiring us to choose an appropriate number of high-quality images from our full dataset and stack them (\S\ref{subsec:makerefim}).
We then extract our initial catalog of stars using PSF photometry, and find astrometric solutions and photometric calibrations for each target by crossmatching with Gaia DR3 \citep{babusiaux2023gaiadr3} and the NOIRLab Source Catalog Data Release 2 \citep[NSC,][]{nidever2018nsc1, nidever2021nsc2} (\S\ref{subsec:refimphot}).
We also determine appropriate proper motion and color selections for cluster membership using the Gaia crossmatch results (\S\ref{subsec:member}).
We then perform several characterization steps to determine which targets will be used to perform transit injections and for use in efficiency estimates (\S\ref{subsec:radii}).
    
    \subsection{Reference Image Construction}\label{subsec:makerefim}
    For each of DECam's CCD chips, we constructed a deep, high-quality reference image in the $r$ and $z$ filters by selecting and combining a small number of the best images from the data taken in 2021, which had the best conditions. 
    To do this, the community pipeline-determined point spread function (PSF), full width at half maximum (FWHM), sky background, zeropoint (magnitude resulting in 1 photon s$^{-1}$), PSF ellipticity, and airmass for all images were compiled from the FITS headers, and sorted by FWHM. 
    Fifteen images were selected in order of increasing FWHM, and any with high sky background, PSF ellipticity, or smaller zeropoint magnitude were skipped.
    Each candidate image was visually inspected to ensure it was a good-quality image. 
    The fifteen selected images were combined using the {\sc ISIS} image subtraction program's {\sc ref.csh} task \citep{alardlupton1998isis, alard2000isis}, which aligns the images; determines a PSF-matching convolution kernel and photometric scaling to match the PSF, zeropoint, and sky background of the input images; and then combines them as a mean stack with 3-$\sigma$ clipping. 
    We refer to the resultant images as the reference image for the relevant chip and filter.
    
    \subsection{Reference Image Photometry \& Calibration}\label{subsec:refimphot}
    We used {\sc DoPHOT} \citep{schechter1993dophot, alonsogarcia2012gcs} to find stars and extract PSF fitting photometry in each $r$ and $z$ reference image.
    For each star, this algorithm measures pixel positions, an instrumental magnitude and uncertainty, a local sky estimate, and a fit type code indicating the type of PSF fit performed.
    We use this code as a data quality indicator.
    An astrometric solution for each {\sc DoPHOT} catalog was found by matching to Gaia DR3's 2016.0 epoch positions.
    Then $r$ and $z$ catalogs were merged to produce an instrumental 2-filter catalog we label ``cmd." 
    
    Following the method of \citet{zang2018cfht}, we calibrate the {\sc DoPhot} photometry to the NOIRLab Source Catalog Data Release 2 \citep[NSC DR2][]{nidever2018nsc1,nidever2021nsc2}, fitting for a zeropoint offset in each chip independently and a color term that is common to all chips in the field. This is done for stars with $17.0 \lesssim r_{\rm NSC}\lesssim 18.5$\footnote{The actual cuts on stars used for calibration are done on the instrumental magnitudes shifted to the NSC magnitude scale only by the mean difference from instrumental to NSC magnitude for all matched stars}, combined reported NSC and instrumental magnitude errors $\sqrt{\sigma_{r,{\rm NSC}}^2+\sigma_{r, {\rm inst}}^2}<0.01$, and instrumental color $(r-z)_{\rm inst}$ color between 0.1 mag less than the mean instrumental color and 0.8 mag greater than the mean instrumental color. This selection includes the 47 Tuc turnoff, much of the SMC giant branch above the clump, and a scattering of field stars. 
    In sparse chips, the median offset between our calibrated magnitude and the NSC DR2 magnitude is typically $\lesssim 5$~mmag, and the standard deviation is typically between 0.01 and 0.02 mag (estimated using a median absolute deviation scaled by 1.48). 
    The densest chips we analyze in this work, N10 and S10, have median offsets of 0.013 and 0.014 mag, respectively, and standard deviations of 0.039 and 0.038 mag, respectively. 
    The worse performance of the photometric calibration in more crowded fields is a result of the aperture photometry used in NSC DR2 photometry, which is more affected by blending than our PSF photometry.

    For the following analysis, we exclude stars located in the N3, N4, S3, and S4 DECam chips, as the dense crowding of the cluster center leads to a number of problems in our image processing and photometric extraction, the solutions of which are left to the second paper.
    This cut brings our sample down to $\sim$1.3 million stars.
    Further reducing the sample to only those stars which have {\sc DoPHOT} ``objtype'' flags of 1 --- indicating that the object is likely a well-distinguished single star that can be fit with the full 7-parameter PSF model in {\sc DoPHOT} --- in both the $r-$ and $z-$bands leaves $\sim$310,000 stars. 

    \subsection{47~Tuc Membership}\label{subsec:member}
\begin{figure}[htb]
        \centering
        \includegraphics[width=\linewidth]{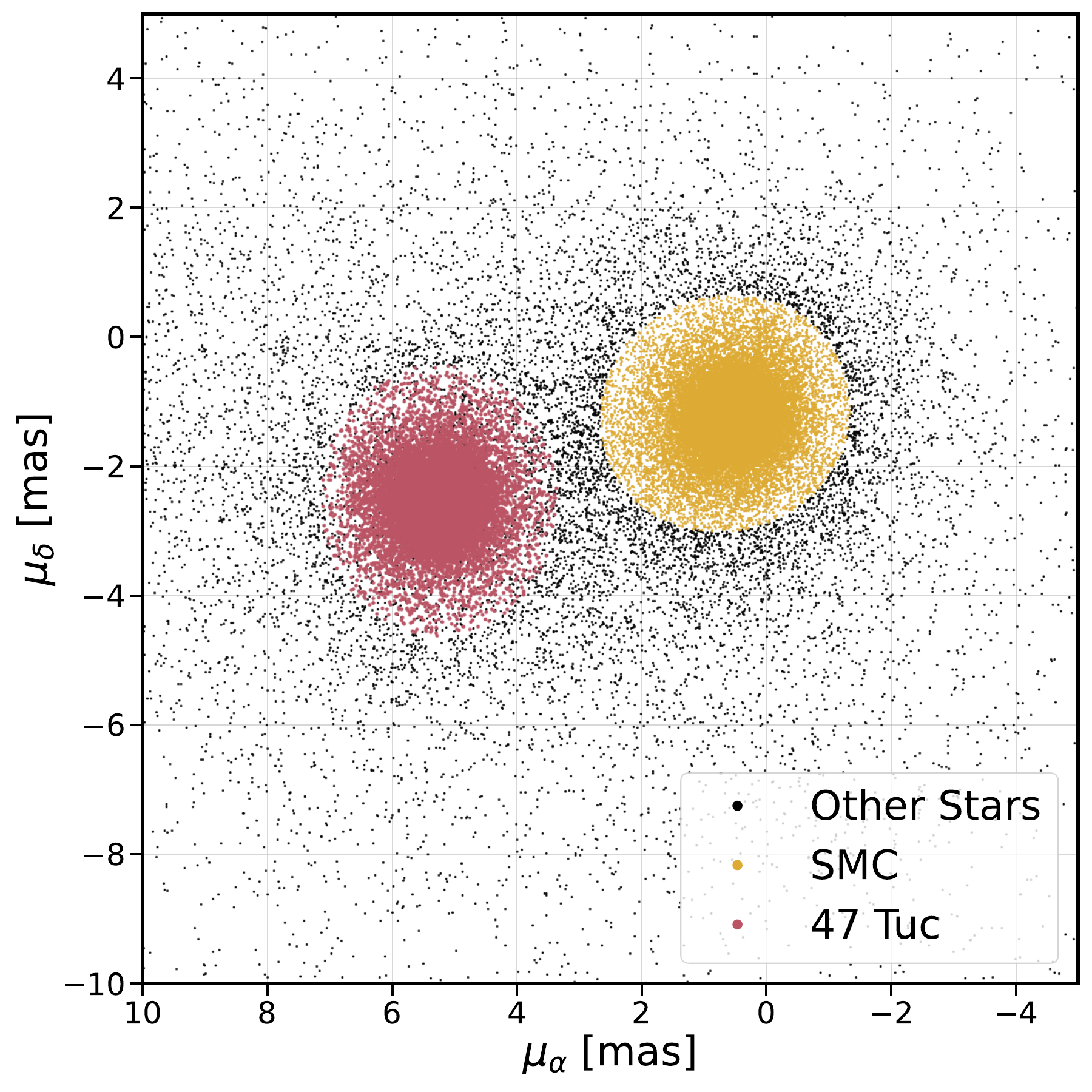}
        \caption{Vector point proper motion diagram plotting Gaia DR3 $\mu_{\rm \alpha}$ and $\mu_{\rm \delta}$. Our color and proper motion-based stellar selection for 47~Tuc is plotted in red. The gold points correspond with likely SMC stars. The black points scattered across the field are other stars, likely foreground Milky Way stars.}
        \label{fig:pmcut}
    \end{figure}

\begin{figure}[htb]
        \centering
        \includegraphics[width=\linewidth]{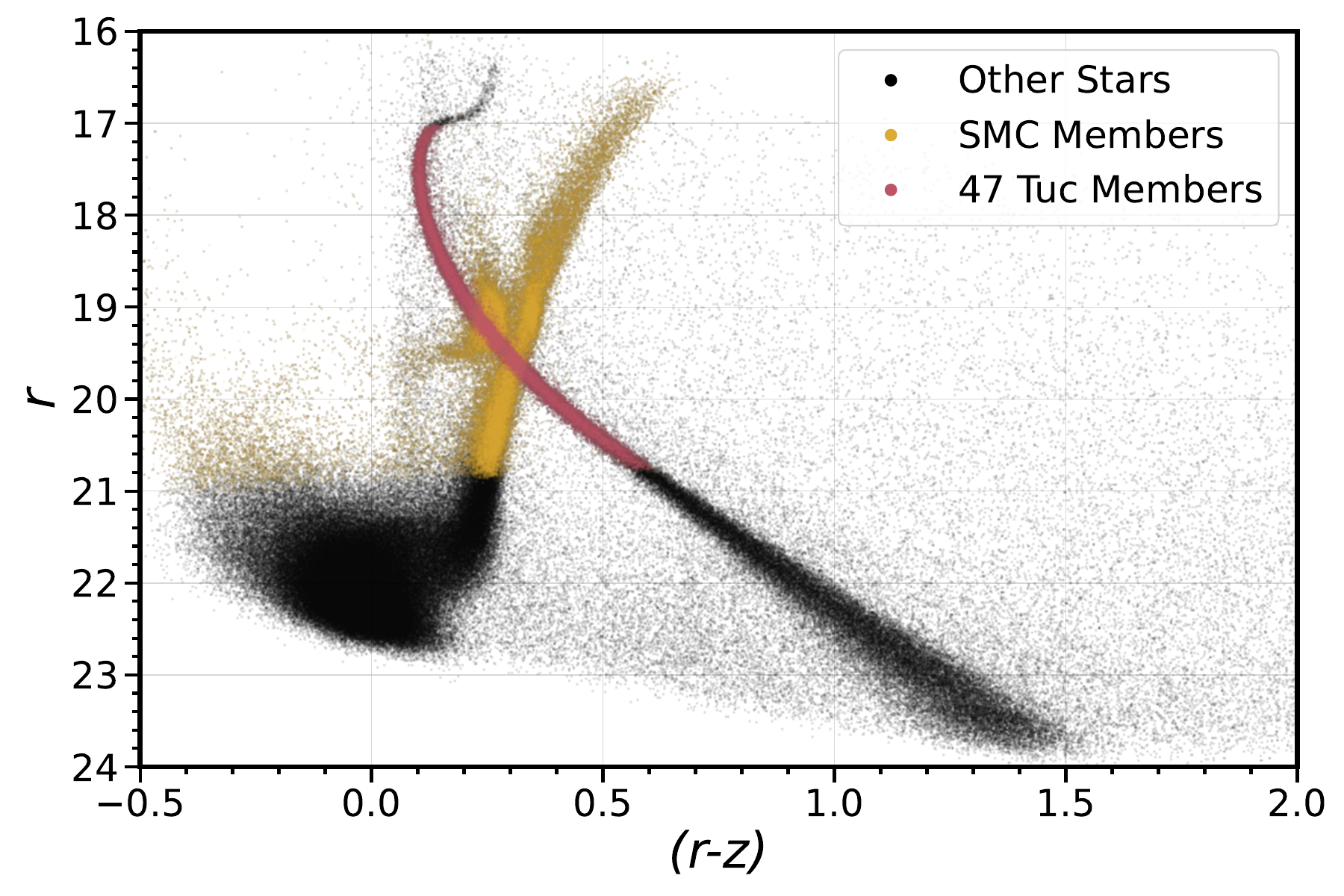}
        \caption{$r$ vs. ($r-z$) color-magnitude diagram of our field, with only the {\sc DoPHOT} type cut applied (black). Our final selection of stars is overplotted in red, and the proper motion-selected SMC stars (selected using the same procedure as for 47~Tuc stars) are overplotted in gold. The remaining black points are targets which either lie outside the 3-$\sigma$ 47~Tuc and SMC proper motion selections, or have no \textit{Gaia} matches.}
        \label{fig:popcmd}
    \end{figure}

    In addition to foreground Milky Way stars, 47~Tuc's visual proximity to the SMC and five other star clusters within it (NGC~121, ESO~28-19, ESO~28-22, HW~5, and Kron~11) requires us to carefully identify true cluster members.
    We crossmatch the sources identified in our images with those in Gaia DR3 \citep{gaia2016gaia, babusiaux2023gaiadr3}, and plot the corresponding proper motions for all sources with matches in Figure \ref{fig:pmcut}. 
    We make an initial wide selection of potential cluster members, using the mean proper motion values for 47~Tuc in \citet{baumgardt2019tucpm}, $\bar{\mu}_{\rm \alpha} = 5.25$ mas/yr and $\bar{\mu}_{\rm \delta} = -2.53$ mas/yr and a selection radius of 
    \begin{equation}\label{eqn:pmselection}
        \sqrt{(\mu_{\rm \alpha} - \bar{\mu}_{\rm \alpha})^{2} + (\mu_{\rm \delta} - \bar{\mu}_{\rm \delta})^{2}} < 3\sigma,
    \end{equation} 
    where $\sigma$=0.94.
    Using the initial selection, we compute new values of $\bar{\mu}_{\rm \alpha}$, $\bar{\mu}_{\rm \delta}$, and their corresponding errors by averaging over the Gaia DR3 values for the selection.
    We make an elliptical selection on ($\bar{\mu}_{\rm \alpha}, \bar{\mu}_{\rm \delta}$) and ($\sigma_{\rm \alpha}, \sigma_{\rm \delta}$), recompute our values, and make another elliptical cut.
    To further constrain the target sample to 47~Tuc stars, we estimate the main sequence locus by plotting the color-magnitude diagram (CMD) for the proper motion selection, computing the median ($r-z$) color in 0.5 mag bins of $r$.
    To the binned median colors, we fit a cubic spline using the {\sc scipy.interpolate} package and select stars within $\pm$0.05 mag of this fit as our final selection.
    This selection leaves us with 19,930 likely cluster members.
    The CMD, with 47 Tuc and SMC populations highlighted, is shown in Figure \ref{fig:popcmd}.
    An overview of these cuts and the resultant numbers of stars can be found in Table \ref{tab:starnums}. 
    Requiring Gaia DR3 matches effectively reduces our faint magnitude limit to $r\sim21.0$, with significant incompleteness below $r\sim20.0$.
    We note that --- although we apply the {\it proper motion} cut before the injections and search are run --- we do not apply the {\it color} cut until after the initial round of vetting has been performed.
    This choice allows us to detect EBs with CMD positions on the binary sequence which would place them outside the range of the color selection.

    \begin{deluxetable*}{lcc}[htb]
        \centering
        \tablecaption{Target star cuts}
        \tablehead{
        \colhead{Cut} & \colhead{Number of Stars Removed} & \colhead{Number of Stars Remaining}
        }
        \startdata
            Full & --- & 1,275,052 \\
            DoPHOT Type Flags & 980,538 & 294,514 \\
            Gaia Cross-Match & 200,742 & 93,772 \\
            $17.0 \leq r \leq 21.0$ & 8,027 & 85,745 \\
            Proper Motion Cut & 71,870 & 21,902 \\
            Color Cut & 1,972 & 19,930 \\
        \enddata
    \tablecomments{Number of stars remaining in our sample after each cut. Our two major limitations are requiring both filters to have DoPHOT type flags of 1, and the requirement that each target have a match with Gaia for proper motion measurements. The Gaia requirement in particular cuts out many of our fainter targets, since its current magnitude limit is $G$=20.7 (corresponding to $r\approx21.0$ for our DECam data).} 
    \label{tab:starnums}
    \end{deluxetable*}

    \subsection{Host Radius Estimates}\label{subsec:radii}
    In order to characterize any transiting or eclipsing systems we find, estimate accurate detection efficiencies, and ultimately achieve an accurate hot Jupiter occurrence rate limit, we must estimate the radii of our target stars.
    To do so, we first estimate the extinctions of our targets and deredden them (\S\ref{subsubsec:extinction}).
    We then transform the dereddened photometry into the PanSTARRS system using spline interpolation functions  (\S\ref{subsubsec:transforms}), so we can use the existing surface brightness relationships in a well-calibrated magnitude system.
    Finally, using the transformed magnitudes, surface brightness relations, and known distance to the cluster, we arrive at our radius estimates (\S\ref{subsubsec:est}).

        \subsubsection{Extinction Estimates}\label{subsubsec:extinction}
        To determine the extinctions, $A_{\lambda}$, of each star, we first obtain estimates of $E(B-V)$ using our cross-match with the NSC, which obtains its $E(B-V)$ values from the SFD maps \citep{schlegel1998dustmaps}.
        However, the catalog only contains $E(B-V)$ estimates for $\sim20\%$ of our targets.
        For those without an estimate, we average the $E(B-V)$ values of stars within 15" which do, and use those averages as their $E(B-V)$. 
        We then use the $E(B-V)$ values, the central wavelengths for the DECam filters, and the \citet{cardelli1989extinction} extinction law, $R_{\rm V}$=3.1, to calculate $A_{r}$ and $A_{z}$ for our sample, with $A_{r}=0.083\pm0.0046$  and $A_{z}=0.045\pm0.0025$ being the mean extinctions.

        \subsubsection{Transformation to PanSTARRS System}\label{subsubsec:transforms}
        To estimate radii, we opt to use the ($g-i$) color-surface brightness relation of \citet{zang2018cfht} in the PanSTARRS DR1 photometric system \citep{chambers2016pssurvey}.
        To use these relations, we transformed our $r$ and $z$ photometry calibrated to the NSC system first into $r_{\rm PS1}$ and $z_{\rm PS1}$, then derived the transforms necessary to convert ($r-z$)$_{\rm PS1}$ to ($g-i$)$_{\rm PS1}$ and $z_{\rm PS1}$ to $i_{\rm PS1}$.
        To derive these transforms, we selected two fields with coverage in both the NSC and PanSTARRS (denoted with NSC and PS1, respectively) --- one relatively reddened field (($\alpha,\delta)=13^{\rm h}06^{\rm m}29^{\rm s},-27^{\circ}02'24''$), and one relatively blue field closer to the disk and containing an open star cluster (($\alpha,\delta)=08^{\rm h}09^{\rm m}22^{\rm s},-26^{\circ}27'36''$).
        From the NSC, we select only stars with photometric errors of $<$0.02 in $g$, $r$, $i$, and $z$.
        From PanSTARRS we select only stars with at least 20 single-epoch detections in all PanSTARRS filters (i.e., {\sc nDetections}$\geq$ 20 in the PanSTARRS flags).
        We then fit splines to the binned medians of the stellar locii, performing iterative sigma-clipping (3 iterations, removing stars outside 3-$\sigma$ from the binned median) to make sure our functions are not biased by outliers.
        We calculate the residuals of the transforms by comparing the values produced by the transform functions to the known PanSTARRS magnitudes of the stars.
        The median offsets and standard deviations of these residuals are shown in \ref{tab:splineparams}, along with the coefficients, knots, and roots required to reproduce the splines. 
        We opt for this path to radius estimates because there are neither published ($r-z$) color-surface brightness relations that we are aware of, nor archival $g$ and $i$ DECam imaging of 47~Tuc with the combination of coverage and depth comparable to our reference images.
            
\begin{figure*}[hbtp!]
            \centering
            \includegraphics[width=\textwidth]{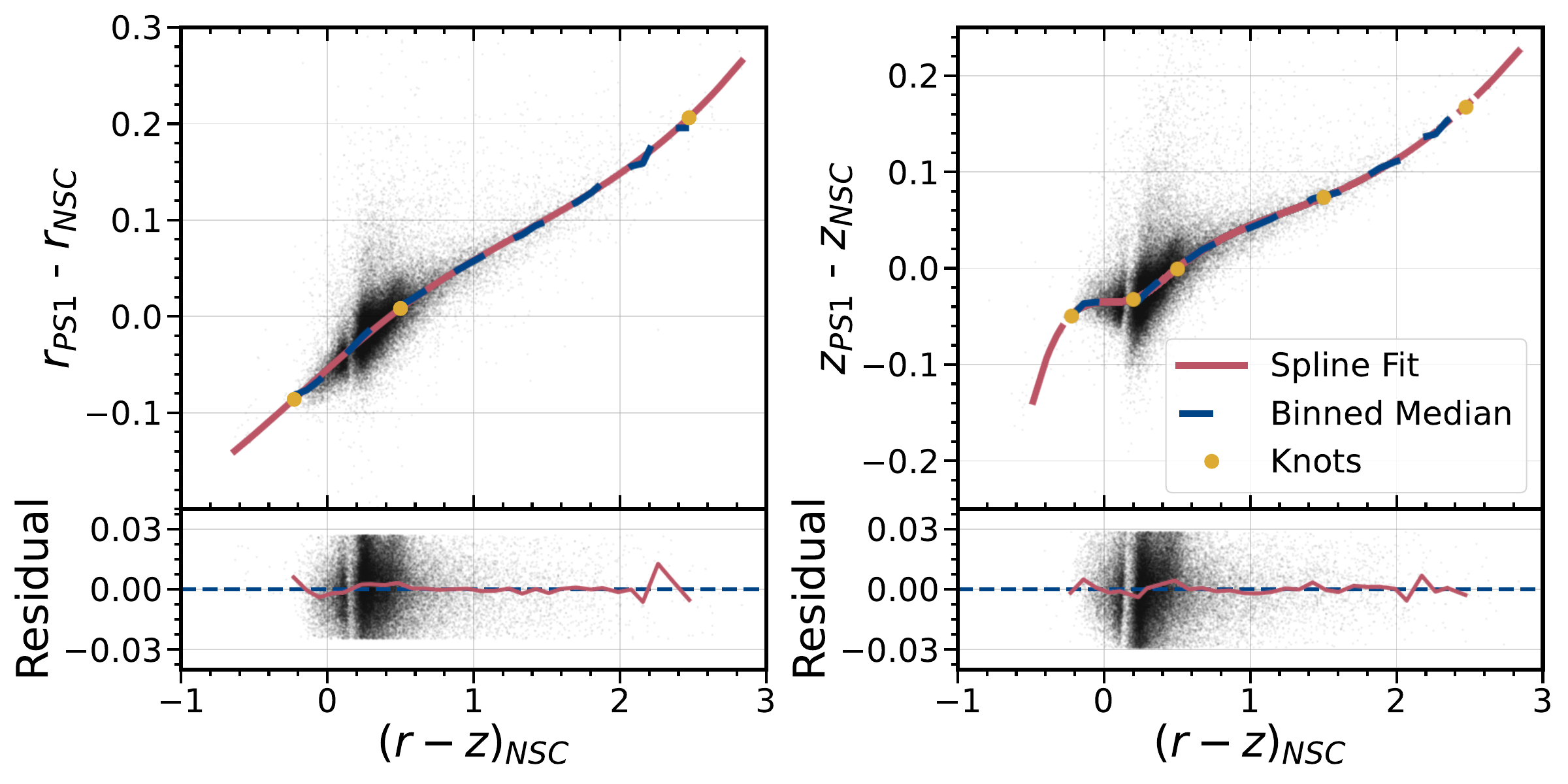}
            \includegraphics[width=\textwidth]{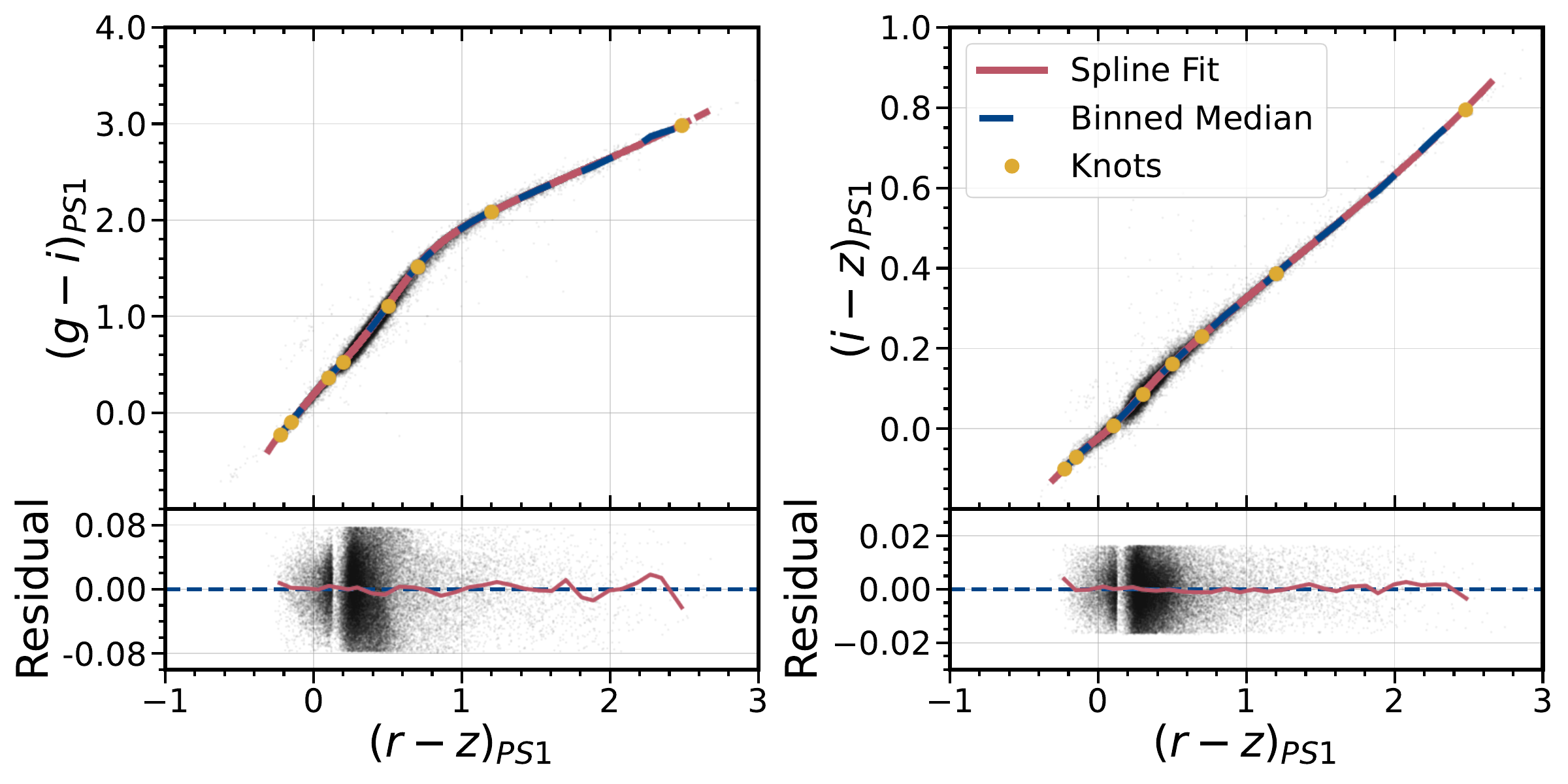}
            \caption{Transformation plots to transform our data from NSC magnitudes to PS1 magnitudes (top row), and between PS1 magnitudes (bottom row). The CMDs of the datasets used to calculate the transforms are shown in black points. The functions generated by the spline fits are shown with the solid red lines in the upper panels. The binned medians of the fit data are shown with the dashed blue lines.  The knots of the splines are indicated by the yellow circles. In the smaller panels, the residuals of sigma-clipped data are plotted as a function of magnitude with black points, their binned medians are plotted with solid red lines, and a residual of 0 is noted with dashed blue lines.}
            \label{fig:combotransform}
        \end{figure*}

        After estimating the transformation functions, we transform our dereddened, NSC-calibrated MISHAPS data to PanSTARRS using the corresponding NSC to PS1 functions.
        We use the resulting $r_{\rm PS1}$ and $z_{\rm PS1}$ values in the $(r-z)_{\rm PS1}$ to $(g-i)_{\rm PS1}$ and $(i-z)_{\rm PS1}$ functions to obtain our $(g-i)_{\rm PS1}$ and $(i-z)_{\rm PS1}$ colors.
        Finally, we add our $z_{\rm PS1}$ to $(i-z)_{\rm PS1}$ to obtain $i_{\rm PS1}$.
        The spline parameters for our fit functions are given in Table \ref{tab:splineparams}, and the functions and their residuals are plotted along with the stars used to determine them in Figure \ref{fig:combotransform}.
        All transforms have residuals better than $\pm0.03$ mag except the ($r-z$)$_{\rm PS1}$ to ($g-i$)$_{\rm PS1}$, which has residuals within $\pm0.08$ mag.
        However, even in the case of this transform, the median residual as a function of ($r-z$) remains close to 0.

        \begin{deluxetable*}{lccccc}[htb]
            \centering 
            \tablecaption{Photometric transform spline parameters}
            \tablehead{
            \colhead{Transform} & \colhead{Coeffs.} & \colhead{Knots} & \colhead{Roots} & \colhead{Median Residual (mmag)} & \colhead{Standard Dev. (mmag)}
            }
            \startdata
                $(r-z)_{NSC}$ to $r_{PS1}-r_{NSC}$ & [-0.086, -0.053, 0.072, & -0.23, 0.50, 2.5 & 0.43 & 1.4 & 12\\
& 0.11, 0.21] & & & & \\
 
$(r-z)_{NSC}$ to $z_{PS1}-z_{NSC}$ & [-0.049, -0.027, -0.048, & [-0.23, 0.20, 0.50,  & 0.50 & 0.19 & 13\\
&  0.037, 0.064, 0.12, & 1.5, 2.5] & & & \\
& 0.17] & & & & \\
 
$(r-z)_{PS1}$ to $(g-i)_{PS1}$ & [-0.22, -0.17, 9.4e-03, & [-0.23, -0.15, 0.10, & -0.10 & 0.28 & 35 \\
& 0.30, 0.63, 1.0, & 0.20, 0.50, 0.70, & & & \\
& 1.8, 2.3, 2.7, & 1.2, 2.5] & & & \\
& 3.0] & & & & \\
 
$(r-z)_{PS1}$ to $(i-z)_{PS1}$ & [-0.098, -0.088, -0.046, & [-0.23, -0.15, -0.10, & 0.077 & 0.037 & 7.4 \\
& -0.0053, 0.089, 0.17, & 0.30, 0.50, 0.70, & & & \\
& 0.26, 0.47, 0.64, & 1.2, 2.5] & & & \\
& 0.80] & & & & \\
            \enddata
            \tablecomments{Table of the coefficients, knots, roots, and residuals of each spline function used to transform our photometry from the DECam system to the PanSTARRS system, to be used with {\sc scipy.interpolate.LSQUnivariateSpline}.}
            \label{tab:splineparams}
        \end{deluxetable*}

        \subsubsection{Radius Estimation}\label{subsubsec:est}
        To estimate the radii of our stars, we use a color-diameter relation to estimate our stars' angular diameters \citep[e.g., ][]{boyajian2014stellardiams}:
        \begin{equation}\label{eqn:zang1}
            \log2\theta_{\rm \star} = \log\theta_{m=0} - 0.2m,
        \end{equation}
        where $m$ is the calibrated and dereddened magnitude in a given filter, $\theta_{m=0}$ is the angular diameter of a star with an apparent magnitude of 0 in the filter, and $\theta_{\rm \star}$ is the target's angular diameter.
        The radius can then be obtained using
        \begin{equation}\label{eqn:zang2}
            R_{\rm \star} = \frac{d}{2}10^{\log2\theta_{\rm \star}},
        \end{equation}
        where $d$ is the estimated distance to 47~Tuc \citep[4.45 kpc, ][]{chen2018gcdistances}.
        We use the relationship obtained in \citet{zang2018cfht}
        \begin{equation}\label{eqn:zang3}
            \log\theta_{i_{\rm PS1}=0} = 0.920 + 0.316((g-i)_{\rm PS1}-0.576)
        \end{equation}
        to determine $\log\theta_{m=0}$ .
        This relation has typical fractional uncertainties of 3.5-6.0\%.
        
\begin{figure}
            \centering
            \includegraphics[width=\linewidth]{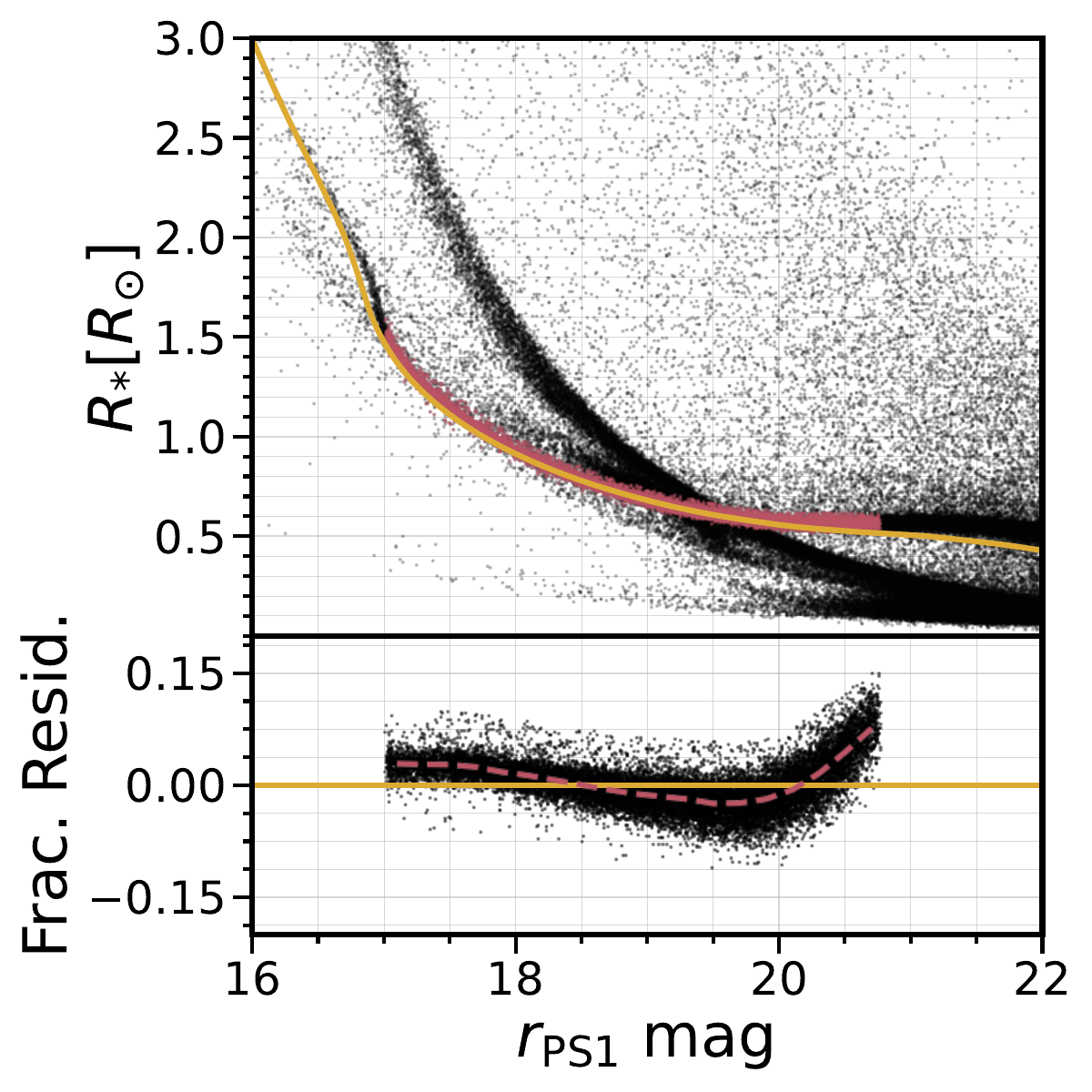}
            \caption{Radius estimation results. \textit{Top panel}: The resultant radius estimates of our full field vs. $r$ magnitude, with our final stellar selection highlighted in red. The radii for a MIST isochrone calculated for the PanSTARRS photometric system are given by the gold line. \textit{Bottom panel}: the residual between our stellar sample's radii and the isochrone's. The median residual is given by the red dashed line.}
            \label{fig:radii}
        \end{figure}

        We use the dereddened and transformed PanSTARRS magnitudes obtained in \S\ref{subsubsec:transforms} to estimate our target radii, with the results shown in Figure \ref{fig:radii}.
        To check the derived radii, we use a MIST isochrone \citep{dotter2016mist0, choi2016mist1} generated for the PanSTARRS photometric system with [Fe/H]=-0.78 \citep{forbes2010mwgcs}, convert the isochrone's absolute magnitudes to apparent magnitudes using the 4.45 kpc distance \citep{chen2018gcdistances}, and estimate the stellar radii using the luminosity and T$_{\rm eff}$.
        We compare the radii obtained for the isochrone stars to the radii obtained for the target stars as a function of $r$ to make sure our results are reasonable.
        Figure \ref{fig:radii} shows the radii obtained for the full field as well as the isochrone, with the radii of our final stellar selection highlighted in red.
        The residuals between the isochrone radii and our estimated radii are shown in the bottom panel.
        We show only the residuals for our 19,930-star selection.
        The radii estimated for the rest of the stars are less likely to be accurate, since determining their radii is dependent on constraining their distances (equation \ref{eqn:zang2}), and thus knowing which population they belong to.

\section{Timeseries Analysis}\label{sec:timeseries}
Our time series analysis of our data involves first extracting difference imaging photometry with {\sc ISIS} (\S\ref{subsec:dia}).
We calibrate the resultant photometry with our PSF photometry from the reference images, and remove systematic trends using the Trend-Filtering Algorithm (\S\ref{subsec:tfa}).

    \subsection{Difference Image Photometry}\label{subsec:dia}
    We use a modified version of the {\sc ISIS} package \citep{alardlupton1998isis, alard2000isis} which replaces the {\sc ISIS} registration task with one based on {\sc Source-Extractor} output \citep{bertin1996tractor, siverd2012kelt1b} to perform difference imaging and extract difference imaging analysis (DIA) photometry.
    The software aligns each image, then convolves the reference image with a PSF of the spatially-varying multi-Gaussian kernel fit to match the reference image to each target image.
    It then subtracts the convolved reference images from the corresponding target images.
    Finally, it produces a photometric table of each source it extracts from the image.

    To convert {\sc ISIS}' difference fluxes to magnitudes, we run the {\sc ISIS} photometry command on the reference image to determine the flux of the object in the reference image, then add this back to the difference flux. The magnitude in the target image $m_t$ is then
    \begin{equation}
    m_t = -m_0 - 2.5\log(F_{\rm ref}-\Delta F_t),
    \label{diffmag}
    \end{equation}
    where $F_{\rm ref}$ is the flux in the reference image, $\Delta F_t$ is the difference flux (computed by {\sc ISIS} as reference minus target) for the target image, and
    \begin{equation}
        m_0 = {\rm med}(-2.5 \log F_{\rm ref} - m_{\rm inst}) 
    \end{equation} is the median difference between the {\sc ISIS} instrumental magnitude of the reference stars $-2.5\log F_{\rm ref}$ and the {\sc DoPHOT} instrumental magnitudes $m_{\rm inst}$. 
    The median is computed using bright stars (PS1 calibrated magnitudes between $16.5$ and $18.5$ for each filter, and {\sc DoPHOT} objtype flag of 1) after 5 iterations of clipping of points further than 2 standard deviations from the median to minimize the impact of outliers due to blending.
    We optimized the configuration parameters for {\sc ISIS} by running it repeatedly on a subset of images of a field in the Galactic bulge while varying the values of one parameter at a time.
    We found the most crucial parameters to adjust to be {\sc rad\_phot} and {\sc rad\_aper}.
    These parameters represent, respectively, the radius around the target center in which the pixels will be used to estimate flux, and the radius used to normalize the flux.
    We found that {\sc rad\_phot}=5.0 px and {\sc rad\_aper}=6.0 px gave us the best results.
    Once all the relevant parameters were optimized, we ran the difference imaging process for each DECam chip.

    \subsection{De-Trending}\label{subsec:tfa}
    Once the difference image analysis (DIA) photometry has been extracted and calibrated, systematic variations are removed using the Trend-Filtering Algorithm (TFA) implemented through {\sc VARTOOLS} \citep{kovacs2005tfa}.
    This process is performed chip-by-chip, as each of DECam's CCDs has slightly different systematics.
    Usage of TFA involves selecting a subset of lightcurves with low variation --- referred to as ``trend stars'' ---which are used to determine the systematic trends in the data.
    To choose the size of this subset, we ran the TFA on lightcurves from the N10 chip multiple times, varying the number of trend stars used in each run.
    We then defined a ``quasi-reduced $\chi^{2}$'' metric
    \begin{equation}
        \text{quasi-}\chi_{\rm red}^{2} = \frac{1}{1.085^{2}} \frac{\sum_{i}^{N_{\rm data}} (RMS_{i}^{2}/\sigma_{i}^{2})}{N_{\rm data} - (N_{\rm trend}+1)}
    \end{equation}
    and used this metric to test the performance of TFA across our magnitude range.
    From the results of this test, we chose to use 15 trend stars, as it gave us the lowest quasi-reduced $\chi^{2}$ without risk of overcorrection.

\begin{figure}[htb]\label{fig:rmsdepth}
        \centering
        \includegraphics[width=\linewidth]{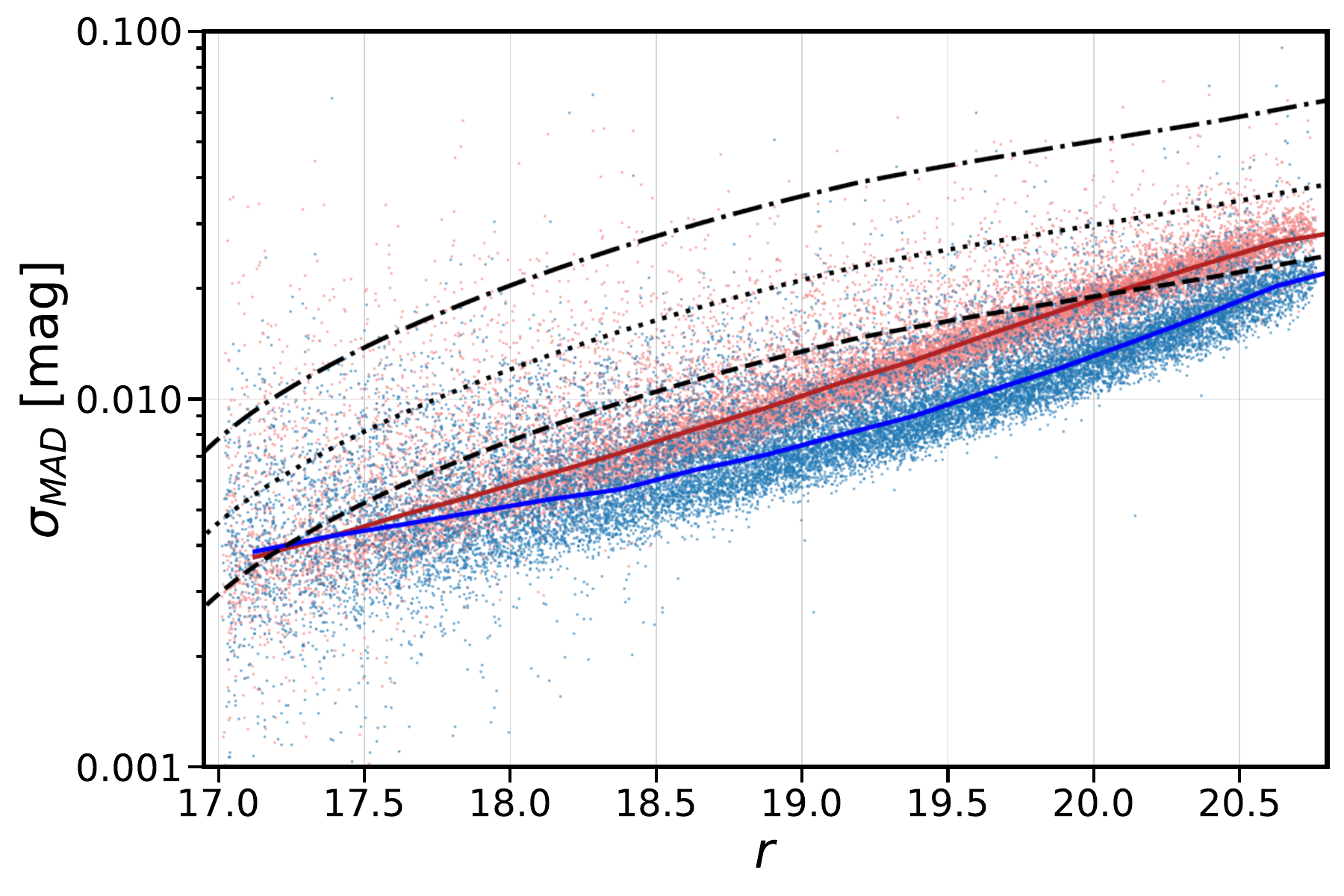}
        \caption{Scaled median absolute deviation (MAD, a robust estimate of RMS) of lightcurve photometry in our observations of 19,930 47~Tuc stars compared to the transit depths of planets orbiting 47~Tuc stars. Data from bad nights are removed prior to calculating this statistic. The black lines show predicted transit depths for $1.3R_{\rm Jup}$ (dash-dotted), $1R_{\rm Jup}$ (dotted), and $0.8R_{\rm Jup}$ (dashed) planets computed using a 47~Tuc compatible isochrone to estimate stellar radii. The points show the MAD scaled to approximate Gaussian $\sigma$ of $r-$band (blue) and $z-$band (red) observations of 47~Tuc. The solid lines show the median MAD versus $r-$magnitude for $r-$band (blue) and $z-$band (red) observations.}
    \end{figure}

    To select the trend stars, we again filtered our full list of stars for those having {\sc DoPHOT} type flags of 1 in both bands and bright magnitudes (within a range of 15.5-17.0 mag). 
    To cut for lightcurves with only low to moderate variation, we sigma-clipped based on
    \begin{equation}\label{eqn:tfaselect}
        \log(RMS) < \rho + 2*MAD(\sigma_{R}),
    \end{equation}
    where $\log(RMS)$ is the logarithm of each measurement's RMS, $\rho$ is the median of $\log(RMS)$, and MAD($\sigma_{R}$) is the median absolute deviation of the scatter scaled to equal $\sigma$ for Gaussian noise.
  From the stars that survive these cuts, we randomly select 15 to serve as our trend stars for TFA.
    To check the resultant photometric precision, we plot the MAD for each target as a function of $r-$magnitude and compare it to the estimated transit depths for planets of various radii (Figure \ref{fig:rmsdepth}).

\section{Transit Search, Efficiency, and Vetting}\label{sec:vetting}
We use the previously determined stellar characteristics to create and inject transit models into our lightcurves for use in our efficiency estimates (\S\ref{subsec:injection}).
After creating our injections, we ran a search algorithm on both the real and synthetic lightcurves simultaneously (\S\ref{subsec:search}), and flagged detections were subjected to a rigorous vetting process.
To estimate our detection efficiency without introducing bias and streamline our analysis, we split our candidate vetting into a two-step process.
First, both real and injected lightcurves flagged by the search were blinded, mixed together, and inspected by-eye (\S\ref{subsec:zooniverse}).
The results of this first round were used to estimate our efficiency in the case of the synthetic lightcurves (\S\ref{subsec:deteff}), and to narrow down the number of candidates that needed to be investigated further in the case of the real lightcurves.
The human vetting left 39 candidates for analysis, which were put through a round of detailed analysis involving checking the targets' locations on the CMD, period searches, comparison of their lightcurves with those of nearby stars, and in-transit difference image stacking (\S\ref{sec:detailed}).
This second round of vetting left 18 objects of interest, which were analyzed more closely.

    \subsection{Transit Injection}\label{subsec:injection}
    To fully characterize the efficiency of our search and calculate an occurrence rate, we need to inject a range of synthetic transits into our lightcurves and analyze how well our search recovers them.
    Using a sample of 19,930 cluster stars with radius estimates, we use the {\sc JKTEBOP} code \citep{southworth2013jktebop1} to calculate transit models and inject them into our data. 

    To obtain the limb-darkening parameters for our transit models, we use a MIST isochrone \citep{dotter2016mist0, choi2016mist1} generated for the DECam photometric system with [Fe/H]=-0.78 dex \citep{forbes2010mwgcs}.
    The isochrone's absolute magnitudes were converted to apparent magnitudes using the distance modulus, with a distance of 4.45 kpc \citep{forbes2010mwgcs}.
    We then matched our stars to the isochrone ``stars" by selecting the isochrone star with the nearest magnitude.
    These results allowed us to estimate \teff and \logg for each of our stars from the isochrone matches, which we then ran through {\sc JKTLD} \citep{southworth2008jktld} to estimate quadratic limb-darkening coefficients in $r$.
    Because {\sc JKTLD} only accepts [Fe/H] values in multiples of 0.5 dex, we calculated limb-darkening models for both [Fe/H]=-1.0 and [Fe/H]=-0.5 dex and interpolated between the two for our desired [Fe/H]=-0.78 dex.
    We note that, in error, we used the same limb-darkening coefficients for $z$ as $r$, which could in principle affect the preliminary vetting, though we expect the effects to be small.
    This error will be addressed in paper 2.

    Orbital periods, $P$, between 0.5--10.0 days and planet radii, $R_{\rm p}$, between 0.5--2.0 $R_{\rm jup}$ were drawn from uniform distributions for each system. 
    These limits are selected to encompass the ranges considered by \citetalias{gilliland2000hstsearch} and \citetalias{masuda2017tucreassessment}, as well as the ranges considered for \textit{Kepler} stars in \citet{fressin2013keplerocc}, allowing us to more carefully compare our occurrence rate results with those of the Milky Way disk.
    Inclinations were randomly selected from the range of $\cos i$ angles which would allow a full or grazing transit.
    The maximum possible $\cos i$ value is also the transit probability for an isotropic distribution of circular orbits,
    
    \begin{equation}\label{eqn:ptr}
       P_{\rm tr} = \cos i_{\rm max} = \frac{R_{\rm \star}+R_{\rm p}}{a}, 
    \end{equation}
    where $a$ is the semimajor axis of the planet's orbit and $R_{\rm \star}$ is the host radius.
    As we draw periods rather than semimajor axes, we compute $a$ given $P$ using Kepler's third law by interpolating the same [Fe/H]=-0.78 dex DECam isochrone to estimate a stellar mass.
    
\begin{figure*}[htb]
        \centering
        \includegraphics[width=\linewidth]{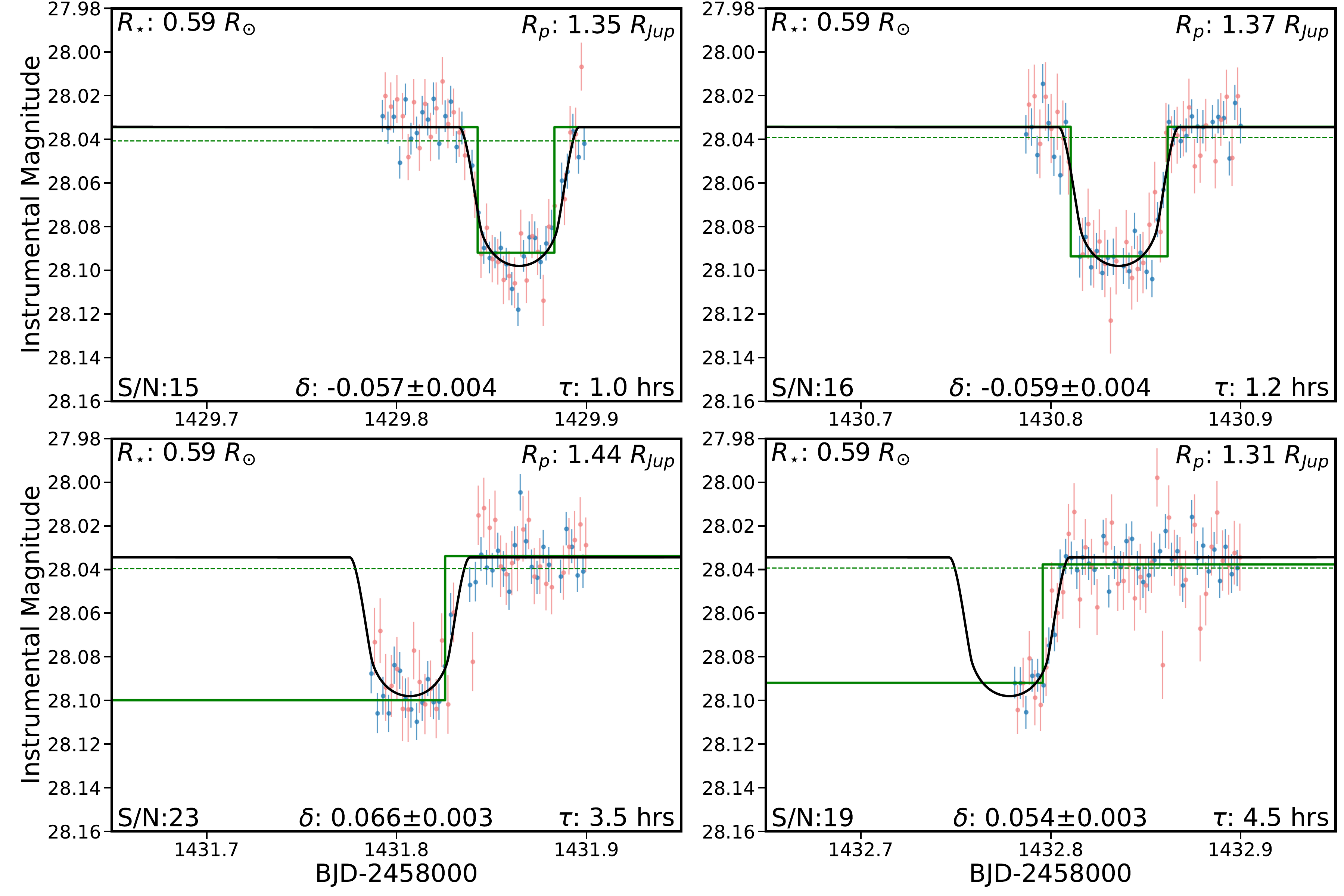}
        \caption{Example of injected full and partial transits of a $1.26 R_{\rm Jup}$ planet orbiting a $0.59 R_{\rm \odot}$, $r=19.9$ star in our data with $P$=0.97 days. Each panel shows a different night, with the top two panels showing nights flagged as having full transits, and the bottom two showing nights flagged as partial. The $r$ band data are shown in  blue, and the $z$ is shown in red. The black line is the injected model, and the green line is the best boxcar model. We note that the top left transit is actually a partial transit since we do not capture the full egress in the data, but the algorithm flags it as a full transit based on its best S/N, $t_{c}$, and $\tau$. Our detection algorithm was able to successfully recover both the full and partial transits with S/N $>$7.}
        \label{fig:injectionlc}
    \end{figure*}

    We injected the resulting {\sc JKTEBOP} models into our de-trended lightcurves.\footnote{Future iterations will inject the transits into the original lightcurves, and the injections will be de-trended along with the real lightcurves.}
    A limited sample ($\sim$2,000 injections) was run and the lightcurves were plotted with parameters and ID number visible to determine whether we were generating models with realistic radii and with depths, limb-darkening effects, and transit times that matched what is expected for systems with similar characteristics.
    These injections were not used in our detection efficiency determination, but the original lightcurves were recycled to create new injections once we were confident in our models.
    We created $\sim$40,000 injection lightcurves from $\sim$20,000 unique lightcurves, by injecting a transiting planet into each original lightcurve twice.
    We chose the total number of injections to achieve binomial uncertainties on the detection efficiency $\sim$5\% in bins of $P$ and $R_{\rm p}$.
    A single lightcurve example with multiple full and partial transits is shown in Figure \ref{fig:injectionlc}.
    Grids of transit injections with varying $P$, $R_{\rm p}$, and $r-$magnitude can be found in Figures \ref{fig:rppgrid}-\ref{fig:rprpartgrid} of Appendix \ref{app:grids}.

    \subsection{Transit Search}\label{subsec:search}
    Our photometric precision is sufficient for detecting single transits of hot Jupiters, so we designed a ``sliding boxcar" algorithm that searches for these without phase-folding our lightcurves.
    The algorithm operates on the lightcurves of individual nights.
    Each night is searched over a grid of potential transit center times, $t_{\rm c}$, and transit durations, $\tau$.
    The boxcar transit model is defined as
    \begin{equation}
    m(t) = \left\{\begin{array}{r l}
    m_{\rm out} \quad {\rm if}&  t < t_{\rm c} - 0.5\tau\\
    m_{\rm out} + \delta \quad {\rm if}& t_{\rm c} - 0.5\tau \leq t \leq t_{\rm c} + 0.5\tau\\
    m_{\rm out} \quad {\rm if}& t_{\rm c} + 0.5\tau < t\\
    \end{array},
    \right.
    \label{eqn:boxcar}
    \end{equation}
    
    \noindent where $t$ is the time, $m(t)$ is the magnitude at time $t$, $m_{\rm out}$ is the out-of-transit magnitude, and $\delta$ is the transit depth (with transits having negative depths). 
    We opt to estimate the out of transit magnitude $m_{\rm out}$ and in-transit magnitude $m_{\rm in}=m_{\rm out}+\delta$ using a median operation to be robust to outliers. 
    This leaves no free parameters aside from the two grid search parameters $t_c$ and $\tau$. 
    To robustly estimate the signal-to-noise of the boxcar transit at each gridpoint, we estimate the in- and out-of-transit RMS --- $\sigma_{\rm in}$ and $\sigma_{\rm out}$, respectively --- using the median absolute deviation scaled by $1.483$ to match $\sigma$ for a Gaussian, then define the transit signal-to-noise ratio as
    \begin{equation}\label{eqn:snr}
        \frac{S}{N} = \frac{\delta}{\sqrt{\sigma_{\rm in}^{2}/n_{\rm in} + \sigma_{\rm out}^{2}/n_{\rm out}}}.
    \end{equation}
    
    The $t_{\rm c}$, $\tau$, $\delta$, $m_{\rm in}$, $\sigma_{\rm in}$, $m_{\rm out}$, and $\sigma_{\rm out}$ of the model that produces the highest S/N are recorded for each night, for each target with at least 6 points in- and out-of-transit, as defined by the model's $t_{\rm c}$ and $\tau$.
    An example is shown in Figure \ref{fig:injectionlc}. 
    Bad nights are searched for transits in case they can be used to confirm or reject candidates, but they are not included when determining whether a lightcurve contains a detection, i.e. a lightcurve with only one detection above our S/N threshold that occurred during a bad night is counted as a non-detection.

\begin{figure}[htb]
        \centering
        \includegraphics[width=\linewidth]{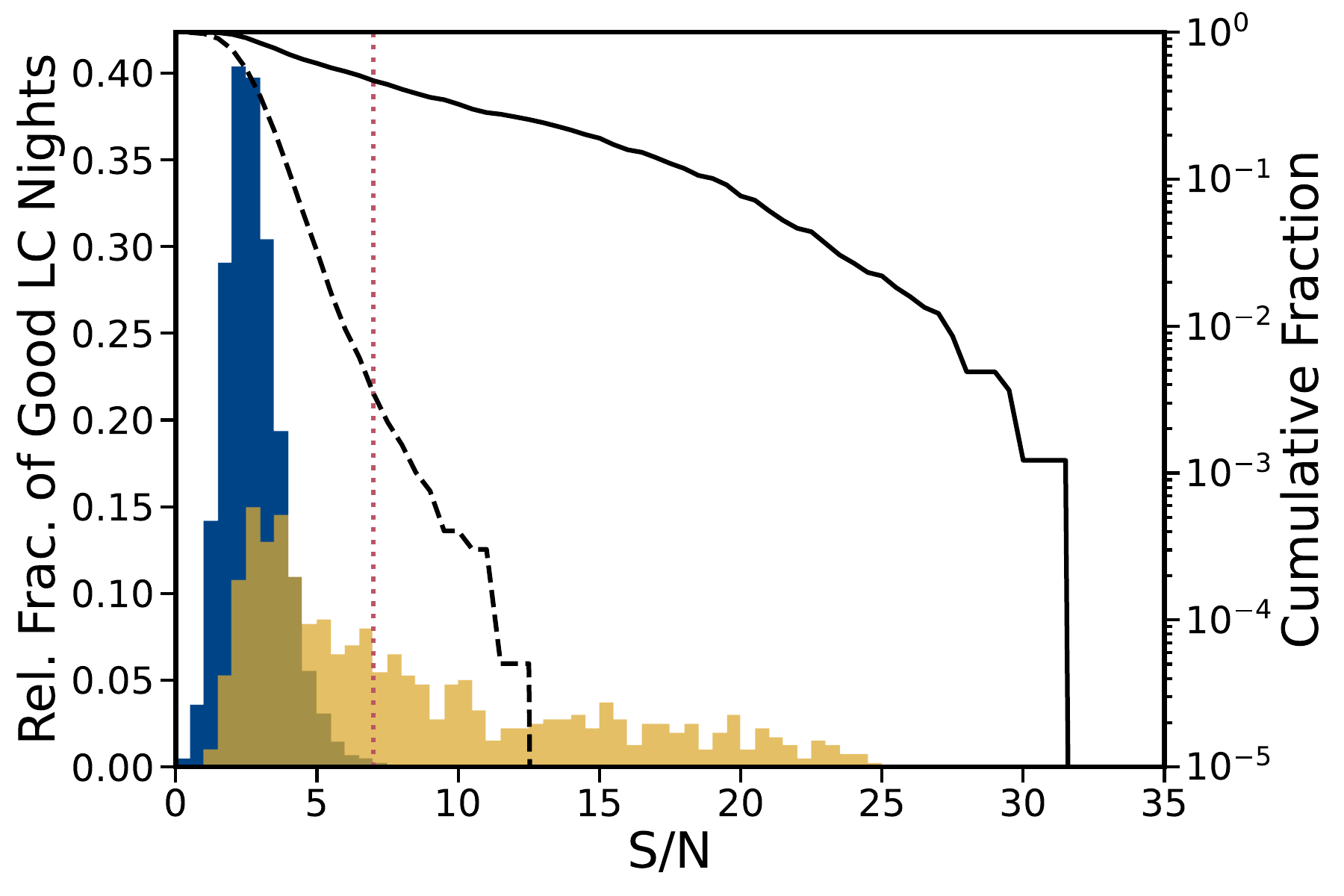}
        \caption{Histogram showing the S/N of nights with (yellow) and without (blue) injected transits. The dotted red line indicates our chosen S/N$\geq7$ threshold. The solid and dashed black lines represent the probability of a lightcurve containing a detection of a given S/N for nights with and without injected transits, respectively.} 
        \label{fig:histall}
    \end{figure}

    To set the detection threshold, we run the search algorithm on the same sample of $\sim$2,000 injections used to test the models.
    We plot the differential and cumulative distributions of S/N values of the best boxcar transit model for nights that do contain injected transits, and for nights that do not, in Figure \ref{fig:histall}.
    We then find the S/N threshold that allows us to detect a reasonable number of transits, while excluding as many false positives as possible, opting ultimately for $S/N\geq7$.
    This choice passes $\sim$47\% of injected transits, while rejecting $\sim$99\% of nights without a transit injection.

    \subsection{Preliminary Zooniverse Vetting}\label{subsec:zooniverse}
\begin{figure*}
        \centering
        \includegraphics[width=\linewidth]{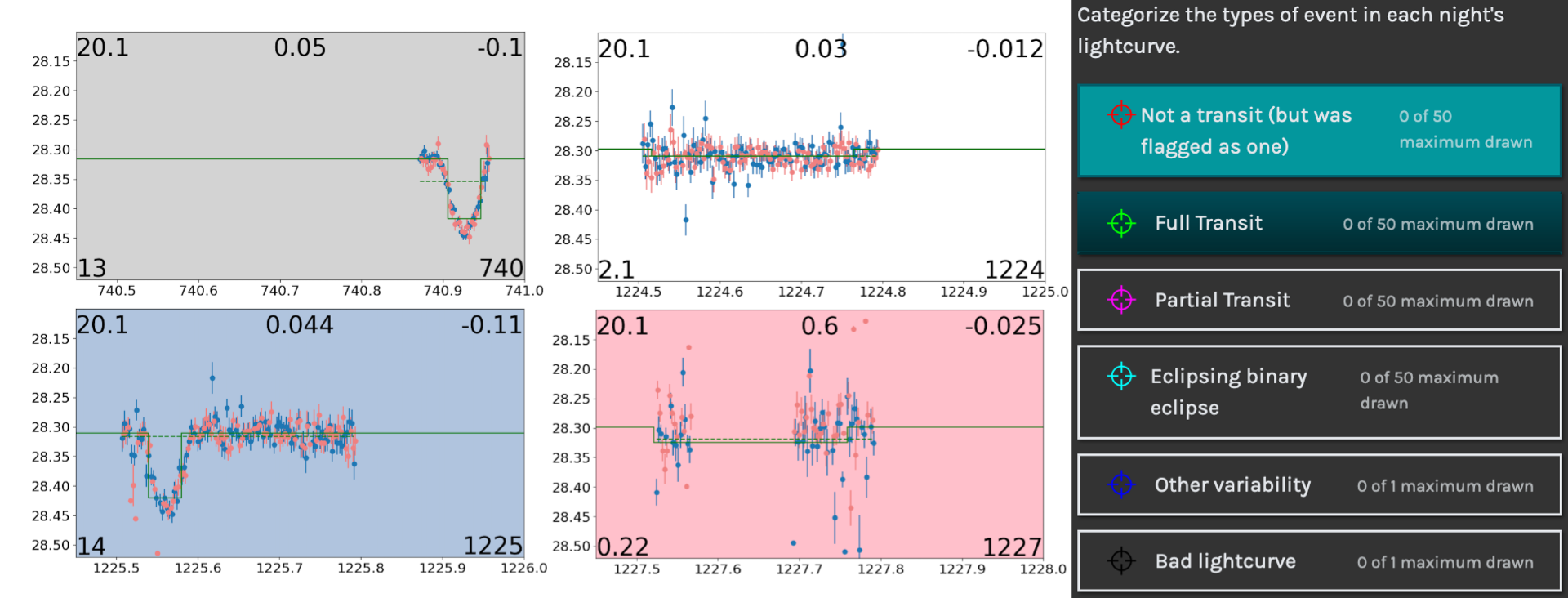}
        \caption{Simplified version of the MISHAPS lightcurve classification workflow. The rightmost panel shows the tools users are given, while the rest show individual nights of the lightcurve. A blue panel indicates a full transit as flagged by the search algorithm, a gray panel indicates a partial transit, and a pink panel indicates a bad night (see \S\ref{subsec:obs} for discussion on how bad nights were selected). In the full Zooniverse workflow, users have the same tools, but are presented with all 24 nights of the lightcurves.}
        \label{fig:zooworkflow}
    \end{figure*}

    Targets with at least one S/N trigger on a good night are passed to preliminary vetting on the Zooniverse platform.
    This vetting step is performed by uploading lightcurve plots of both injected and real candidates into a Zooniverse workflow.
    The lightcurves have all identifying information removed, including what chip of the DECam CCD they lie on, their MISHAPS pipeline ID numbers, and whether they are real or injected lightcurves.
    Lightcurves are inspected and classified in the workflow, event by event.
    A simplified version of the Zooniverse workflow is shown in Figure \ref{fig:zooworkflow}.

    The workflow presents the user with a complete lightcurve, split into subplots of each survey night so the features are easier to analyze.
    Nights we have deemed to have bad data quality are highlighted in pink, while detections the algorithm considers full transits are highlighted in blue and detections it considers partial transits are highlighted in gray.
    The plots show the de-trended instrumental $r$ and $z$ lightcurves aligned to a common baseline on each night, and the boxcar model with the highest $S/N$.
    Each panel also provides the median instrumental magnitude, the $\sigma_{\rm MAD}$ of the night's data, and the best $S/N$ of the event calculated by the search algorithm.
    Users place a mark on the nights in which they think a real transit event may have occurred, with the available marks being spurious detection, full transit, partial transit, EB, variable star, and bad-quality lightcurve.
    They are encouraged to classify features in nights that were not flagged by the search, in case there are additional transit-like events that are recognizable in bad nights or below our $S/N$ threshold.
    Two members of our team inspected all $\sim$11,000 Zooniverse lightcurves --- which contained $\sim$1,500 lightcurves without injections --- and 39 targets had at least one flagged night where both agreed it was a possible transit.
    We performed more detailed vetting on each of these (\S\ref{sec:detailed}), though the results of that detailed vetting were not incorporated into the analysis of our recovery rate (\S\ref{subsec:deteff}).
    This step was performed prior to the color cut so EB candidates could be retained.

    \subsection{Detection Efficiency}\label{subsec:deteff}
    To trace the relative importance of each step in our two-step search and vetting process on the recovery rate calculation, we define two efficiencies: the detection efficiency of our search algorithm, $\epsilon_{\rm det}$, and the detection efficiency of our Zooniverse vetting, $\epsilon_{\rm Zoo}$. 
    
    $\epsilon_{\rm det}$ is the fraction of injected transits that are successfully detected by our search algorithm, computed as
    \begin{equation}
        \epsilon_{\rm det} = \frac{\sum_j^{N_{\rm inj}} [S/N_{{\rm max},j} \ge 7] [S_j=1]}{\sum_j^{N_{\rm inj}} [S_j=1]},
    \end{equation}
    where $N_{\rm inj}$ is the number of injected lightcurves, square brackets are the Iverson bracket taking the value 1 if the contained condition is true and 0 if false, and $S/N_{{\rm max},j}$ is the largest nightly transit search signal-to-noise from a good night for the $j^{\rm th}$ lightcurve. 
    We have also included a selection function Iverson bracket to allow binning or the computation of detection efficiencies across different ranges of parameters, where the selection is defined as
    \begin{align}
        S_j = {} & [R_{\rm p,min}\le R_{{\rm p},j} < R_{\rm p,max}] \times \\ & [P_{\rm min}\le P_j < P_{\rm max}]  [r_{\rm min}\le r_j < r_{\rm max}],
    \end{align}
    with $R_{\rm p,min}$, $R_{\rm p,max}$, and $R_{{\rm p},j}$ the minimum, maximum, and $j^{\rm th}$ planet radius, respectively, and similarly for the period $P$ and calibrated $r$ magnitude.
    
    $\epsilon_{\rm Zoo}$ is the fraction of detected injections that are confirmed by human checks to be possible transits. 
    To compute it, we first define $C_j$ as the count of good nights with a transit that passes the signal-to-noise threshold and which have a Zooniverse categorization of either a partial or full transit unanimously for all users
    \begin{equation}
    C_j = \sum_k^{N_{\rm nt}}[S/N_k>7][N_{\rm pf}=N_{\rm user}],
    \end{equation}
    where $N_{nt}$ is the number of good nights, $S/N_k$ is the transit signal-to-noise of the $k^{\rm th}$ night, $N_{pf}$ is the number of partial or full transit classifications, and $N_{\rm user}$=2 is the number of classifying Zooniverse users. 
    This allows us to write
    \begin{equation}
    \epsilon_{\rm Zoo} = \frac{\sum_j^{N_{\rm inj}}[C_j\ge 1][S_j=1]}{\sum_j^{N_{\rm inj}} [S/N_{{\rm max},j}>7][S_j=1]}.
    \end{equation}
    
    The total detection efficiency must also include the geometric transit probability to account for the fact that only a small range of orbits will cross the face of the star from our point of view. 
    This depends on the stellar radius, planet radius and period, so to accurately weight the contribution of different stars to the average detection efficiency, we compute the total detection efficiency as 
    \begin{equation}
    \epsilon_{\rm total} = \frac{\sum_j^{N_{\rm inj}} P_{{\rm tr},j} [C_j\ge 1] [S_j=1]}{\sum_j^{N_{\rm inj}} [S_j=1]},
    \end{equation}
    where $P_{{\rm tr},j}$ is the transit probability of the $j^{\rm th}$ injection (computed using equation~\ref{eqn:ptr}). 
    As we have injected transiting planets into each lightcurve twice, the above evaluation appropriately weights the detection efficiency across the properties of our stellar sample. 
    The weighting would also be appropriate if we had chosen to select stars for injection randomly with replacement.
 
\begin{figure*}[htb]
        \centering
        \includegraphics[width=\textwidth]{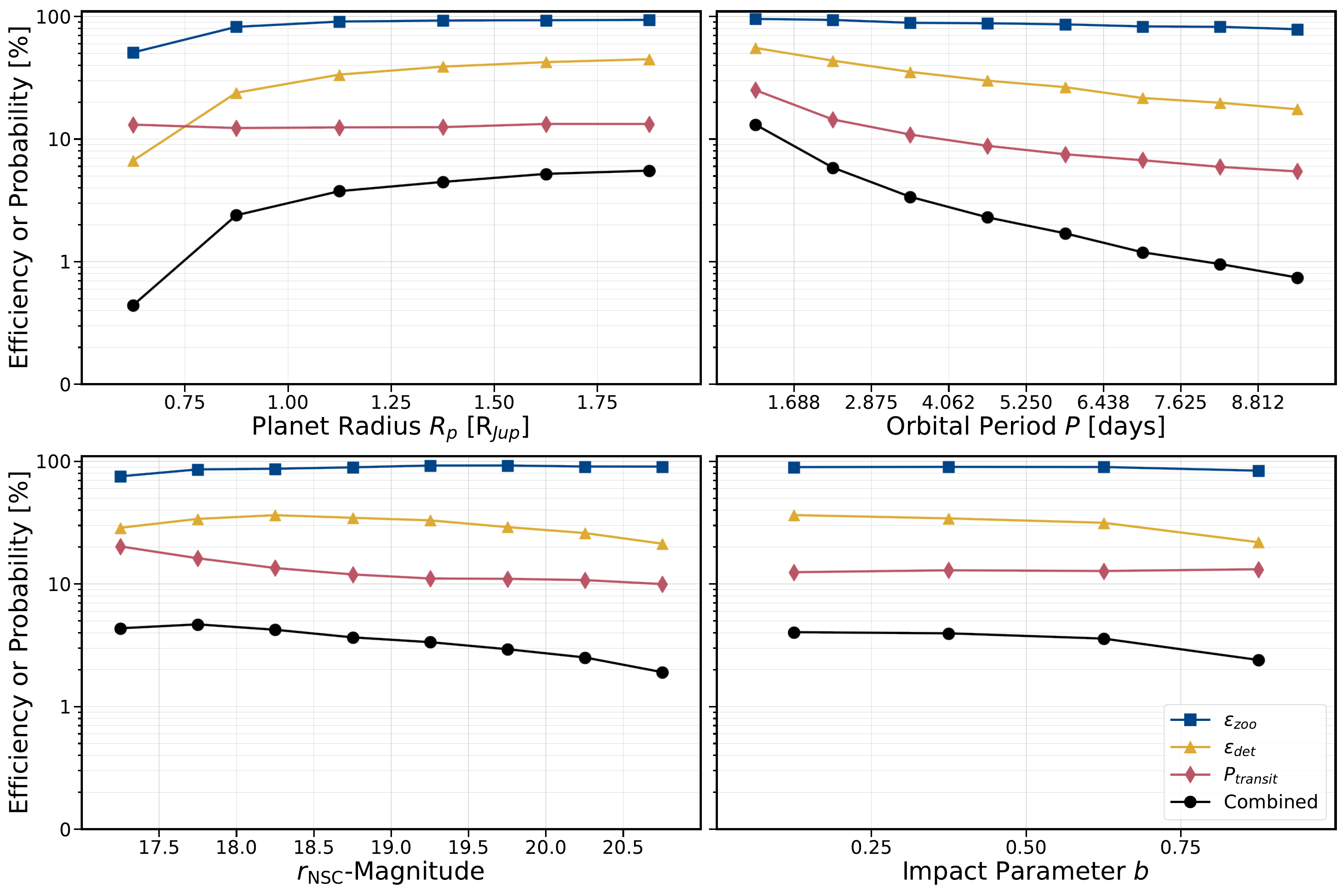}
        \caption{Efficiencies of our transit search and Zooniverse vetting, binned as a function of planet radius $R_{\rm p}$ (top left), orbital period $P$ (top right), $r-$magnitude (bottom left), and impact parameter $b$ (bottom right). The blue-triangle lines show our detection efficiency $\epsilon_{\rm det}$, the red-square lines show our Zooniverse vetting efficiency $\epsilon_{\rm Zoo}$, the green-diamond lines show our transit probability $P_{\rm transit}$, and the black-circle lines show the combined efficiencies.}
        \label{fig:1d_eff}
    \end{figure*}

\begin{figure}[htb]
        \centering
        \includegraphics[width=\linewidth]{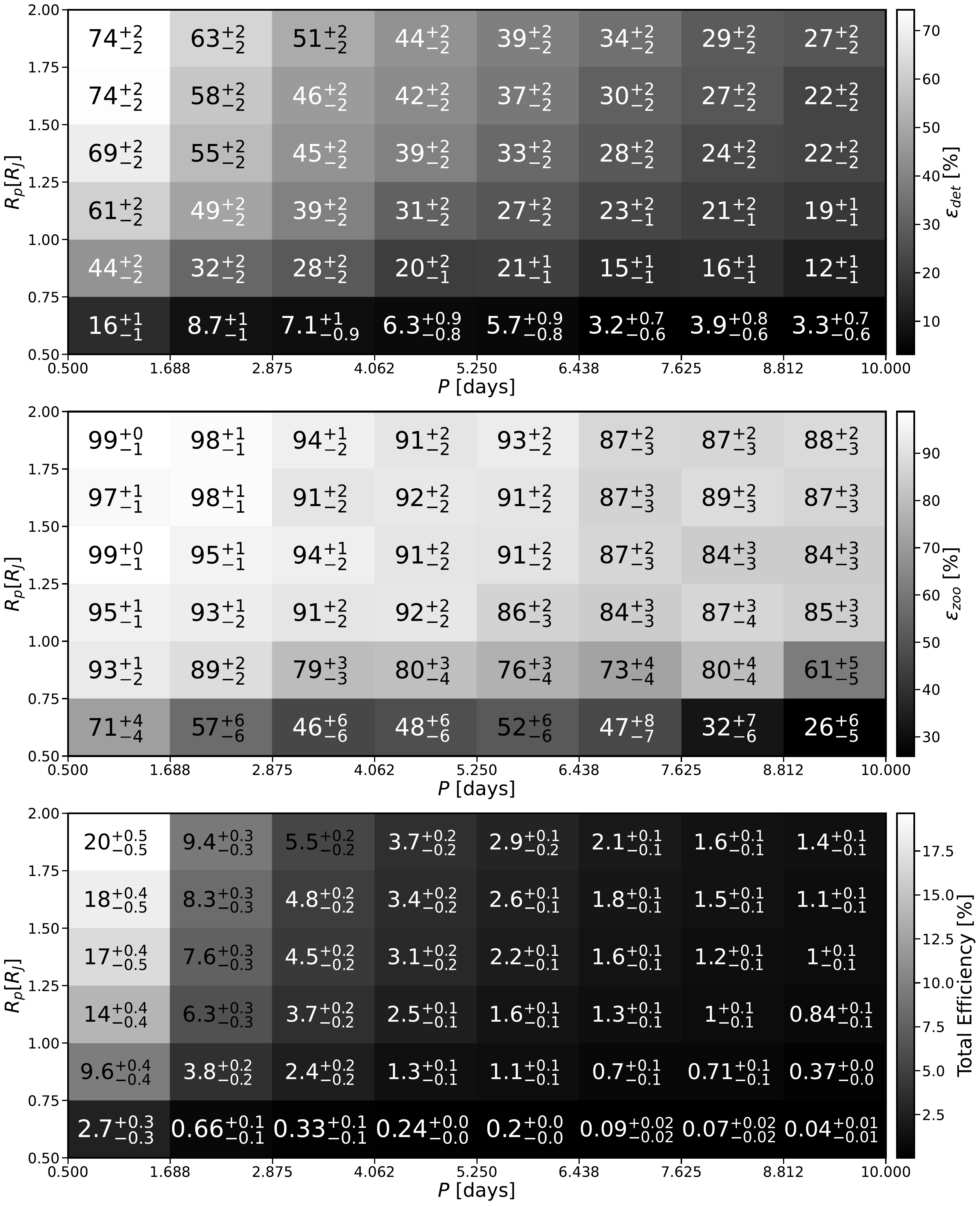}
        \caption{2D plots of detection efficiencies of our pipeline, estimated in bins of $R_{\rm p}$ and $P$. The top panel shows $\epsilon_{\rm det}$ for the search algorithm, and middle panel shows $\epsilon_{\rm Zoo}$ for the Zooniverse vetting, and the bottom panel shows the overall efficiency incorporating the transit probability (e.g., $P_{\rm tr}\epsilon_{\rm det}\epsilon_{\rm Zoo}$). The bins are color-coded by efficiency, and the numbers labeling each bin give the respective detection efficiency and $1-\sigma$ errors in percent.}
        \label{fig:2d_eff_Rp_P}
    \end{figure}

\begin{figure}[htb]
        \centering
        \includegraphics[width=\linewidth]{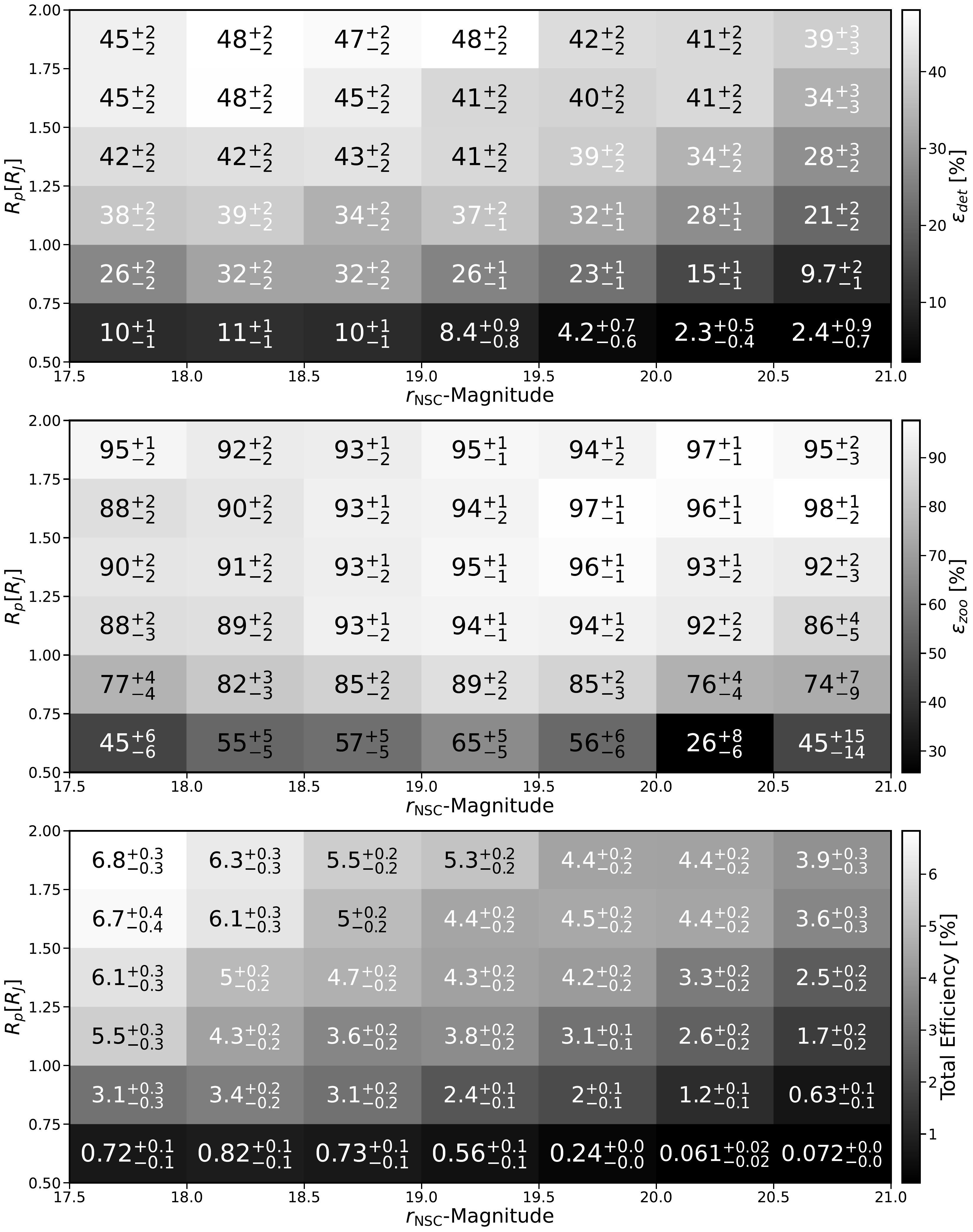}
        \caption{Same as Figure \ref{fig:2d_eff_Rp_P}, but plotting detection efficiencies in bins of $R_{\rm p}$ and $r-$magnitude.}
        \label{fig:2d_eff_Rp_rmag}
    \end{figure}

    We compute these efficiencies in bins of $R_{\rm p}$, $P$, and $r-$magnitude after applying the color cut, using parameter ranges that encompass those considered by \citetalias{gilliland2000hstsearch}, \citetalias{weldrake2005tuc}, \citetalias{fressin2013keplerocc}, and \citetalias{masuda2017tucreassessment} ($0.5 R_{\rm Jup} \leq R_{\rm p} \leq 2.0 R_{\rm Jup}$, 0.5 days $\leq P_{\rm orb} \leq$ 10.0 days).
    Figure \ref{fig:1d_eff} shows the efficiencies as a function of $R_{\rm p}$, $P$, $r-$magnitude, and impact parameter $b$, averaged over the other parameters.
    Figures \ref{fig:2d_eff_Rp_P} and \ref{fig:2d_eff_Rp_rmag} show combined efficiencies over 2D grids of $R_{\rm p}$ vs. $P$ and $R_{\rm p}$ vs. $r-$magnitude for $\epsilon_{\rm det}$ and $\epsilon_{\rm Zoo}$.
    For the 2D efficiency plots, errors were computed using binomial confidence intervals at a confidence level of 68\%, or 1-$\sigma$.

    We notice a significant decrease in $\epsilon_{\rm det}$ for smaller $R_{\rm p}$ and longer $P$.
    Additionally, though our $\epsilon_{\rm Zoo}$ shows no large trends in $r-$magnitude, it does show a steady decrease over our range of $P$ and a large increase over $R_{\rm p}$ = 0.5-1.0 $\rjup$.
    However, this behavior is expected, since smaller planets will have transits with lower S/N and shallower depths, and longer-period planets are less likely to have a transit fall in our observing windows.

\section{Detailed Candidate Vetting}
\label{sec:detailed}
A major potential challenge for MISHAPS is the rejection of false positives caused by systematic, time-correlated variations in the lightcurves, either astrophysical, instrumental, or otherwise.
In some instances, these false positives are simply caused by contamination from nearby bright stars, and can be easily eliminated.
Others are caused by stellar blending in our difference images, which may either be astrophysical or due to problems with our difference imaging.
EB false positives can appear in three main forms: M-dwarf secondaries which cause eclipses of similar depths to hot Jupiter transits; EBs blended with target stars, causing diluted eclipses of similar depths to hot Jupiter transits; and hierarchical triple systems with EB components having similarly diluted eclipses. 
We therefore designed our detailed vetting process to rule out these false positives.

The detailed vetting was completed in two passes. In the first pass we:
\begin{enumerate}
    \item Stacked the in-transit difference images for each night with a detection to see if the in-transit light lost, averaged over a transit, was displaced from the target on the reference image, indicating the source of the signal was a blended, likely unrelated star. 
    Difference images to be stacked were chosen to be those that produce in-transit datapoints in the  boxcar model for full or partial events. 
    The stacked difference images are compared to the reference image, and are prepared for both $r$ and $z$ filters.
    \item Selected all stars within 3 aperture radii of the target and plotted their lightcurves to see if they display similar trends, potentially indicating contamination from either that star or a nearby very bright star.
    \item Used {\sc VARTOOLS}' BLS, AOV, and Lomb-Scargle period searches \citep{kovacs2002bls, schwarzenberg-czerny1989aov, devor2005aov, press1992nr, zechmeister2009ls} to obtain preliminary estimates of the target's period, then folded the lightcurves on those periods to see if a strong transit, EB, or other variable signal emerged.
\end{enumerate}

\begin{figure}[htb!]
    \centering
    \includegraphics[width=\linewidth]{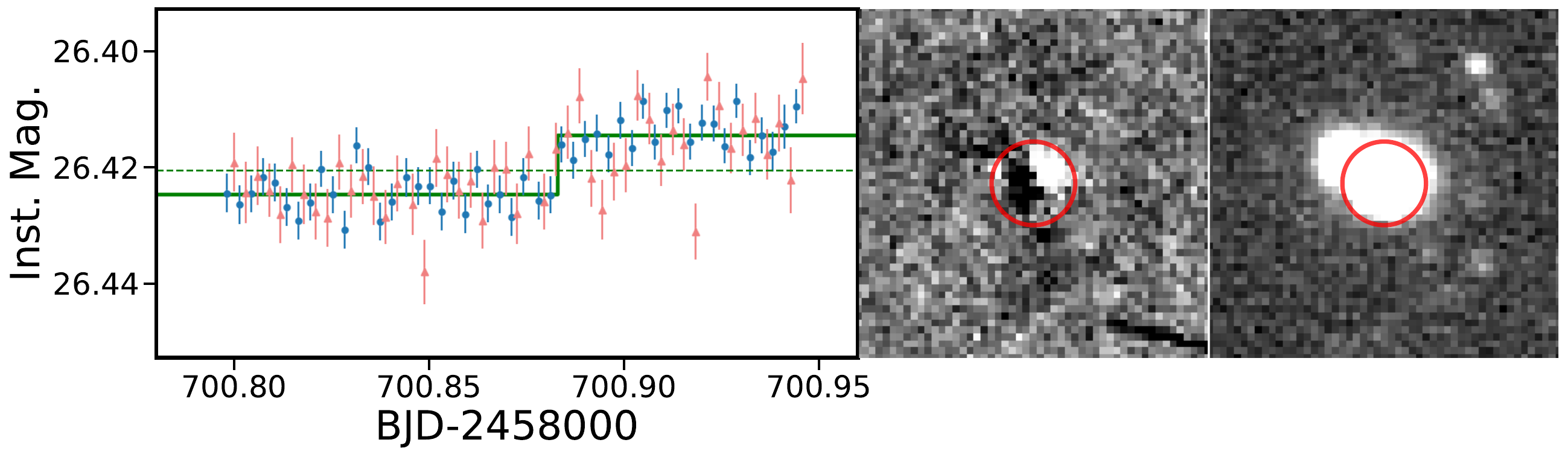}
    \caption{Example of a candidate rejected for a clear blend. The lightcurve of the detection is plotted on the left with the search's model. The stacked in-transit difference image and reference image are shown in the middle and right panels, respectively. Though the lightcurves seem to indicate a clear longer duration eclipse, inspection of the images leads us to reject this candidate.}
    \label{fig:blend}
\end{figure}
    
Of the 39 candidates that made it through the Zooniverse vetting prior to the color cut, only 18 survived this round of analysis. 
The majority of those rejected --- 14 out of 21 --- were found to have messy difference images near the target, often with a dipole feature, with many of the dipoles belonging to a very close star within ${\sim}1$~PSF FWHM of the target (i.e., blended in the reference image). 
An example of a dipole is shown in Figure~\ref{fig:blend}. 
The dipoles are caused by a quirk of the growth of DECam's CCD chips. 
Image artefacts referred to as ``tree rings'' --- visible in flat fields and bright sky backgrounds --- are caused by photoelectric imperfections in the DECam CCDs, and result in the effective area of each pixel varying \citep{plazas2014decamccds, bernstein2017decamastrom, bernstein2018decamphotom}.
When the science images are flat-fielded, the tree rings introduce small astrometric shifts in stars positioned on the rings~\citep{baumer2017flatfielding}. 
The image alignment step of our difference imaging only accounts for large-scale astrometric distortions, and thus cannot account for these localized shifts on the scale of the entire chip. 
This shift results in dipoles in our difference imaging because the star may be positioned differently in the reference image than in the individual images relative to an assumed regular pixel grid. 
Visually, these appear as a positive and negative area in the image where the misaligned star is over-subtracted on one side and under-subtracted on the other. 
As the aperture is centered on the reference image star, a small shift in the target image can result in some of the positive dipole landing outside the aperture and more of the negative dipole landing inside, which --- if this happens in the same sense for a sequence of images --- can result in a transit-like systematic in the lightcurve. 
The reverse case causing an apparent brightening can also look like a partial transit. 
The chance of such systematics is greatly increased when there is a nearby neighbor with a dipole that might only have the positive or negative side landing in the aperture. 
We did not find any blended eclipsing binaries through the stacked difference images.

Another 5 candidates were rejected because they shared similar lightcurve features with nearby stars, indicating either systematic errors or contamination from bright stars or diffraction spikes.
The remaining 2 were determined to be inappropriately classified in Zooniverse, as they showed no clear indications of eclipse-like signals in their lightcurves or difference images.
Given the number of lightcurves inspected in Zooniverse (${\sim}11,000$), this appears to be a reasonable simultaneous error rate for two users.

We did not find any significant periodic transiting or eclipse signals in any candidates using the period finding tools, though this is likely because we included data from bad nights in the period search on the first pass. 
On the second pass, we reran the periodogram search only with data from good nights.

To confirm or reject the remaining 18 candidates, we performed a second round of detailed analysis. 
For these candidates, we produced a second set of difference imaging photometry, this time using smaller 500x500 pixel image cutouts centered on each target.\footnote{Unless the target is within 500 pixels of the image edge, in which case the 500x500 pixel region that puts the target as close to center as possible is chosen.} 
On these cutouts, we reran the difference imaging with {\sc ISIS}, and produced new lightcurves to see if clearer photometry could be obtained when aligning and subtracting a smaller image centered on the target. In this way the image alignment and spatially variable kernel of the difference imaging should be optimized at the center of the image cutout. To minimize the impact of image misalignment and dipoles on photometry, we increased the {\sc ISIS} {\sc rad\_phot} and {\sc rad\_aper} parameters to 10.0 px and 11.0 px respectively to better contain the entire dipole in the aperture for the target star, though at the cost of adding sky background noise.

As an extra check on the photometry, we also performed aperture photometry on the cutout images using {\sc AstroImageJ} \citep{collins2017aij}.
We compare the original lightcurves, cutout lightcurves, and aperture photometry lightcurves to see whether the eclipse features seen in the lightcurves are consistent or if they disappear, indicating false positives from, e.g., poor initial difference imaging and photometry, or systematic effects caused by the analysis.
Finally, we make new in- and out-of-transit image stacks using the target-centered difference images.

\begin{figure*}[htb!]
    \centering
    \includegraphics[width=\textwidth]{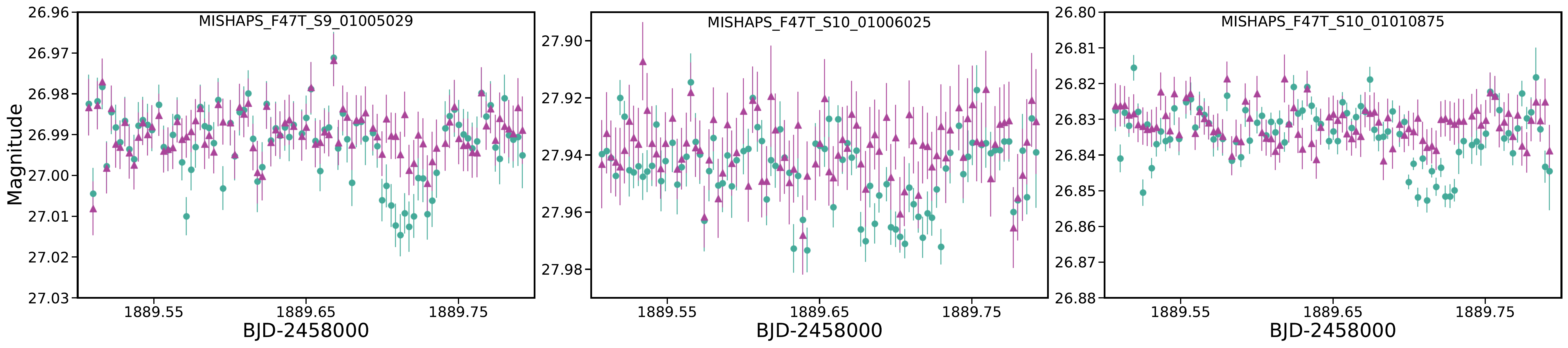}
    \caption{$r-$band lightcurves from survey night 1225 for MISHAPS\_F47T\_S9\_01005029 (left), MISHAPS\_F47T\_S10\_01006025 (middle), and MISHAPS\_F47T\_S10\_01010875 (right). The de-trended lightcurves are plotted in teal circles, and the target-centered lightcurves are plotted in purple triangles. Each lightcurve showed a similar dip at roughly the same time in the original de-trended $r-$band lightcurves, but the target-centered photometry was able to remove this signal, indicating a likely systemic error in the difference images for this time period.}
    \label{fig:sys1225}
\end{figure*}

Viewing these lightcurves and the rest of the vetting products holistically, we are able to conclusively rule out each of the remaining 18 candidates. In some cases there is a singular clear reason for rejecting the candidate based on the second pass of detailed vetting. For example, the target-centered difference imaging lightcurve for some candidates did not show the same transit-like signal as the original pipeline lightcurve. Three examples of these rejected candidates are shown in Figure~\ref{fig:sys1225}, which show transit-like signals in the original pipeline lightcurve at roughly the same time at $BJD{\sim}2459889.7$; the three targets are located on two neighboring DECam chips, S9 and S10. In contrast, the target-centered difference image lightcurve for each candidate at the same time is flat within the photometric uncertainties. We reject all three of these candidates, and others for the same reason.

Other cases are less clear cut, but still lead to rejection. Rather than providing a repetetive accounting of the process of ruling out each one, we opt to present the analysis that we conducted for a single candidate with the most promising transit-like event in its original pipeline lightcurve. All but four of the candidates are eliminated in a similar fashion, though with minor differences in the analysis of each. 

\begin{figure}[htb]
    \centering
    \includegraphics[width=\linewidth]{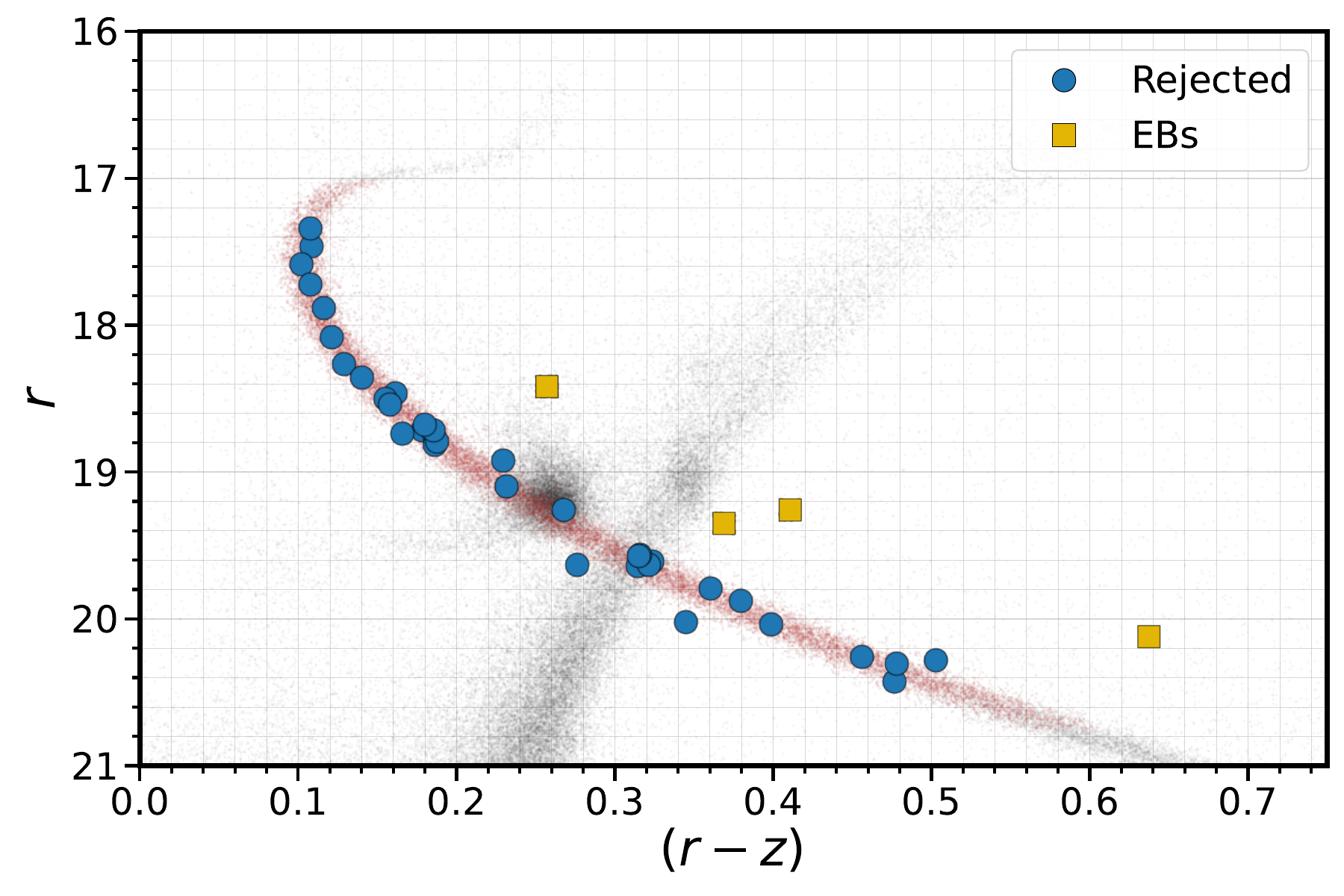}
    \caption{CMD for our field (black points) with our targets (red points) and candidates highlighted. The rejected candidates are given by the blue circles, and the EBs are given by the yellow squares.}
    \label{fig:candcmd}
\end{figure}

\begin{figure}[htb]
    \centering
    \includegraphics[width=\linewidth]{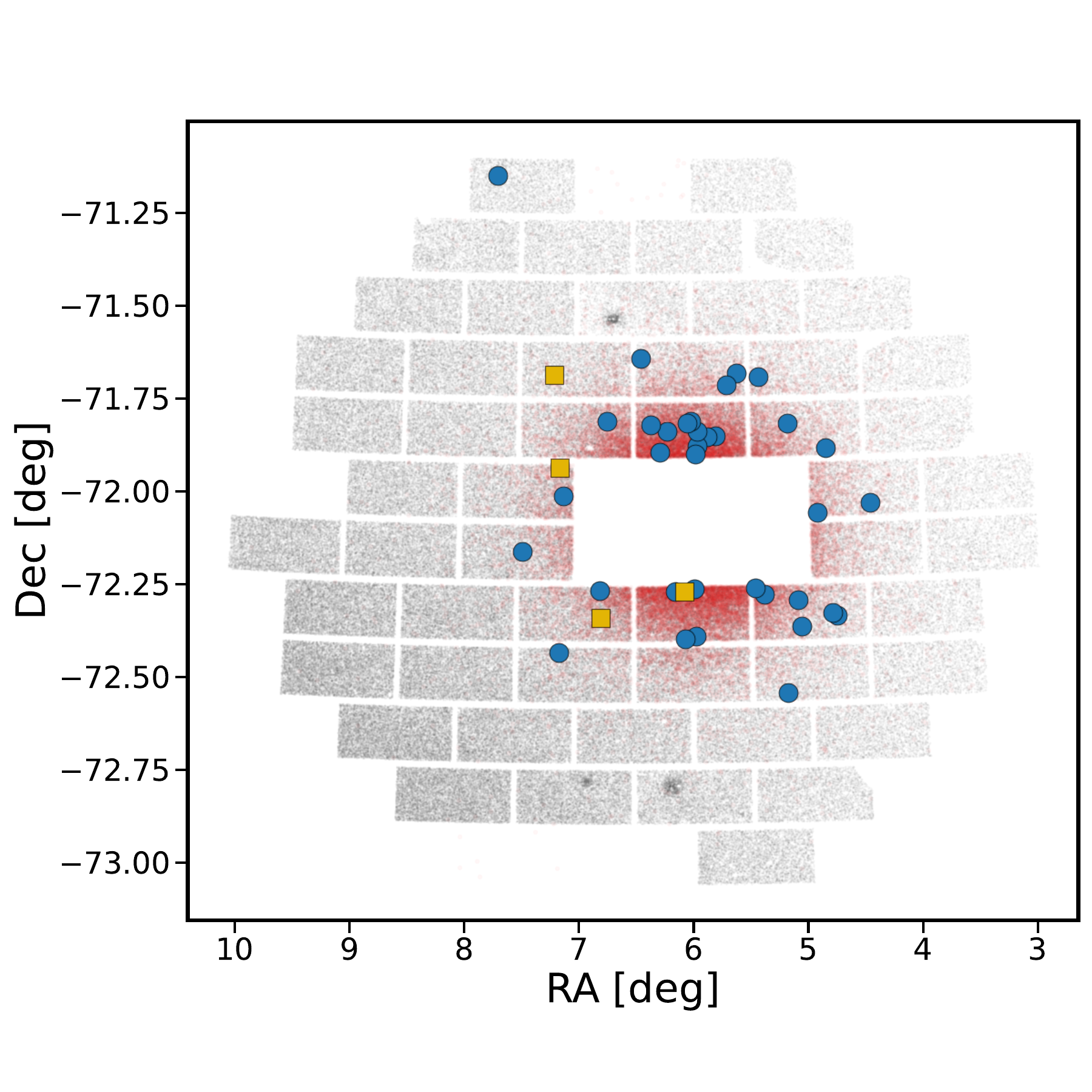}
    \caption{Positions of all high-quality ({\sc DoPHOT} flag=1) star detections in our field (black points) and our target stars (red points), with the candidates highlighted. The rejected candidates are given by the blue circles, and the EBs are given by the yellow squares. The large missing section in the center is due to our removing the central CCDs from our dataset. The rest of the seemingly-missing sections are due to either faulty CCDs (middle left, bottom middle) or not having enough stars meeting our requirements to have those sections appear populated.}
    \label{fig:candpos}
\end{figure}

The vetting stage or cause of rejection for all the original 39 candidates is listed in Table~\ref{tab:cands}, along with the signal-to-noise ratio of the detection and the times of detected full or partial eclipses reported by the transit search.\footnote{For detections of full eclipses, the search's best $t_{\rm c}$ is given.
For detections of partial eclipses, the times of ingress and egress are instead calculated using $t_{\rm c} \pm \frac{1}{2}T$, and the one which lies within the data is included in the table.} 
The positions of each candidate on 47~Tuc's color-magnitude diagram are shown in Figure \ref{fig:candcmd}, and their physical locations in the cluster are shown in Figure \ref{fig:candpos}.
The lightcurves, reference images, and in-transit stacked difference images for each candidate's detection nights can be found in Appendix \ref{app:candinfo}. 
In the next two subsections, we give an example of the analysis that led to the rejection of the 35 transit candidates (\S\ref{subsec:case}) and details of the 4 remaining candidates that we classify as EBs (\S\ref{subsec:rejeb}).  

\begin{longtable*}{cccccccccc}
    \caption{Candidates} \\
     & $R_{\star}$ & $r_{\rm PS1}$ & $\delta$ & & $t_{c} | t_{i,e}^{1}$ & T & Transit & Final & Rejection \\
    ID & (\rsol) & (mag) & (mag) & $S/N$ & (days) & (hrs) & Type & Classification & Stage \\
    \hline
    \hline
    N5\_01007377 & 1.18 & 17.5 & -0.003 & 8.25 & 1432.81 & 3.72 & Partial & Rejected & Second pass \\ 
    
    N9\_01001945 & 0.79 & 18.6 & 0.012 & 7.76 & 700.87 & 1.72 & Partial & Rejected & Second pass \\ 
    
    N9\_01002359 & 0.90 & 18.2 & -0.01 & 10.78 & 700.88 & 2.97 & Partial & Rejected & Second pass \\ 
    
    N9\_01005022 & 0.62 & 20.4 & -0.048 & 7.00 & 1403.90 & 1.47 & Partial & Rejected & First pass \\ 
      &  &  & -0.035 & 8.37 & 1404.92 & 0.72 & Partial &   &   \\ 
      &  &  & 0.051 & 9.40 & 1430.92 & 3.47 & Partial &   & \\ 
      &  &  & -0.047 & 9.56 & 1431.85 & 1.47 & Full &   &  \\ 
      &  &  & -0.054 & 7.08 & 1432.9 & 0.72 & Full &   &   \\ 
    
    N9\_01005409 & 0.66 & 19.3 & -0.023 & 7.03 & 1885.59 & 0.72 & Full & Rejected & First pass \\ 
    
    N9\_01010459 & 0.72 & 18.8 & -0.015 & 7.16 & 1404.92 & 1.22 & Partial & Rejected & First pass \\ 
    
    N9\_01012179 & 0.55 & 20.5 & 0.257 & 8.47 & 1433.81 & 3.22 & Partial & Rejected & First pass \\ 
    
    N10\_01015356 & 0.84 & 18.4 & -0.011 & 8.67 & 1403.91 & 1.47 & Partial & Rejected & Second pass \\
      &  &  & 0.006 & 7.33 & 1405.88 & 1.97 & Partial &   &   \\ 
    
    N10\_01016157 & 0.57 & 20.3 & -0.12 & 8.07 & 1431.87 & 0.72 & Full & Rejected & First pass \\
      &  &  & -0.085 & 8.03 & 1432.82 & 0.97 & Partial &   &   \\ 
    
    N10\_01018928 & 0.74 & 18.8 & -0.045 & 10.06 & 697.90 & 0.97 & Partial & Rejected & First pass \\ 
      &  &  & -0.025 & 8.31 & 1432.9 & 0.97 & Full &   &   \\ 
    
    N10\_01019290 & 0.84 & 20.2 & 0.062 & 15.74 & 698.83 & 3.72 & Partial & EB & --- \\ 
      &  &  & 0.062 & 15.90 & 1430.83 & 2.97 & Partial &   &   \\ 
      &  &  & -0.072 & 11.19 & 1889.59 & 0.97 & Full &   &   \\ 
    
    N10\_01021994 & 0.74 & 18.8 & -0.018 & 10.93 & 1403.91 & 1.47 & Partial & Rejected & First pass \\ 
    
    N11\_01005481 & 0.95 & 18.5 & -0.018 & 8.59 & 1431.84 & 0.97 & Full & EB & --- \\ 
    
    N11\_01005644 & 0.62 & 19.7 & -0.017 & 7.13 & 698.90 & 1.97 & Partial & Rejected & Second pass \\ 
      &  &  & 0.03 & 9.47 & 1402.94 & 0.72 & Partial &   &   \\ 
      &  &  & -0.032 & 8.20 & 1405.92 & 1.22 & Partial &   &   \\ 
      &  &  & -0.037 & 10.74 & 1431.85 & 1.47 & Full &   &   \\ 
      &  &  & -0.024 & 7.23 & 1432.83 & 1.22 & Partial &   &   \\ 
      &  &  & 0.027 & 9.73 & 1433.83 & 2.97 & Partial &   &   \\ 
      &  &  & -0.032 & 9.47 & 1885.57 & 0.72 & Partial &   &   \\ 
      &  &  & -0.028 & 8.75 & 1886.56 & 0.97 & Partial &   &   \\ 
    
    N15\_01005656 & 0.82 & 18.4 & 0.014 & 8.33 & 1431.92 & 3.22 & Partial & Rejected & Second pass \\
      &  &  & -0.015 & 7.14 & 1889.71 & 0.72 & Full &   &   \\
    
    N17\_01009501 & 0.81 & 18.5 & 0.008 & 7.71 & 1403.90 & 1.22 & Partial & Rejected & Second pass \\ 

    S2\_01002825 & 0.6 & 19.9 & -0.034 & 7.92 & 1406.90 & 1.22 & Partial & Rejected & Second pass \\ 
      &  &  & -0.025 & 7.11 & 1407.93 & 0.72 & Partial &   &   \\ 
      &  &  & -0.02 & 8.36 & 1431.84 & 2.97 & Partial &   &   \\ 
    
    S2\_01006929 & 0.67 & 19.2 & -0.013 & 8.82 & 1432.89 & 2.22 & Partial & Rejected & First pass \\ 

    S5\_01001384 & 0.62 & 19.7 & -0.04 & 9.13 & 1430.81 & 3.47 & Partial & Rejected & First pass \\ 
      &  &  & -0.04 & 13.31 & 1431.83 & 3.22 & Partial &   &   \\ 
    
    S5\_01002002 & 0.84 & 19.3 & 0.44 & 25.18 & --- & 2.97 & Partial & EB & --- \\ 
    
    S9\_01001873 & 0.74 & 18.8 & 0.015 & 8.08 & 700.93 & 4.22 & Partial & Rejected & First pass \\ 
    
    S9\_01005029 & 0.75 & 18.8 & -0.018 & 9.54 & 1889.72 & 0.97 & Full & Rejected & Second pass \\    
    
    S10\_01006025 & 0.6 & 19.7 & -0.022 & 7.76 & 1889.71 & 1.47 & Full & Rejected & Second pass \\ 
    
    S10\_01008184  & 0.56 & 19.7 & -0.018 & 8.17 & 1403.90 & 2.22 & Partial & Rejected &  First pass \\
     &  &  & 0.024 & 10.37 & 1430.93 & 3.72 & Partial &   &   \\
    
    S10\_01010875 & 0.78 & 18.6 & -0.016 & 8.60 & 1889.72 & 0.97 & Full & Rejected & Second pass \\ 
    
    S10\_01011133 & 0.71 & 18.9 & -0.014 & 8.67 & 699.90 & 1.47 & Partial & Rejected & First pass \\ 
    
    S10\_01011503 & 0.58 & 20.4 & -0.042 & 9.12 & 1432.83 & 2.97 & Partial & Rejected & First pass \\ 
      &  &  & -0.066 & 7.54 & 1889.69 & 0.72 & Full &   &   \\ 
    
    S10\_01012927 & 0.61 & 19.7 & 0.041 & 8.28 & 1889.55 & 5.97 & Partial & Rejected & First pass \\ 
    
    S10\_01013998 & 1.10 & 17.7 & -0.007 & 7.31 & 700.9 & 1.47 & Full & Rejected & First pass \\ 
      &  &  & -0.008 & 8.85 & 1429.91 & 0.72 & Partial &   &   \\
    
    S10\_01020146 & 0.62 & 19.6 & 0.015 & 7.59 & 1432.93 & 4.47 & Partial & Rejected & First pass \\ 
    
    S10\_01022206 & 0.98 & 18.0 & -0.01 & 7.52 & 1432.93 & 3.72 & Partial & Rejected & First pass \\ 
    
    S10\_01024610 & 0.57 & 20.1 & -0.018 & 7.58 & 1403.94 & 1.72 & Partial & Rejected & Second pass \\ 
    
    S11\_01004730 & 1.04 & 17.8 & -0.009 & 8.75 & 1403.89 & 1.72 & Partial & Rejected & Second pass \\ 
    
    S15\_01005434 & 0.72 & 18.9 & -0.008 & 7.29 & 700.88 & 1.97 & Partial & Rejected & Second pass \\ 
    
    S16\_01000575 & 0.60 & 20.0 & -0.022 & 7.88 & 700.90 & 2.22 & Partial & Rejected & First pass \\ 
    
    S16\_01001403 & 1.24 & 17.4 & 0.009 & 7.87 & 1404.89 & 1.22 & Partial & Rejected & First pass \\ 
    
    S16\_01009327 & 0.73 & 19.0 & -0.009 & 7.16 & 1403.90 & 1.47 & Partial & Rejected & First pass \\ 
      &  &  & -0.009 & 9.11 & 1430.93 & 3.47 & Partial &   &   \\ 
    
    S17\_01006408 & 0.75 & 19.4 & -0.051 & 13.57 & 1432.88 & 1.22 & Partial & EB & --- \\ 
      &  &  & -0.051 & 14.67 & 1886.63 & 1.22 & Full &  &  \\ 
    
    S31\_01003863 & 0.53 & 20.1 & -0.019 & 7.43 & 700.88 & 1.22 & Partial & Rejected & First pass \\ 
    \hline
    \multicolumn{10}{c}
    a NOTE---For full transits, the $t_{\rm c}$ returned by the transit search algorithm is given. For partial transits, \\
    \multicolumn{10}{c}
    either the time of ingress $t_{i}$ or time of egress $t_{e}$ is given, depending on what part of the transit we capture.
\end{longtable*}\label{tab:cands}

    \subsection{Case Study: MISHAPS\_F47T\_S11\_01004730}\label{subsec:case}
        \begin{figure}
        \includegraphics[width=\columnwidth]{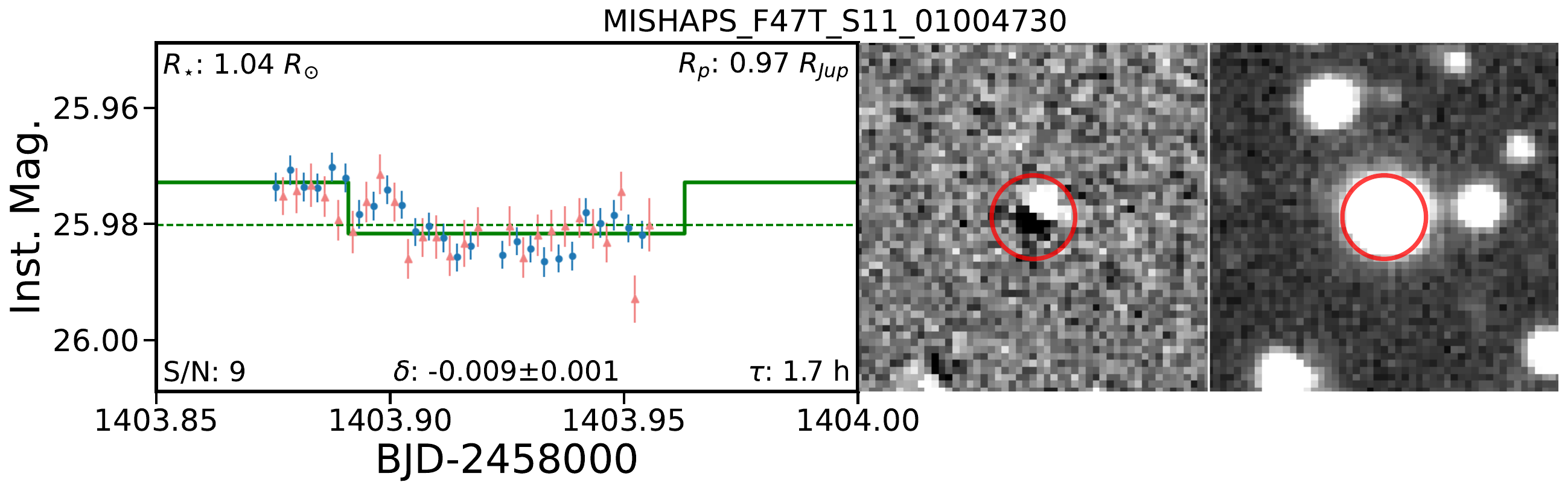}
        \includegraphics[width=\columnwidth]{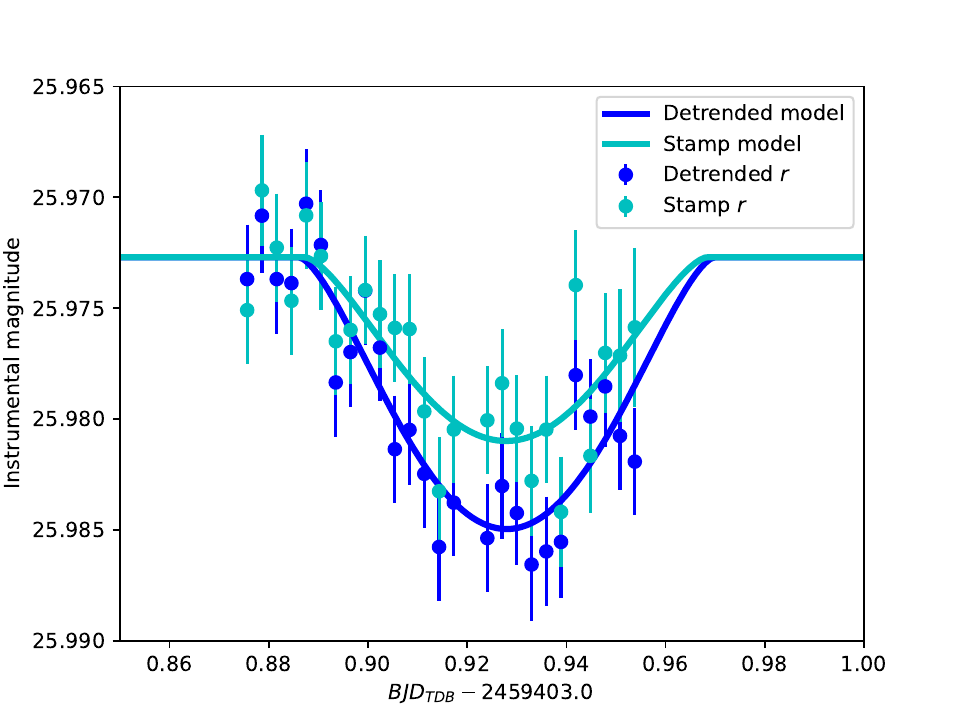}\\
        \includegraphics[width=\columnwidth]{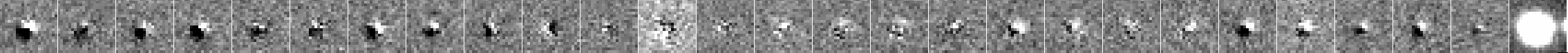}
        \caption{\emph{Top row}: Transit search lightcurve plot for MISHAPS\_F47T\_S11\_01004730 from night 1403 showing a $S/N=8.7$ detection. The images in the middle and right of the row are the in-transit stacked difference image and the reference image. \emph{Middle}: Detrended (blue) and target-centered (cyan) lightcurves of the night 1403 transit. The lines show transiting planet fits to each lightcurve. The target-centered lightcurve is shifted so that the model baselines match. \emph{Bottom}: Time series of target-centered difference images of the target on the same night showing the evolution of dipoles. The last image in the sequence is the reference image.}
        \label{fig:varying_dipole}
    \end{figure}
     
    MISHAPS\_F47T\_S11\_01004730 is an $r=17.8$ turnoff star with a partial event flagged in night 1403 with $S/N=8.7$ and a boxcar depth of $8.7$~mmag. The $z-$band lightcurve follows the $r-$band lightcurve, which shows ${\sim}5$ points at baseline, a long ingress, possibly a short flat bottom, and the beginning of the egress, as shown in the top left panel of Figure~\ref{fig:varying_dipole}. 
    Our radius estimate for the target star is $1.04 R_{\sun}$.
    If the transit depth were the same as the boxcar fit's estimate and it was flat-bottomed, this would correspond to an eclipser or transiting planet of radius ${\sim}1.0\rjup$. 
    The stacked difference image shows a dipole (top-center panel of the plot), slightly more pronounced in its negative (white) parts than its positive (black) parts, with no evidence of a blend in the stacked difference image or reference image (top right panel). 
    There were no similar signals in the lightcurves of nearby targets. Given the strong signal and lack of clear issues in the first pass of detailed vetting, we conducted a second pass of vetting.
    
    The target-centered difference imaging lightcurve and the {\sc AstroImageJ} aperture photometry lightcurve show a similar event, which lends some credence to the signal being real. 
    However, notably, the $r$-band target-centered lightcurve transit is significantly shallower than the detrended lightcurve's transit (shown in the middle panel of Figure~\ref{fig:varying_dipole}). 
    We fit both the original pipeline lightcurve and the target-centered difference image lightcurve with transiting planet models calculated using Batman~\citep{kreidberg2015batman} fits to yield best fit transits with depths of $-12.3$~mmag and $-8.3$~mmag, respectively. 
    As the target-centered photometric extraction uses a larger aperture, it is possible that the inclusion of flux from another star entering the aperture on the reference image could cause the depth to be shallower, but we rule that out as a possibility --- the target-centered lightcurve is only ${\sim}0.01$ mag brighter, whereas to cause enough dilution to decrease the transit depth by ${\sim}33$\% would require a neighboring star to increase the flux in the aperture by ${\sim}0.31$~mag. 
    This depth difference between the two lightcurves measured with the same data but slightly different techniques was the first piece of evidence to reject the candidate, but we did not deem it conclusive enough on its own. 
    We therefore inspected the individual difference images in a timeseries, as shown in the thin lower panel of Figure~\ref{fig:varying_dipole}, and found that the variations of strength of the dipole are correlated with the flux, with the dipole nearly disappearing at the bottom of the apparent transit.

    In addition to the flagged transit, the lightcurve of this object shows variability at a comparable level on some nights, with some of this variability having the appearance of ingresses, egresses or full transits.
    Boxcar depths in the detrended lightcurves range from 3 to 9 mmag, with one of these having a boxcar $S/N=6.3$ and the rest lower. 
    Lomb-Scargle, box least squared, and analysis of variance periodograms \citep{hartman2016vartools,zechmeister2009ls,press1992nr,kovacs2002bls,schwarzenberg-czerny1989aov,devor2005aov} do not reveal any clear periods, which, if the single transit-like signals were periodic and detectable by eye, we would expect to phase together into a clear folded signal.

    The combination of the variations of the dipole correlating with the transit lightcurve, the unexplained difference in transit depth between the detrended and target-centered difference images, and our inability to phase the lightcurve variability to a clear transit leads us to reject this candidate.

    \subsection{Eclipsing Binaries}\label{subsec:rejeb}
\begin{figure}[h]
            \centering
            \includegraphics[width=\linewidth]{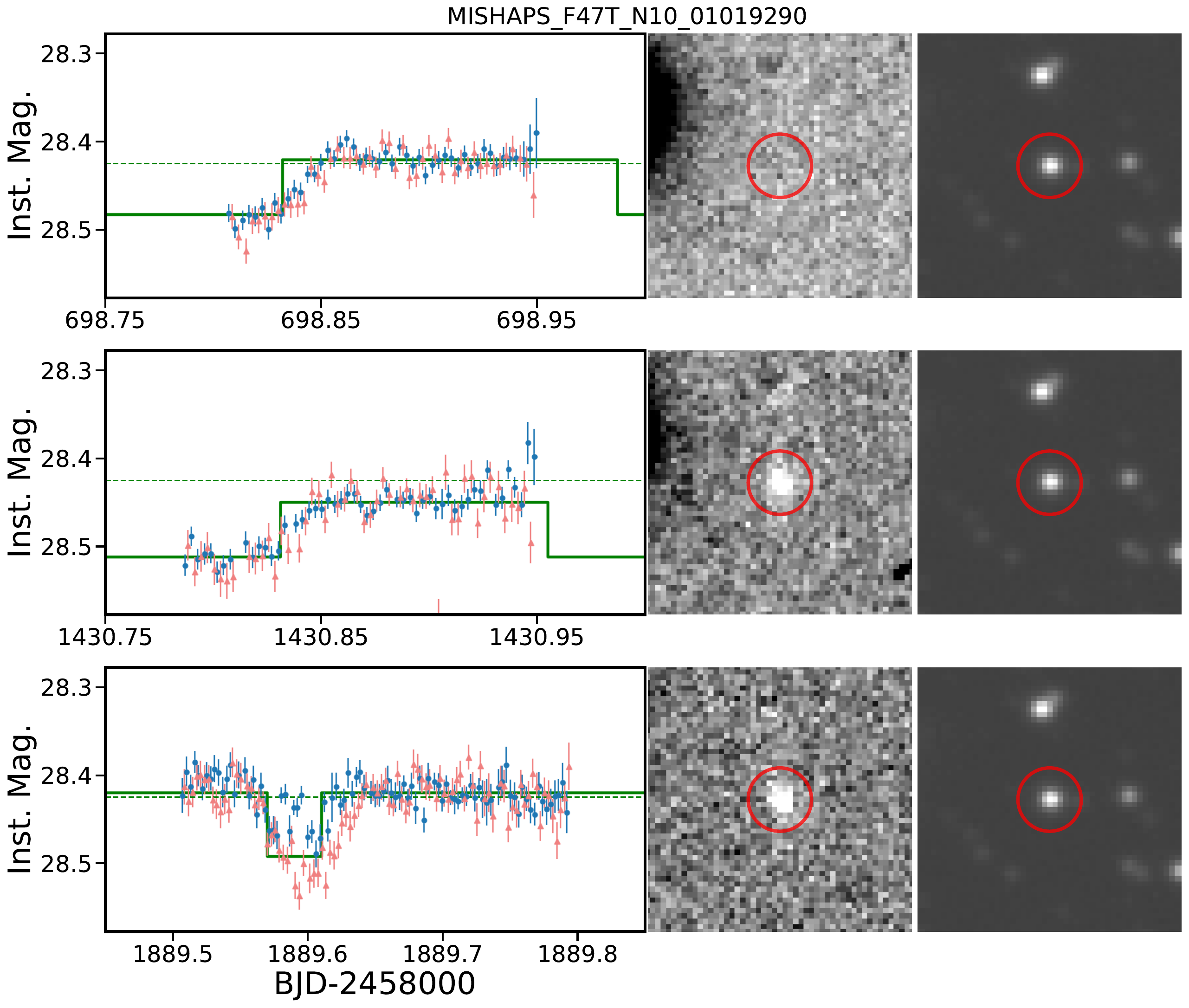}
            \includegraphics[width=\linewidth]{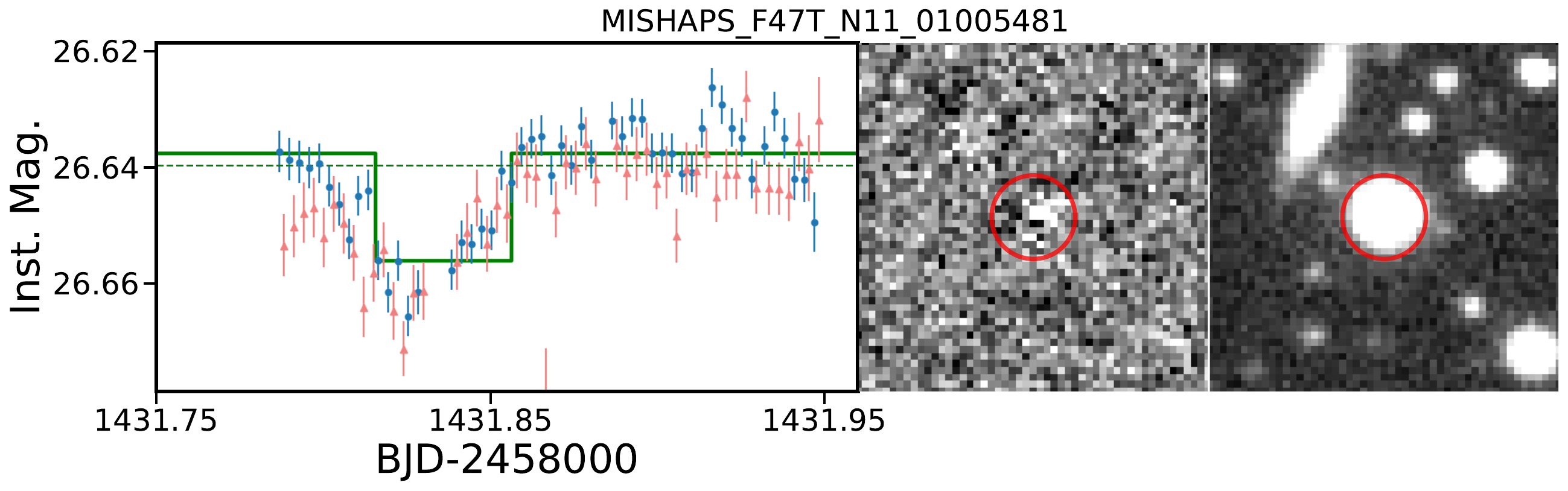}
            \includegraphics[width=\linewidth]{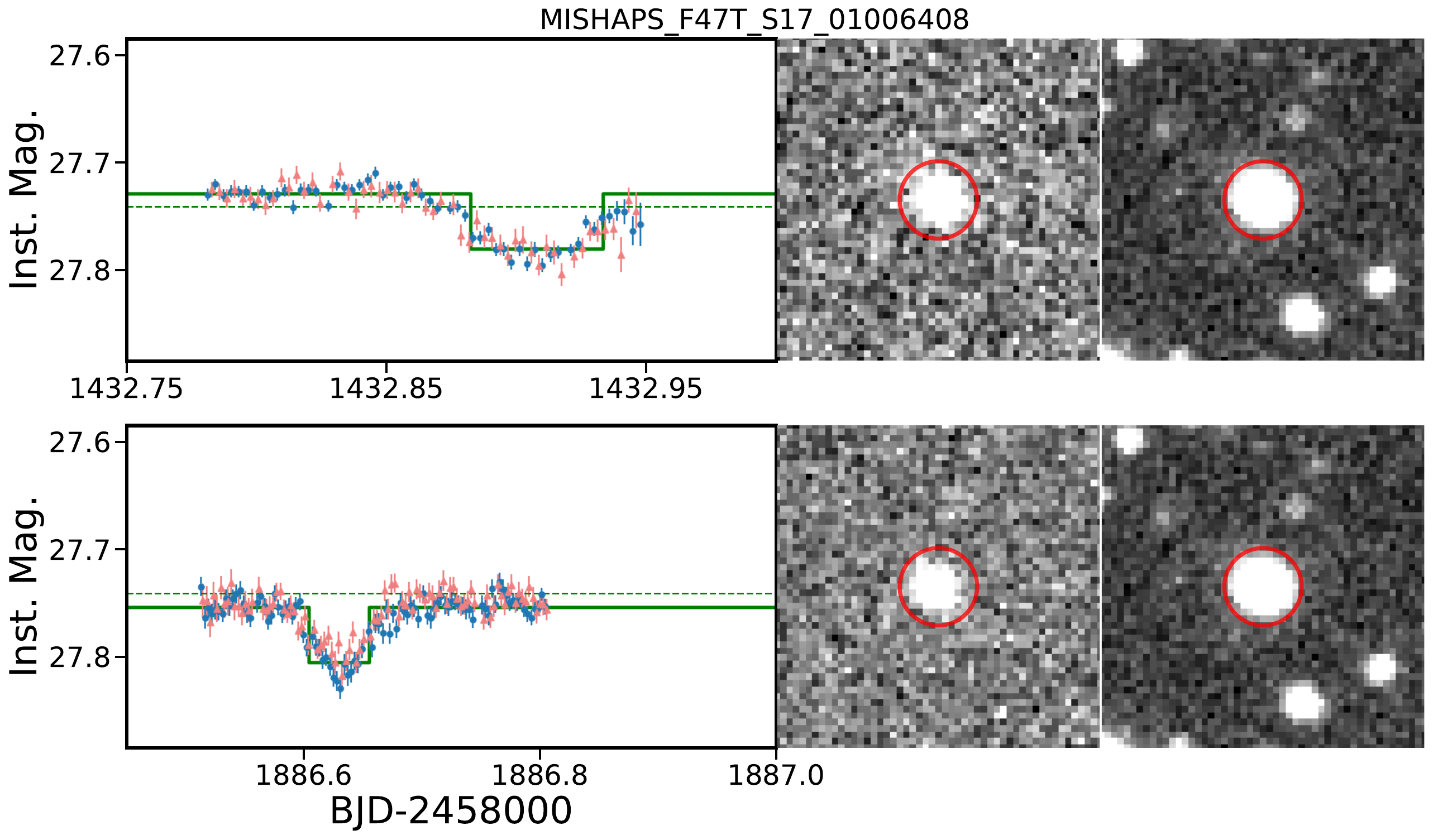}
            \includegraphics[width=\linewidth]{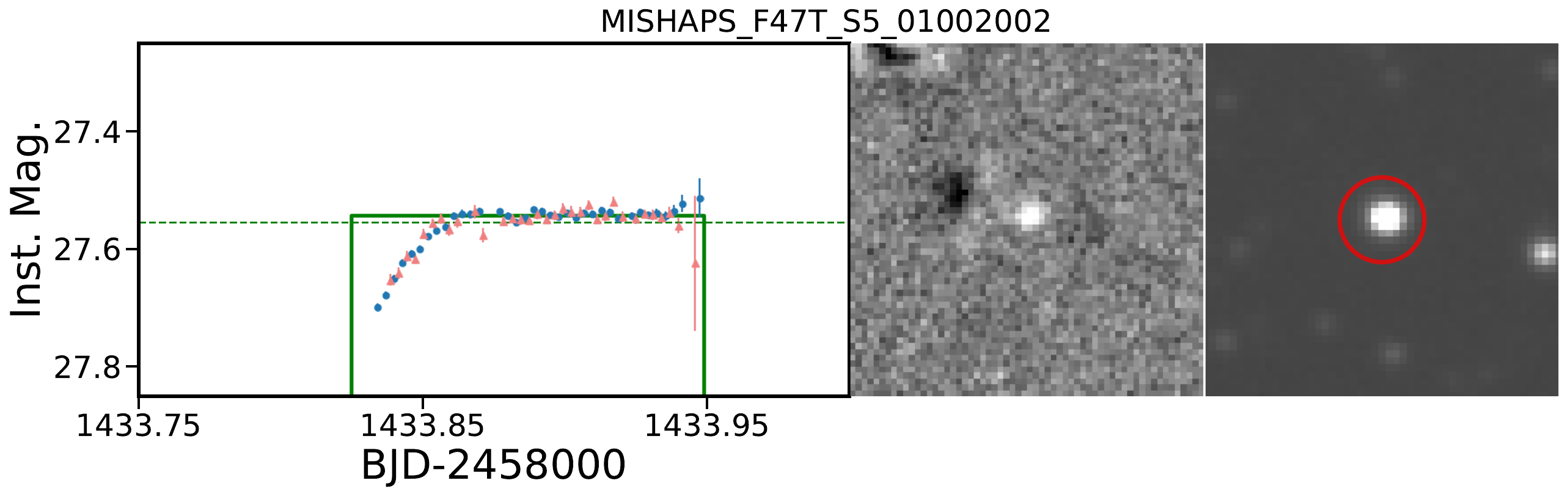}
            \caption{Lightcurves and stacked in-eclipse difference images for EB candidates. The lightcurves of each detection are shown in the left column, with the $r-$band data plotted in blue, and the $z-$band data plotted in red. The middle column shows the stacked in-eclipse difference images for each detection, and the right column shows the reference image for side-by-side comparison.}
            \label{fig:ebims}
        \end{figure}

    Our search and 2-step vetting leave us with 4 EB candidates.
    We cross-referenced the EB candidates with \citet{weldrake2004vars} and to Vizier catalogs containing the targets.
    One target --- MISHAPS\_F47T\_S5\_01002002 --- is contained in the OGLE catalog of EBs \citep{pawlak2016ogleeb} and is also flagged as an EB in Gaia~DR3 \citep{babusiaux2023gaiadr3}, but we find no existing literature matches for the other 3.
    Other catalogs searched included \citet{roman1987cat}, the Guide Star Catalog \citep{lasker2008gsc}, \citet{kaluzny2013case}, ATLAS-REFCAT 2 \citep{tonry2018atlasrefcat2}, \citet{stetson2019gccat}, and \citet{jimenezarranz2023cat}.
    All four targets lie on 47~Tuc's binary sequence and show clear indications of eclipses both in their lightcurves and in their stacked in-``transit" difference images.
    These candidates are all rejected by our color cuts that exclude stars on the binary sequence, but demonstrate the value of MISHAPS' high-cadence color observations.
    The flagged portions of their lightcurves, stacked difference images, and reference images are shown in Figure \ref{fig:ebims}

        \subsubsection{MISHAPS\_F47T\_N10\_01019290}\label{subsubsec:ebn10}
        MISHAPS\_F47T\_N10\_01019290 is located at 00$^{\rm h}$24$^{\rm m}$19.43$^{\rm s}$, -72$^{\circ}$16'14.00''.
        For this target, we observe partial eclipses in nights 698 and 1430, each with an estimated depth of 0.062 mag.
        In night 1889, we observe a full eclipse with an estimated depth of 0.072 and search-estimated duration of $\sim$1.0 hr.
        The target also displays long-term variability.
        The target lies above 47~Tuc's main sequence near the equal luminosity line.

        \subsubsection{MISHAPS\_F47T\_N11\_01005481}\label{subsubsec:ebn11}
        MISHAPS\_F47T\_N11\_01005481 is located at 00$^{\rm h}$27$^{\rm m}$13.93$^{\rm s}$, -72$^{\circ}$20'30.78''.
        For this target, we observe a single full eclipse on night 1431.
        The search returns an estimated depth of 0.018 and duration of 1.0 hr for this eclipse for this eclipse --- though visual inspection suggests the depth is closer to 0.028 --- with roughly equal depths in $z$ and $r$.
        Though the equal depths could indicate a planetary transit, the sharp, V-shaped bottom of the eclipse is indicative of an EB or grazing transit.
        This target exhibits long-term variability, though we are not able to recover a well-phased period for it.
        This target falls on 47~Tuc's binary sequence in the CMD, and is ultimately rejected from consideration in our sample by the final color cut in target selection that excludes 47 Tuc's binary sequence.

        \subsubsection{MISHAPS\_F47T\_S17\_01006408}\label{subsubsec:ebs17}
        MISHAPS\_F47T\_S17\_01006408 is located at 00$^{\rm h}$28$^{\rm m}$51.47$^{\rm s}$, -71$^{\circ}$41'07.20''.
        For this target, we observe a partial eclipse in night 1432 and a full eclipse in night 1886.
        Both eclipses have estimated depths of 0.051 and estimated durations of 1.2 hrs, and have V-shaped bottoms indicative of EBs or grazing transits.
        The $r-$band eclipses are slightly deeper than the $z-$band eclipses, and the target falls on 47~Tuc's binary sequence in the CMD and is rejected by the color cut.

        \subsubsection{MISHAPS\_F47T\_S5\_01002002}\label{subsubsec:ebs5}
        When cross-matching with other catalogs, this candidate is identified as OGLE~SMC-ECL-6273, and is flagged as an EB in Gaia (DR3 4689635452915711616).
        It is located at 00$^{\rm h}$28$^{\rm m}$40.45$^{\rm s}$, -71$^{\circ}$56'12.79''.
        In our data, we observe one partial eclipse in night 1433.
        The search estimates the depth of our observed eclipse to be 0.44, but our observations do not capture the full egress and thus do not reflect an accurate eclipse depth.
        This target also exhibits long-period variations, but we are not able to recover our own period estimate.  

\section{Occurrence Rate Limit}\label{sec:occurrence}
Our goal is to constrain the occurrence rate of hot Jupiters in 47 Tuc.
For the purpose of this paper, we define our occurrence rate $f_{\rm HJ}$ as the average number of planets per star, assuming that period and radius are distributed uniformly in each parameter across a specified range, i.e., 
\begin{equation}
    f_{\rm HJ} = \int_{R_{\rm p, min}}^{R_{\rm p, max}} \int_{P_{\rm min}}^{P_{\rm max}} \frac{d^{2}N}{d R_{\rm p} d P} d P d R_{\rm p},
\end{equation}
where $R_{\rm p}$ is the planet radius and $P$ is the orbital period, and $d^{2}N/dR_{\rm p}dP$ is the joint period-radius distribution.

As we have not found any convincing candidates, we can only place an upper limit on their occurrence. 
We estimate our 95\% upper limit on the occurrence rate by assuming the number of candidate planets we expect to detect, $N_{\rm exp}$, is Poisson distributed
\begin{equation}
P(N_{\rm cand}) = \frac{N_{\rm exp}^{N_{\rm cand}}e^{-N_{\rm exp}}}{N_{\rm cand}!},
\label{eqn:poisson}
\end{equation}
where $N_{\rm cand}$ is the number of surviving candidates.
Setting the probability of finding no planets equal to $(1-0.95)$ and solving for $N_{\rm exp}$ when $N_{\rm cand}=0$, we get $N_{\rm exp}=3$. 
The number of candidates we expect to detect is set by $f_{\rm HJ}$, $N_{\star}$, and $\epsilon_{\rm total}$ as
\begin{equation}
N_{\rm exp} = N_{\star} f_{\rm HJ} \epsilon_{\rm total}.
\label{eqn:nexp}
\end{equation}
Previous searches for hot Jupiters in globular clusters have expressed their sensitivity as $N_{\rm exp}$ for an assumed $f_{\rm HJ}$, but we find it useful to define a further quantity, $N_1$, the expected number of planets to be found in a survey if the occurrence rate is 1 planet per star
\begin{equation}
N_1 = N_{\star} \epsilon_{\rm total}.
\label{eqn:n-one}
\end{equation}
The upper limit on the occurrence rate can then be written
\begin{equation}
f_{\rm HJ} < \frac{3}{N_1}.
\label{eqn:fhjlimit}
\end{equation}

It is worth noting two points about $N_1$ and its corresponding limit on the occurrence rate $f_{\rm HJ}$. 
Both are dependent on the range of planet parameters that are considered. 
Somewhat counter-intuitively, if the parameter range is increased into areas that have lower detection efficiency, $N_1$ will fall, and the occurrence rate limit will weaken.
This is because the $N_1$ definition assumes one planet per star in the parameter range under consideration, so increasing the range reduces the density of planets per unit parameter space in regions where the survey is most sensitive and distributes them to areas that are less sensitive. 
However, despite the occurrence rate limit weakening with increased parameter space, the number of actual planets that can be found will increase, because the number of ``available'' planets will increase with increasing parameter space, even if it may be at a slower rate than the parameter space increases. 
Therefore, it is important to only compare occurrence rates over the same range of parameters. 
Additionally, one should not seek to achieve the lowest possible occurrence rate limit by restricting the parameter space and instead attempt a measurement in as wide a parameter space that still retains reasonable sensitivity and likelihood of planet occurrence.

For these reasons we opt to present our main occurrence rate limit in nearly the full range over which we calculate detection efficiencies, taking the full range of periods $0.5\le P < 10$~days, and only a slightly restricted radius range of $0.75\le R_{\rm p} < 2.0~R_{\rm Jup}$. 
We decide to ignore the $0.5$--$0.75$~$R_{\rm Jup}$ range of planet radii because our average detection efficiency drops precipitously by a factor of ${\sim} 5.5$ relative to the $0.75$--$1.0$~$R_{\rm Jup}$ bin, and because we expect very few hot Jupiters to have radii below $0.75$~$R_{\rm Jup}$ if their incident flux is above ${\sim}2\times 10^{8}$~erg~s$^{-1}$~cm$^{-2}$~\citep[e.g.,][]{demory2011inflated,thorngren2018ohmic}. 
For a planet in a 10~day orbit around an $r=21$ star in 47 Tuc, we compute a minimum incident flux of $2.5\times 10^{8}$~erg~s$^{-1}$~cm$^{-2}$ using the MIST version 1.2 isochrone with [Fe/H]=-0.75 dex at a distance of 4.45~kpc and with no extinction~\citep{dotter2016mist0,choi2016mist1}. 

To accurately account for the variations in detection efficiency and transit probability as a function of period and target brightness as shown in Figure~\ref{fig:2d_eff_Rp_rmag}, and to properly weight the sample by the cluster mass function, we compute $N_1$ by computing $\epsilon_{\rm total}$ in equation~\ref{eqn:n-one} directly as the weighted average over all injection-recovery simulations meeting the radius and period limits. 
As we have injected transiting planets into each lightcurve twice, the application of equation \ref{eqn:nexp} appropriately weights the detection efficiency across the properties of our stellar sample.

Using our detection efficiencies for the 19,930 stars passing our target selection cuts, we estimate $N_1 = 693$ for $0.75 \le R_{\rm p} < 2.0~R_{\rm Jup}$ and $0.5\le P<10$~days, which yields the upper limit on the occurrence rate of hot Jupiters in 47 Tuc  
\begin{equation}
f_{\rm HJ} < 0.43\% \quad {\rm for} \quad\begin{array}{l}0.75 \le R_{\rm p} < 2.0~R_{\rm Jup}\\0.5\le P<10~{\rm days}.\end{array}
\label{fn:main_result}
\end{equation}

To compare our occurrence rate limit to previous limits and measurements, we need to compute the limit for the same range of period and planet radius as the other studies.
We list our estimates and those from previous studies in Table~\ref{tab:occratecomp}.

\citetalias{gilliland2000hstsearch} used injection-recovery simulations of their \textit{Hubble} data for planets with $0.8\le R_{\rm p} < 2.0~R_{\rm Jup}$ and $0.5\le P<8.3$~days to predict their expected yield under the assumption of an occurrence rate $f_{\rm HJ}=0.8$--$1.0$\%, $N_{\rm exp}=17$. 
Taking the center of their occurrence rate range, substituting equation~\ref{eqn:n-one} into equation~\ref{eqn:nexp}, and rearranging, we can compute their survey sensitivity
\begin{equation}
N_1^{\rm G00} = \frac{N_{\rm exp}}{f_{\rm HJ}} = 1889.
\end{equation}
The parameter range is more restrictive than we used for equation~\ref{eqn:fhjlimit}. 
When we use the same range of $R_{\rm p}$ and $P$ as \citetalias{gilliland2000hstsearch} we compute $N_1=830$. 
With neither \citetalias{gilliland2000hstsearch} or our survey finding any planets, the occurrence rate limits are $f_{\rm HJ}<0.16$\% and $f_{\rm HJ}<0.36$\%, respectively.

\citetalias{gilliland2000hstsearch}'s survey searched stars near 47 Tuc's center with Hubble's field of view, while our survey has so far ignored the central regions of the cluster. Our two occurrence rate limits are  therefore completely independent, and we can combine them simply by summing $N_1$ for each survey ($N_1^{{\rm C24}+{\rm G00}}=1889+830=2719$) to yield a stronger occurrence rate constraint
\begin{equation}
f_{\rm HJ} < 0.11\% \quad {\rm for} \quad\begin{array}{l}0.8 \le R_{\rm p} < 2.0~R_{\rm Jup}\\0.5\le P<8.3~{\rm days}.\end{array}
\label{eqn:comb-limit}
\end{equation}

\citetalias{weldrake2005tuc} searched stars throughout 47 Tuc for hot Jupiters and also found none. 
Because our target star sample likely overlaps significantly, we can not combine our limits without a star-by-star accounting of injection and recovery that is outside the scope of this paper. 
However, we can compare results by computing $N_1$ on the same range of $R_{\rm p}$ and $P$. \citetalias{weldrake2005tuc} search out to 16 days, beyond our injections, but present results as a function of period. 
We can therefore compute \citetalias{weldrake2005tuc}'s $N_1$ for a range of parameters enclosed by our injection-recovery simulations. 
From Table 2 of \citetalias{weldrake2005tuc}, we gather that in the range $2.042\le P< 8.672$ days, they expect to find $N_{\rm exp}=6.7$ planets of radius $R_{\rm p}=1.3~\rjup$ and assuming an occurrence rate of $f_{\rm HJ}=0.8$\%, implying $N_1^{\rm W05}=378$ and $f_{HJ}<0.79\%$.  
Over the same range of $R_{\rm p}$ (we take the finite range $1.2$--$1.4~\rjup$) and $P$ we find $N_1=565$ and $f_{\rm HJ}<0.53\%$.

It is also useful to compare our limits to the occurrence rate in the solar neighborhood. 
\citet{fressin2013keplerocc} present occurrence rates using data from the {\it Kepler} spacecraft corrected for false positives. 
In the period ranges $0.8$––$3.4$~days and $0.8$--$10$~ days, for planets with radius $0.53\le R_{\rm p}<1.96\rjup$, they measure hot Jupiter occurrence rates of $0.082\pm0.019$\% and $0.43\pm0.05$\%. 
Over these period and $R_{\rm p}$ ranges we compute $N_1=1178$ and $515$, and place limits of $f_{\rm HJ}<0.25\%$ and $<0.58\%$, respectively.

\begin{figure}[htb!]
    \centering
    \includegraphics[width=\linewidth]{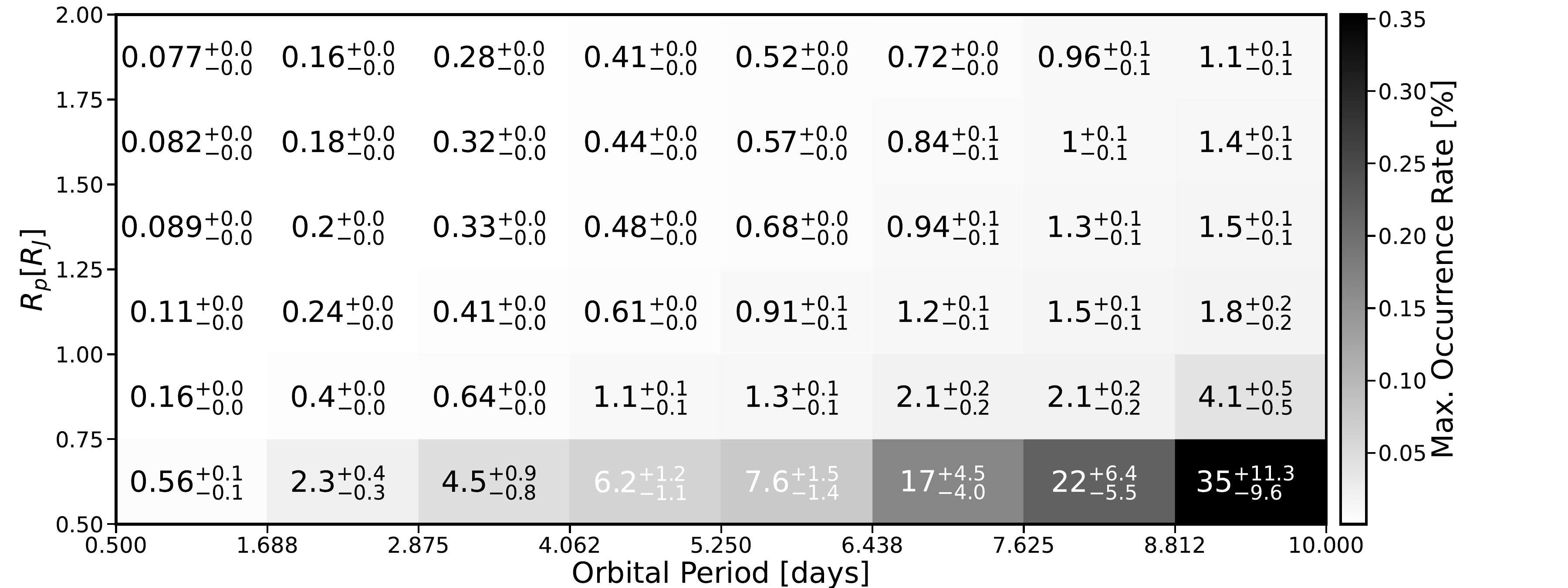}
    \caption{2D occurrence rate upper limits, estimated in bins of $R_{\rm p}$ and $P$. The labels on each bin give the occurrence rate in percent.}
    \label{fig:occrpp}
\end{figure}

To enable more detailed comparisons with future work, we provide with this article the results of our injection-recovery tests for all injections at the NOIRLab Astro Data Lab\footnote{https://datalab.noirlab.edu/data/mishaps}, and a sample dataset on Zenodo\footnote{https://doi.org/10.5281/zenodo.15857086}.
For calculations that do not require data processing, we also present our occurrence rate limits for bins of planet radius and period in Figure~\ref{fig:occrpp}. 
To combine cells in this plot, the individual limits can be converted to $N_1=3/f_{\rm HJ,max}$, averaged, and then converted back to a limit with equation~\ref{eqn:fhjlimit}.

\begin{deluxetable*}{lccccc}
    \centering
    \tablecaption{Hot Jupiter occurrence rate limits for 47 Tuc and measurements in the {\it Kepler} field}
    \tablehead{
         \colhead{Comparison} & \colhead{$R_{\rm p}$ Range ($\rjup$)} & \colhead{$P$ Range (days)} & \colhead{$N_{\rm \star}$} & 
         \colhead{$N_1$} & \colhead{$f_{\rm HJ}$}
    }
    \startdata
	\multicolumn{6}{l}{{\bf 47 Tuc estimates}}\\
	This work & $0.75$--$2.0$ & $0.5$--$10.0$ & $19,930$ & $692.8$ & $<0.43$\% \\[4pt]
	\citet{gilliland2000hstsearch} & $0.8$--$2.0$ & $0.5$--$8.3$ & $34,091$ & $1888.9$ & $<0.16$\% \\
	\citetalias{masuda2017tucreassessment} & \multicolumn{2}{c}{{\it Kepler} hot Jupiters} & $34,091$ & $1222.2$ & $<0.25$\% \\
	This work & $0.8$--$2.0$ & $0.5$--$8.3$ & $19,930$ & $830.1$ & $<0.36$\% \\
	{\bf This work + \citetalias{gilliland2000hstsearch}} & $0.8$--$2.0$ & $0.5$--$8.3$ & $54,021$ & $2719.0$ & {\bf $<0.11$\%} \\[4pt]
	\citet{weldrake2005tuc} & $1.3$ & $1.0$--$10.0$ & $21,920$ & $377.7$ & $<0.79$\% \\
	This work  & $1.2$--$1.4$ & $1.0$--$10.0$ & $19,930$  & $564.6$ & $<0.53$\% \\[4pt]
	This work & $0.53$--$1.96$ & $0.8$--$3.4$ & $19,930$ & $1177.8$ & $<0.25$\% \\
	This work & $0.53$--$1.96$ & $0.8$--$10.0$ & $19,930$ & $514.6$ & $<0.58$\% \\[6pt]
	\multicolumn{6}{l}{\bf Kepler field estimates}\\
	\citet{fressin2013keplerocc} & $0.53$--$1.96$ & $0.8$--$3.4$ & $132,756$ & --- & $0.082\pm0.019$\% \\
	\citet{fressin2013keplerocc} & $0.53$--$1.96$ & $0.8$--$10.0$ & $132,756$ & --- & $0.43\pm0.05$\% \\[6pt]
	\multicolumn{6}{l}{\bf 47 Tuc predictions}\\
	\citet{masuda2017tucreassessment} $0.57\le 
	M_{\star}<0.88$ & $0.8$--$2.0$ & $0.5$--$8.3$ & --- & --- & $0.18$\% \\
	\citet{masuda2017tucreassessment} $0.57\le M_{\star}<0.88$ & $0.8$--$2.0$ & $0.68$--$10.0$ & --- & --- & $0.24_{-0.09}^{+0.10}$\% \\
	{\citet{fressin2013keplerocc} $0.57\le M_{\star}<0.88$ \& [Fe/H]$=-0.78$} & $0.53$--$1.96$ & $0.8$--$10.0$ & --- & --- & $0.028$\% \\
	{\citet{fressin2013keplerocc} $0.57\le M_{\star}<0.88$ \& [$\alpha$/H]$=-0.48$} & $0.53$--$1.96$ & $0.8$--$10.0$ & --- & --- & $0.055$\%\\
    \enddata
    \tablecomments{Occurrence rates for this work are computed for different ranges of planet radius and period to match other studies. The 47 Tuc predictions list two different estimates from \citetalias{masuda2017tucreassessment} that are their estimates of the {\it Kepler}-field occurrence rates for masses matching 47 Tuc's main sequence stars searched by \citetalias{gilliland2000hstsearch} but with different period ranges, and assume no metallicity dependence. The {\it Kepler} field predictions scale the occurrence rates measured by \citetalias{fressin2013keplerocc} for the same host mass range by the joint mass-metallicity planet occurrence dependence estimated by \citet{johnson2010gpoccurrence} for 47~Tuc's [Fe/H], and assuming that [$\alpha$/H] can be substituted for [Fe/H] in the relation.}
    \label{tab:occratecomp}
\end{deluxetable*}

\section{Discussion}\label{sec:Discussion}
The primary goal of our survey is to improve the constraints on the hot Jupiter occurrence rate in 47 Tuc, inspired by the re-evaluation of previous searches by \citetalias{masuda2017tucreassessment}. 
Our occurrence rate limit by itself does not achieve this, but when added to the \citetalias{gilliland2000hstsearch} limit estimated for a completely separate sample of 47~Tuc stars, our combined 95\%-confidence upper limit of 0.11\% conclusively rules out the equivalent 0.18\% occurrence rate \citetalias{masuda2017tucreassessment} estimated for {\it Kepler}-field stars with the same range of masses as the 47~Tuc stars \citetalias{gilliland2000hstsearch} surveyed. 
However, it is worth noting that when \citetalias{gilliland2000hstsearch}'s non-detection is converted into a 95\% occurrence rate limit over a range of planetary parameters ($f_{\rm HJ}<0.16$\%), it just about rules out \citetalias{masuda2017tucreassessment}'s estimate of the {\it Kepler} low-mass host occurrence rate ($f_{\rm HJ}=0.18$), not accounting for the large but unquoted Poisson uncertainties. 
\citetalias{masuda2017tucreassessment}'s conclusion that the \citetalias{gilliland2000hstsearch} data set does not rule out the {\it Kepler}-field low-mass occurrence rate is likely due to their use of the actual planet radii and periods of the {\it Kepler} planets to estimate the detection efficiency rather than a uniform distribution of $R_{\rm p}$ and $P$.
This choice has the effect of weighting the distribution of injections to longer periods, which are harder to detect. 
With this sample of injections, the effective constraining power of the search, quantified by $N_1$, is about two thirds of that over a uniform distribution of planet radius and period.

We have not reproduced \citet{masuda2017tucreassessment}'s methodology, but if we assume a similar two thirds detection efficiency for an appropriately weighted distribution, our limit combined with \citet{masuda2017tucreassessment} would yield a combined $N_1=1776$ and an occurrence rate limit of $f_{\rm HJ}<0.17$\%, which would just about rule out \citetalias{masuda2017tucreassessment}'s prediction of $f_{\rm HJ}=0.18$\% that assumes no metallicity dependence.

If we assume that the strong dependence of giant planet occurrence rate on metallicity holds for globular cluster planet formation, then our new limits do not place particularly interesting constraints on 47 Tuc's hot Jupiter occurrence rate.
\citet{johnson2010gpoccurrence} fit for a giant planet occurrence rate that depends on host mass and metallicity of the form
\begin{equation}
f_{\rm GP} \propto M^{\alpha} 10^{\beta [{\rm Fe}/{\rm H}]},
\label{eqn:fehdependence}
\end{equation}
finding $\alpha=1.0 \pm 0.3$ and $\beta=1.2\pm0.2$ for giant planets with semimajor axis $<2.5$~AU. 
Similar estimates for hot Jupiters yield an even stronger metallicity dependence, with \citet{guo2017hjmet} finding a metallicity dependence of $f_{\rm HJ}\propto 10^{2.1\pm0.7 [{\rm Fe}/{\rm H}]}$, and \citet{petigura2018cksiv} found $f_{\rm HJ}\propto 10^{3.4_{-0.8}^{+0.9} [{\rm Fe}/{\rm H}]}$, both studies having relatively small numbers of hot Jupiters and do not account for a possible host mass dependence. 
If we are optimistic and adopt the \citet{johnson2010gpoccurrence} metallicity relation and apply it to 47~Tuc's metallicity $[{\rm Fe}/{\rm H}]=-0.78$~dex~\citep{cordero2014tucabundances} and use \citetalias{masuda2017tucreassessment}'s $f_{\rm HJ}=0.24$\% occurrence rate for hot Jupiters with $P<10$ days, we can predict the hot Jupiter occurrence rate in 47~Tuc to be $0.028$\%. 
However, if we can assume that $\alpha$-elements actually set the planet occurrence rate and not iron, then by replacing [Fe/H] with $[\alpha/{\rm H}] = [{\rm Fe}/{\rm H}] + [\alpha/{\rm Fe}]$ in equation~\ref{eqn:fehdependence}, we can predict $f_{\rm HJ}=0.055$\% by adopting $[\alpha/{\rm Fe}] = 0.3$~dex~ \citep{cordero2014tucabundances}. 
While our occurrence limits apply to slightly different planet radius and period ranges, the ones that combine our search with \citetalias{gilliland2000hstsearch} are within a factor of $2$--$3$ of the $\alpha$ prediction. 
A 95\% upper limit that rules out the $\alpha$ prediction will require a survey sensitivity of $N_1=5450$.

This work is only the first step toward our goal of providing improved occurrence rate limits in 47~Tuc. 
It is already 50\% more sensitive than the previous ground-based search of 47~Tuc~\citepalias{weldrake2005tuc}, and per star our sensitivity is within 10\% of \citetalias{gilliland2000hstsearch}'s.
We note that our analysis is the first globular cluster planet search to quantify the human vetting detection efficiency, which we find is high (${\sim}90$\%) but significantly lower than 1, especially at longer periods. 
However, we have only determined this efficiency for the first vetting step; it is likely that the remaining vetting steps we take also are not 100\% efficient. 
Additionally, our analysis is the first 47~Tuc transit search to exclude Milky Way field stars and stars in the SMC via proper motion cuts, which provide a small but significant contribution to the stellar sample in the magnitude range $r\approx 18$--$20$ due to the 47 Tuc main sequence crossing the SMC giant branch in the regions of the red clump and red giant branch bump. 
In our next paper, we will improve our occurrence rate limits by incorporating the central cluster chips into our analysis, including faint stars missed by {\it Gaia} but on which we can effectively detect transits, by upgrading our search to incorporate more realistic transit shapes, and to include data from nights that were not as bad as originally assumed, or that were gathered after the inclusion cutoff for this paper. 
Combined, we anticipate these additions to increase $N_1$ to the point where we can challenge the most optimistic predictions of planet occurrence rate if we again include \citetalias{gilliland2000hstsearch}.

\section{Conclusions}\label{sec:conclusions}
Using the Dark Energy Camera on CTIO's 4-m Blanco telescope, we perform an extensive survey of the globular cluster 47~Tucanae for transiting hot Jupiters.
Our observing strategy allows us to obtain the coverage, photometric depth, and precision necessary to detect single full and partial transits of Jupiter radius planets.
Using a purpose-built data pipeline involving difference imaging, photometric calibration and detrending, transit injections, automated transit searching, and vetting, we carefully search 19,930 lightcurves for transit-like eclipses.

Though we find no planets in 47~Tuc, we are able to constrain the giant planet occurrence rate to an upper limit of $<$0.43\% at 95\% confidence.
When combined with the results of \citet{gilliland2000hstsearch}, we can place an even tighter limit of $<$0.11\%. 
We present detailed analysis of 39 interesting targets, consisting of 4 eclipsing binaries and 35 transit candidates.
Using our two-step vetting process, we are able to reject all 35 of these transit candidates as false positives without follow-up observations.

\section*{Acknowledgments}
The authors would like to thank our support astronomer Dr. Guillermo Damke; our observing support staff Claudio Aguilera, Alberto Alvarez, Manuel Hern\'andez, Rodrigo Hern\'andez, Jacqueline Ser\'on, and Hernan Tirado; and the rest of the staff at CTIO for their assistance and hospitality over the course of our observations.

We would also like to thank the referee for their helpful comments and suggestions.

Work by AC \& MP was supported by Louisiana Board of Regents Research Competitiveness Subaward LEQSF(2020-23)-RD-A-10. MP \& JK were supported by NASA award NNG16PJ32C.

This project used data obtained with the Dark Energy Camera (DECam), which was constructed by the Dark Energy Survey (DES) collaboration. Funding for the DES Projects has been provided by the US Department of Energy, the US National Science Foundation, the Ministry of Science and Education of Spain, the Science and Technology Facilities Council of the United Kingdom, the Higher Education Funding Council for England, the National Center for Supercomputing Applications at the University of Illinois at Urbana-Champaign, the Kavli Institute for Cosmological Physics at the University of Chicago, Center for Cosmology and Astro-Particle Physics at the Ohio State University, the Mitchell Institute for Fundamental Physics and Astronomy at Texas A\&M University, Financiadora de Estudos e Projetos, Fundação Carlos Chagas Filho de Amparo à Pesquisa do Estado do Rio de Janeiro, Conselho Nacional de Desenvolvimento Científico e Tecnológico and the Ministério da Ciência, Tecnologia e Inovação, the Deutsche Forschungsgemeinschaft and the Collaborating Institutions in the Dark Energy Survey.

The Collaborating Institutions are Argonne National Laboratory, the University of California at Santa Cruz, the University of Cambridge, Centro de Investigaciones Enérgeticas, Medioambientales y Tecnológicas–Madrid, the University of Chicago, University College London, the DES-Brazil Consortium, the University of Edinburgh, the Eidgenössische Technische Hochschule (ETH) Zürich, Fermi National Accelerator Laboratory, the University of Illinois at Urbana-Champaign, the Institut de Ciències de l’Espai (IEEC/CSIC), the Institut de Física d’Altes Energies, Lawrence Berkeley National Laboratory, the Ludwig-Maximilians Universität München and the associated Excellence Cluster Universe, the University of Michigan, NSF’s NOIRLab, the University of Nottingham, the Ohio State University, the OzDES Membership Consortium, the University of Pennsylvania, the University of Portsmouth, SLAC National Accelerator Laboratory, Stanford University, the University of Sussex, and Texas A\&M University.

Based on observations at Cerro Tololo Inter-American Observatory, NSF’s NOIRLab (NOIRLab Prop. IDs 2019A-0315, PI: M. Penny; 2022B-511337, PI: A. Crisp). This research also uses services or data provided by the Astro Data Lab, which is part of the Community Science and Data Center (CSDC) Program of NSF NOIRLab. NOIRLab is operated by the Association of Universities for Research in Astronomy (AURA), Inc. under a cooperative agreement with the U.S. National Science Foundation.

Raw imaging data were processed with the DECam Community Pipeline \citep{valdes2014decamcp}.

This work has made use of data from the European Space Agency (ESA) mission
{\it Gaia} (\url{https://www.cosmos.esa.int/gaia}), processed by the {\it Gaia}
Data Processing and Analysis Consortium (DPAC,
\url{https://www.cosmos.esa.int/web/gaia/dpac/consortium}). Funding for the DPAC
has been provided by national institutions, in particular the institutions
participating in the {\it Gaia} Multilateral Agreement.

The Pan-STARRS1 Surveys (PS1) and the PS1 public science archive have been made possible through contributions by the Institute for Astronomy, the University of Hawaii, the Pan-STARRS Project Office, the Max-Planck Society and its participating institutes, the Max Planck Institute for Astronomy, Heidelberg and the Max Planck Institute for Extraterrestrial Physics, Garching, The Johns Hopkins University, Durham University, the University of Edinburgh, the Queen's University Belfast, the Harvard-Smithsonian Center for Astrophysics, the Las Cumbres Observatory Global Telescope Network Incorporated, the National Central University of Taiwan, the Space Telescope Science Institute, the National Aeronautics and Space Administration under Grant No. NNX08AR22G issued through the Planetary Science Division of the NASA Science Mission Directorate, the National Science Foundation Grant No. AST-1238877, the University of Maryland, Eotvos Lorand University (ELTE), the Los Alamos National Laboratory, and the Gordon and Betty Moore Foundation.

This publication uses data generated via the \href{https://www.zooniverse.org/}{Zooniverse} platform, development of which is funded by generous support, including a Global Impact Award from Google, and by a grant from the Alfred P. Sloan Foundation.

\facilities{Blanco (DECam)}

\software{Astropy \citep{astropy:2013, astropy:2018, astropy:2022}, CFITSIO \citep{pence1999cfitsio}, DoPHOT \citep{mateo1989dophot, schechter1993dophot}, FITSH \citep{pal2012fitsh}, ISIS \citep{alardlupton1998isis, alard2000isis}, JKTEBOP \citep{southworth2013jktebop1}, JKTLD \citep{southworth2008jktld}, matplotlib \citep{Hunter2007matplotlib}, NumPy \citep{harris2020numpy}, pandas \citep{reback2020pandas}, Source Extractor \citep{bertin1996tractor}, VARTOOLS \citep{hartman2016vartools}, WCSTOOLS \citep{mink1998wcstools1, mink2019wcstools2}}

\bibliography{47Tuc}{}
\bibliographystyle{aasjournalv7}

\appendix
\begin{turnpage}
\section{Injection Grids}\label{app:grids}
We performed transit injections into detrended lightcurves with planet parameters in the ranges $0.5~R_{\rm Jup} \leq R_{\rm p}  \leq 2.0~R_{\rm Jup}$ and $0.5 \leq P \leq 10.0$~days. The following plots show examples of our injections for bins \rplan over $P$ and $r$. For the $P$ grid, shown in Figure \ref{fig:rppgrid}, only full transits are shown. For the $r$ grids, shown in Figures \ref{fig:rprfullgrid} \& \ref{fig:rprpartgrid}, respectively, we show both full and partial transits.

    \begin{figure*}[htbp]
        \centering
        \includegraphics[width=\linewidth]{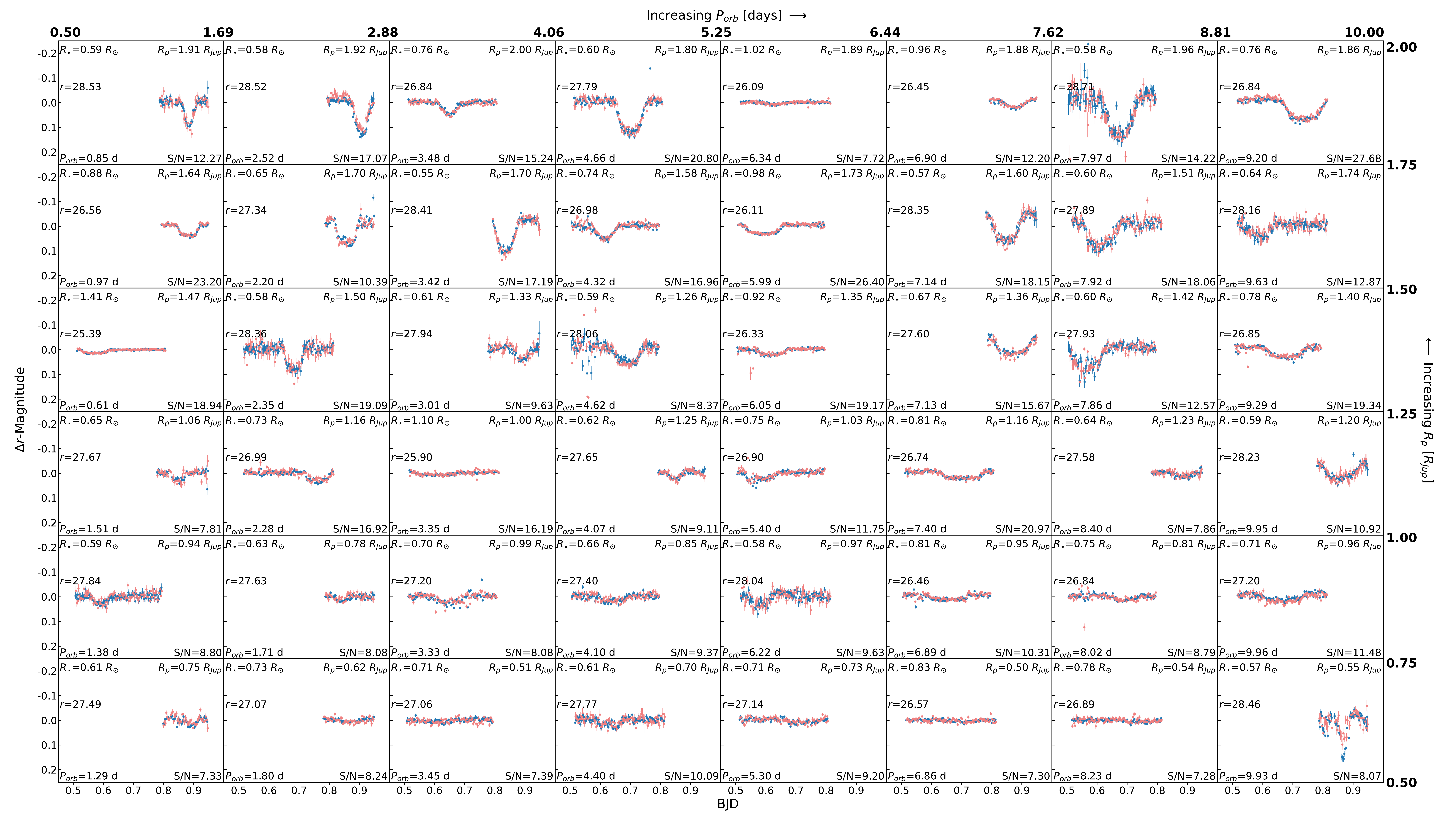}
        \caption{Examples of injected transits that pass our S/N=7 detection threshold, spanning the range of $R_{\rm p}$ and $P$ that we explore. Each plot falls within one of our $P$ and $R_{\rm p}$ bins, with binned $P$ increasing from left to right and binned $R_{\rm p}$ increasing from bottom to top. The bin edges are denoted in bold along the top and right. $r-$band data are shown in blue, while $z-$band data are shown in red.}
        \label{fig:rppgrid}
    \end{figure*}

    \begin{figure*}[htbp]
        \centering
        \includegraphics[width=\linewidth]{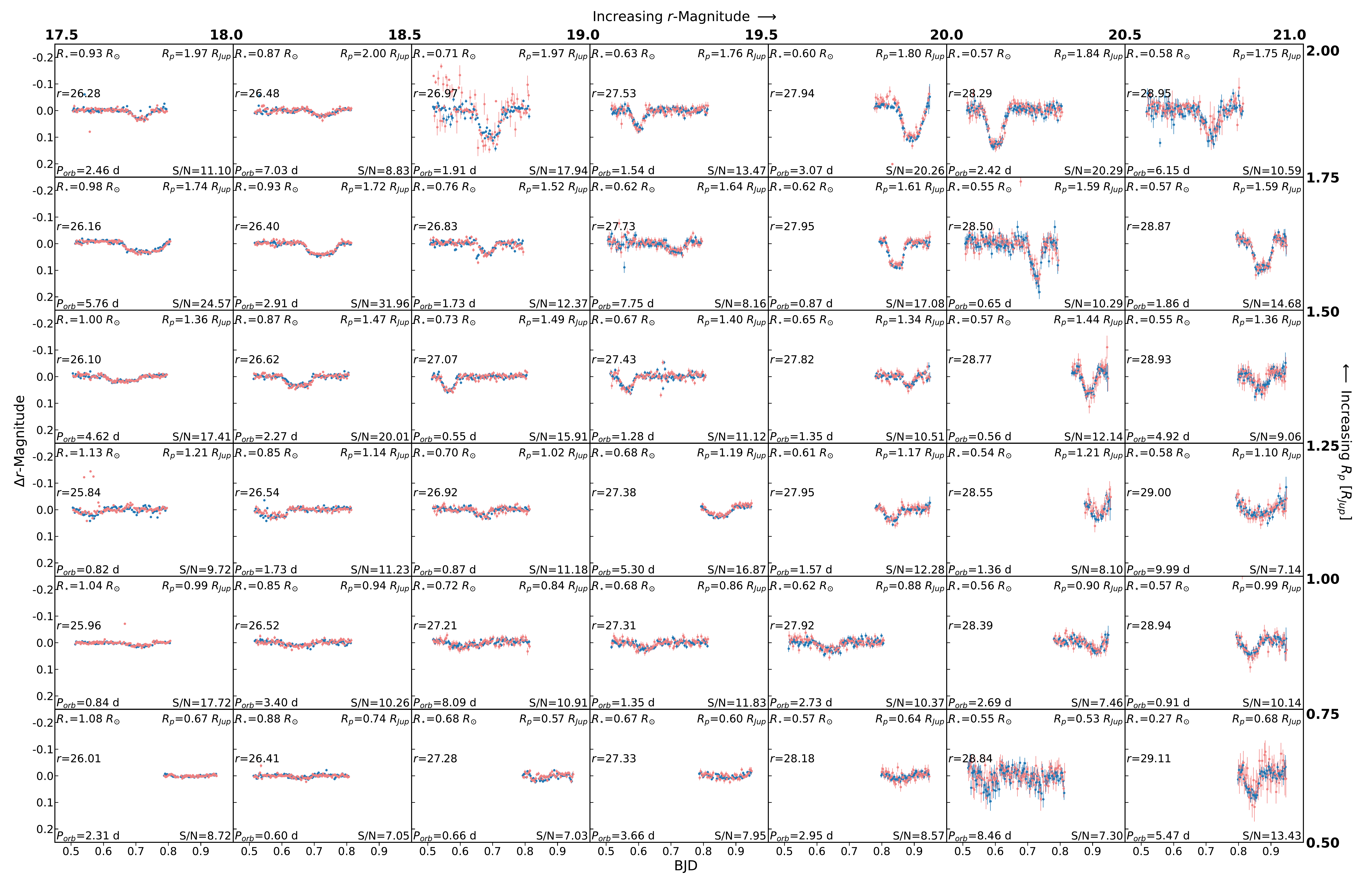}
        \caption{Examples of injected full transits that pass our S/N=7 detection threshold, spanning the range of $R_{\rm p}$ and $r$ that we explore. Each plot falls within one of our $r$ and $R_{\rm p}$ bins, with binned $r$ increasing from left to right and binned $R_{\rm p}$ increasing from bottom to top. The bin edges are denoted in bold along the top and right. $r-$band data are shown in blue, while $z-$band data are shown in red.}
        \label{fig:rprfullgrid}
    \end{figure*}

    \begin{figure*}[htbp]
        \centering
        \includegraphics[width=\linewidth]{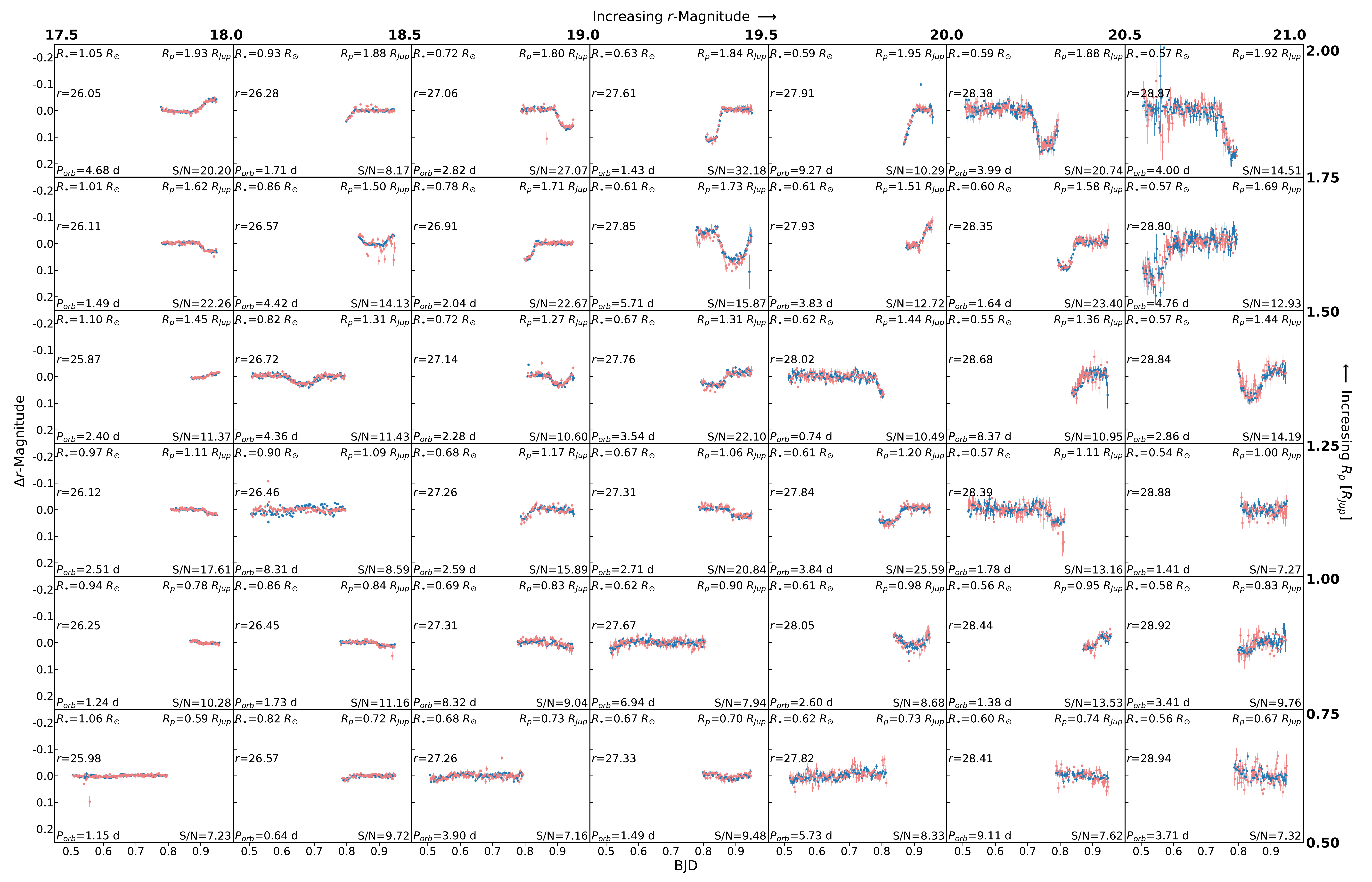}
        \caption{Examples of injected partial transits that pass our S/N=7 detection threshold, spanning the range of $R_{\rm p}$ and $r$ that we explore. Each plot falls within one of our $r$ and $R_{\rm p}$ bins, with binned $r$ increasing from left to right and binned $R_{\rm p}$ increasing from bottom to top. The bin edges are denoted in bold along the top and right. $r-$band data are shown in blue, while $z-$band data are shown in red.}
        \label{fig:rprpartgrid}
    \end{figure*}
\end{turnpage}

\clearpage
\section{Additional Candidate Information}\label{app:candinfo}
The following plots show the lightcurves and images for each of our 35 non-EB candidates.
In each plot, the left panel is the lightcurve of a night with a detection, the middle panel is the in-transit difference image stack, and the right panel is the reference image.
In the lightcurve panels, the $r-$band data are plotted in blue, the $z-$band data are plotted in red,  the boxcar model from the search is plotted in solid green, and the median magnitude is plotted in dashed green.
The values displayed on the plot are --- clockwise from top left --- the estimated stellar radius, estimated planet radius, transit duration, transit depth, and S/N of the detection.
For candidates with multiple detections, there is one row of plots per detection.

 \begin{figure*}[b]
	\centering
	\includegraphics[width=0.49\textwidth]{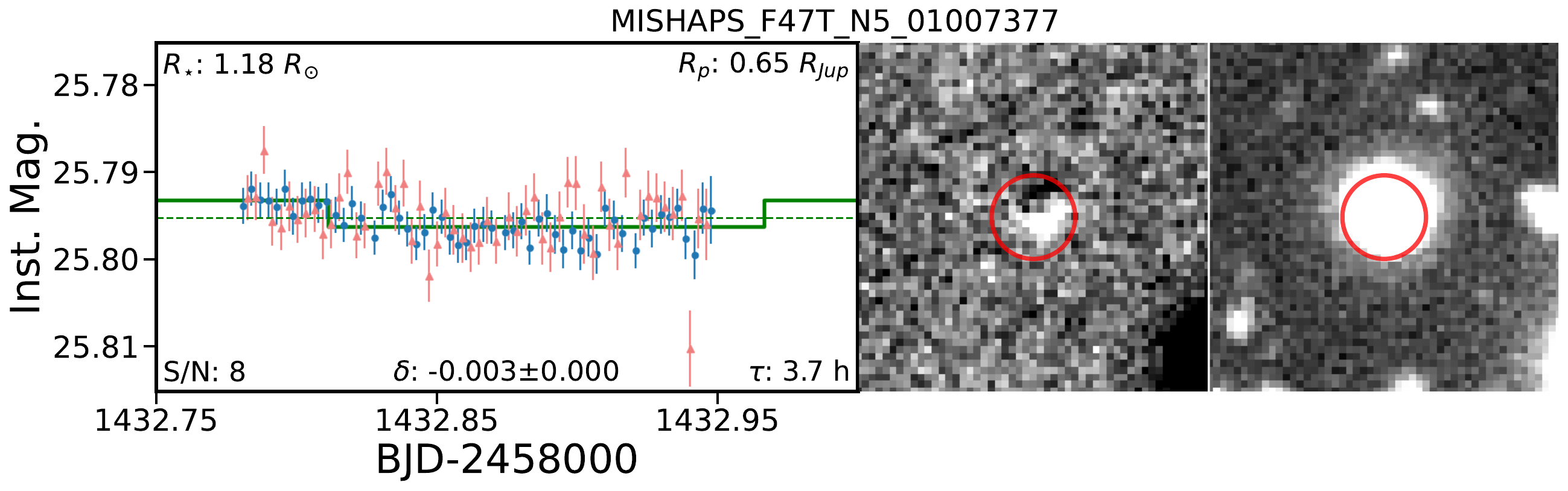}
	\includegraphics[width=0.49\textwidth]{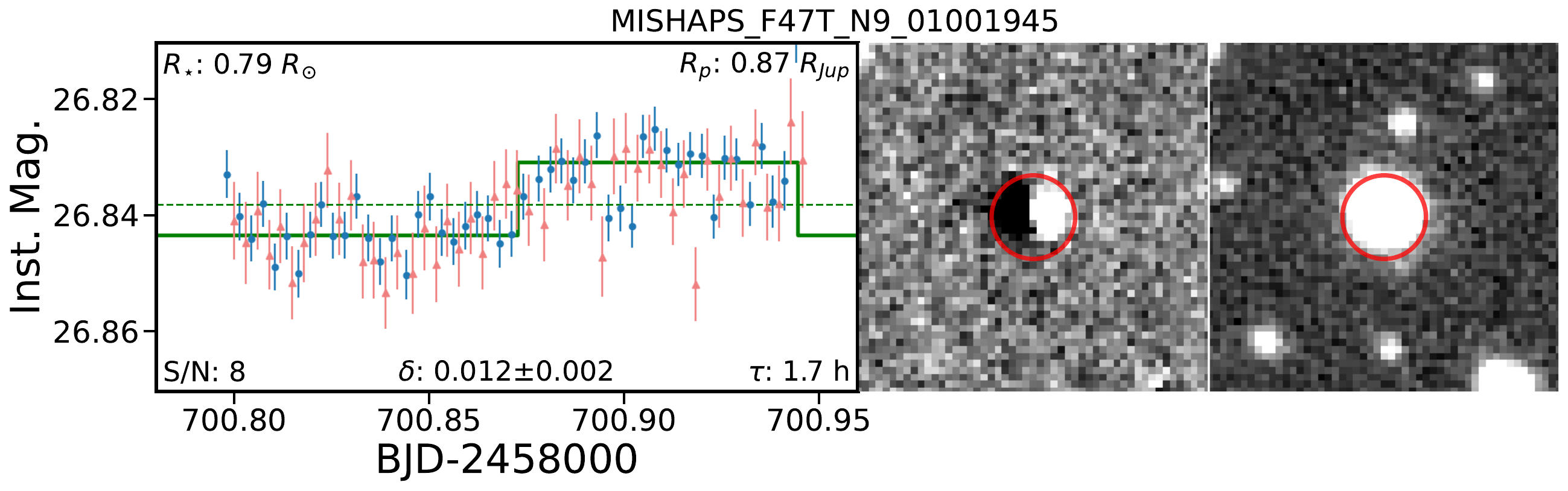}\\
	\includegraphics[width=0.49\textwidth]{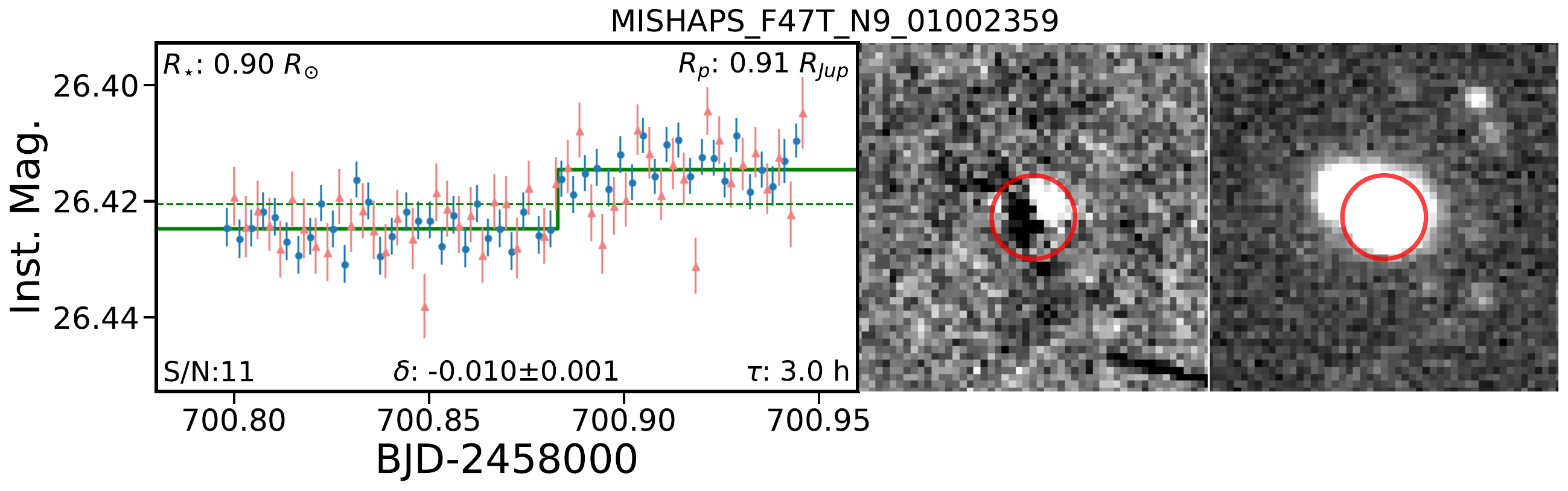}
	\includegraphics[width=0.49\textwidth]{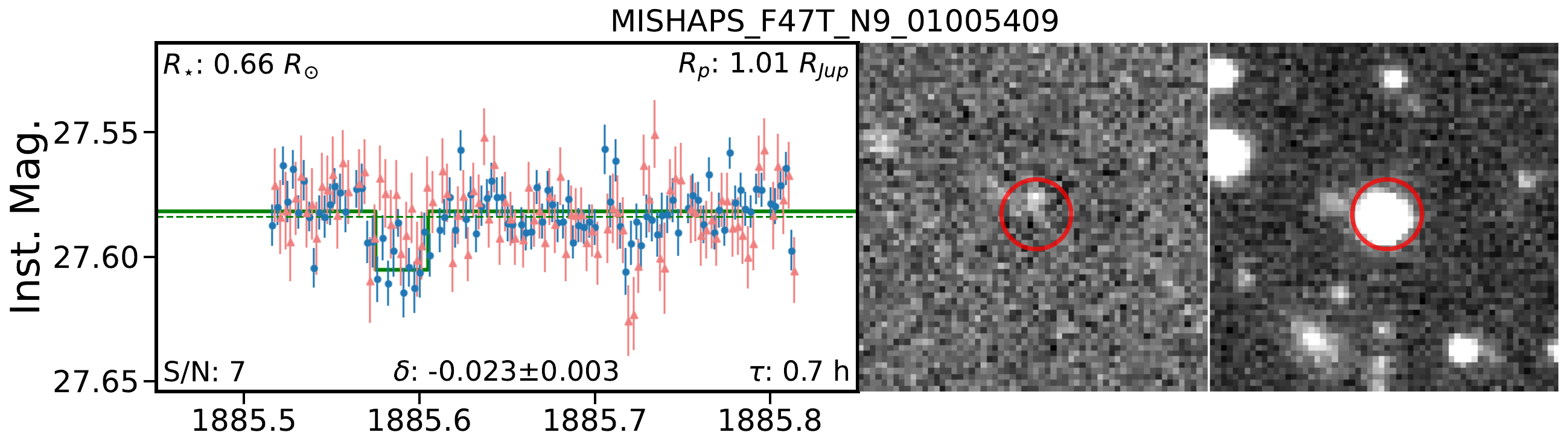}\\
	\includegraphics[width=0.49\textwidth]{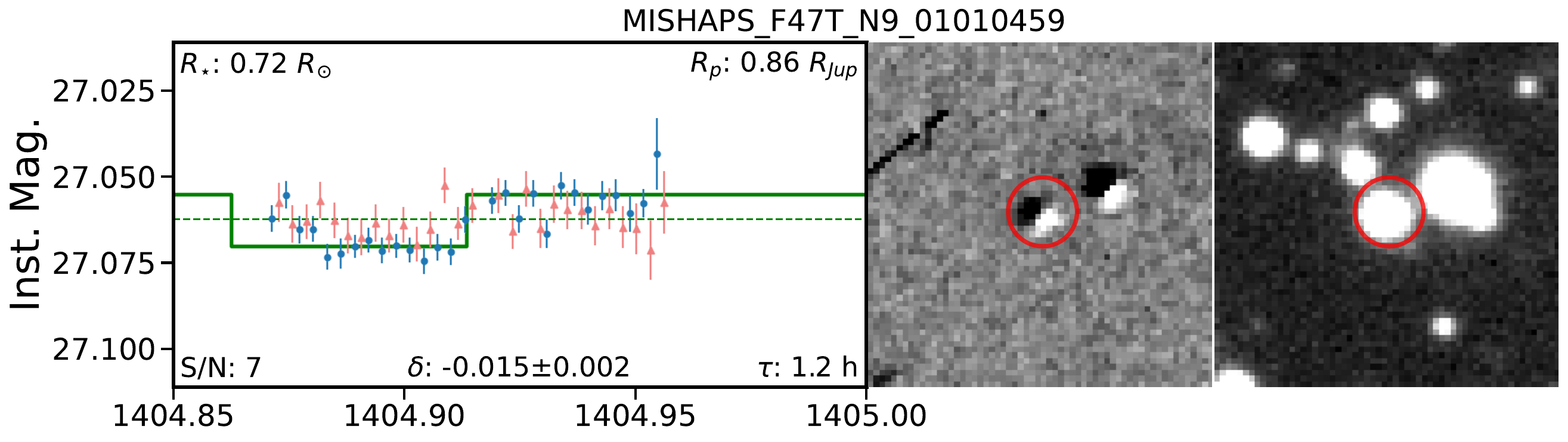}
	\includegraphics[width=0.49\textwidth]{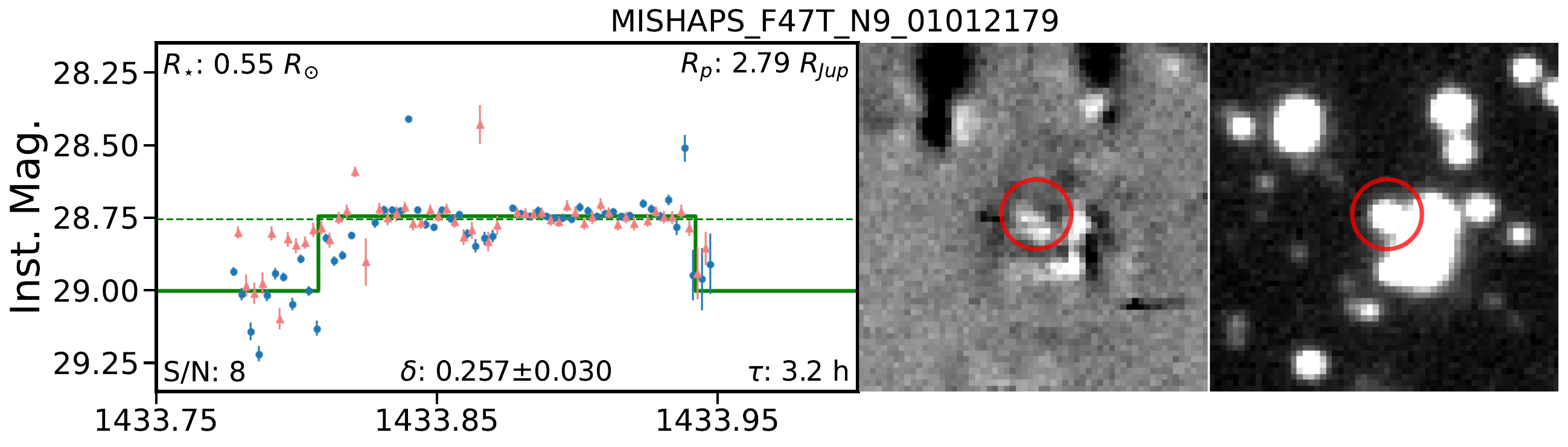}\\
	\includegraphics[width=0.49\textwidth]{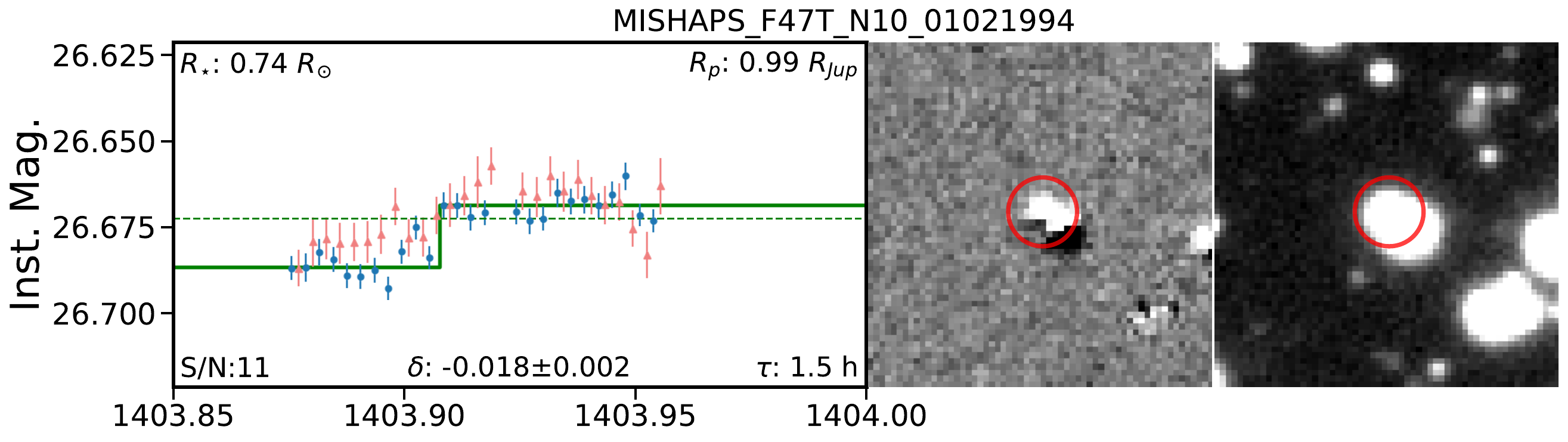}
	\includegraphics[width=0.49\textwidth]{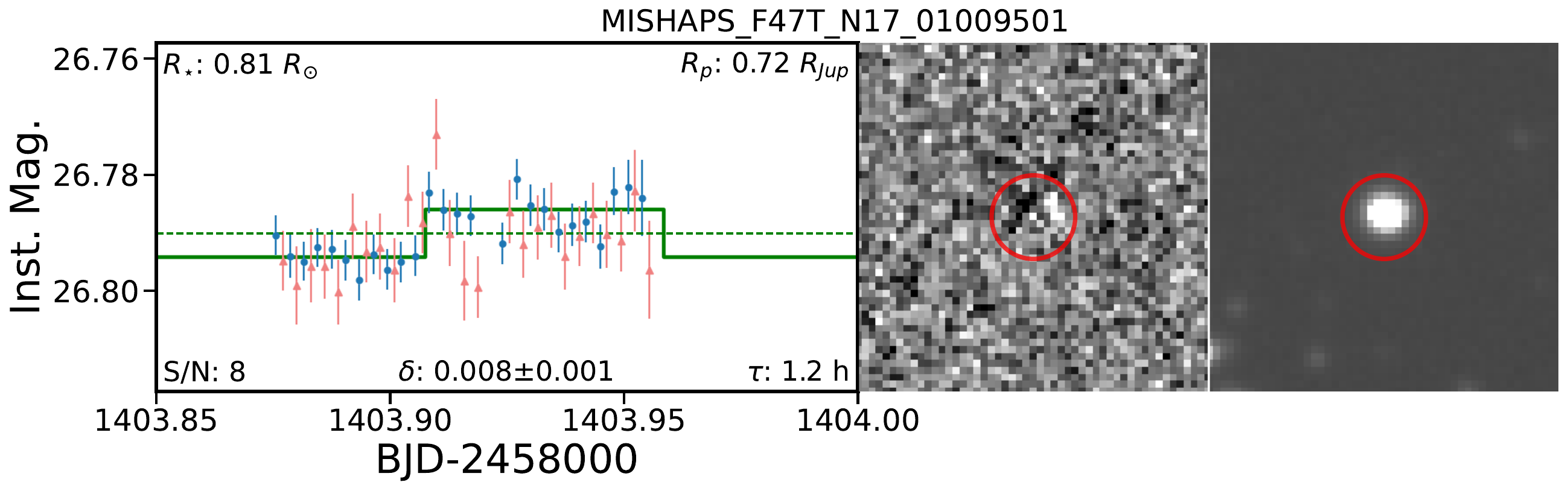}\\
	\includegraphics[width=0.49\textwidth]{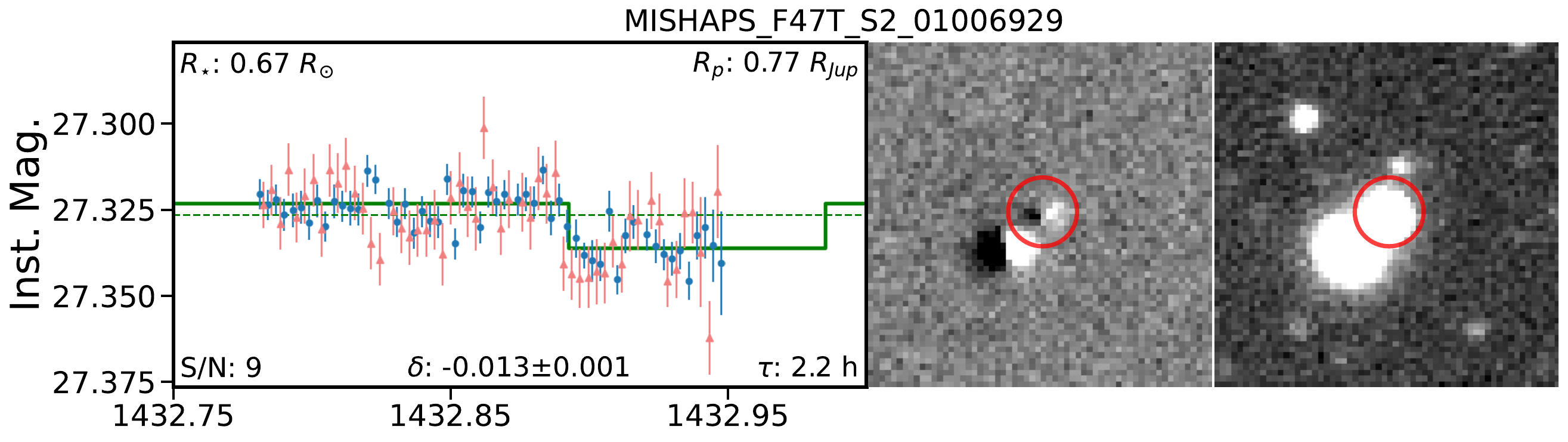}
	\includegraphics[width=0.49\textwidth]{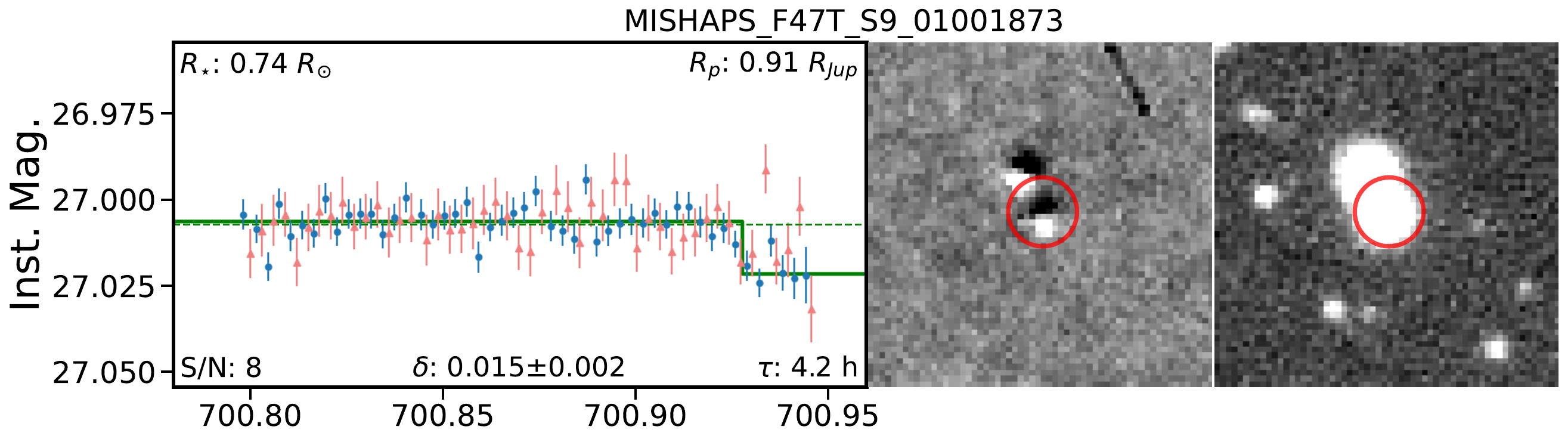}\\
	\includegraphics[width=0.49\textwidth]{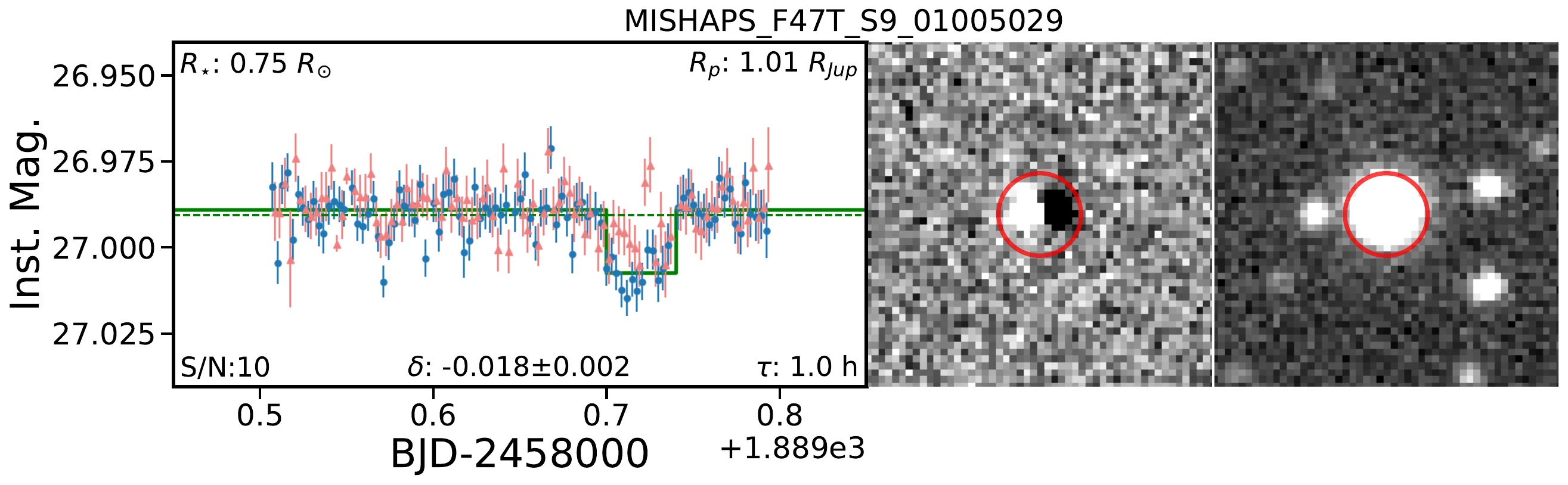}
	\includegraphics[width=0.49\textwidth]{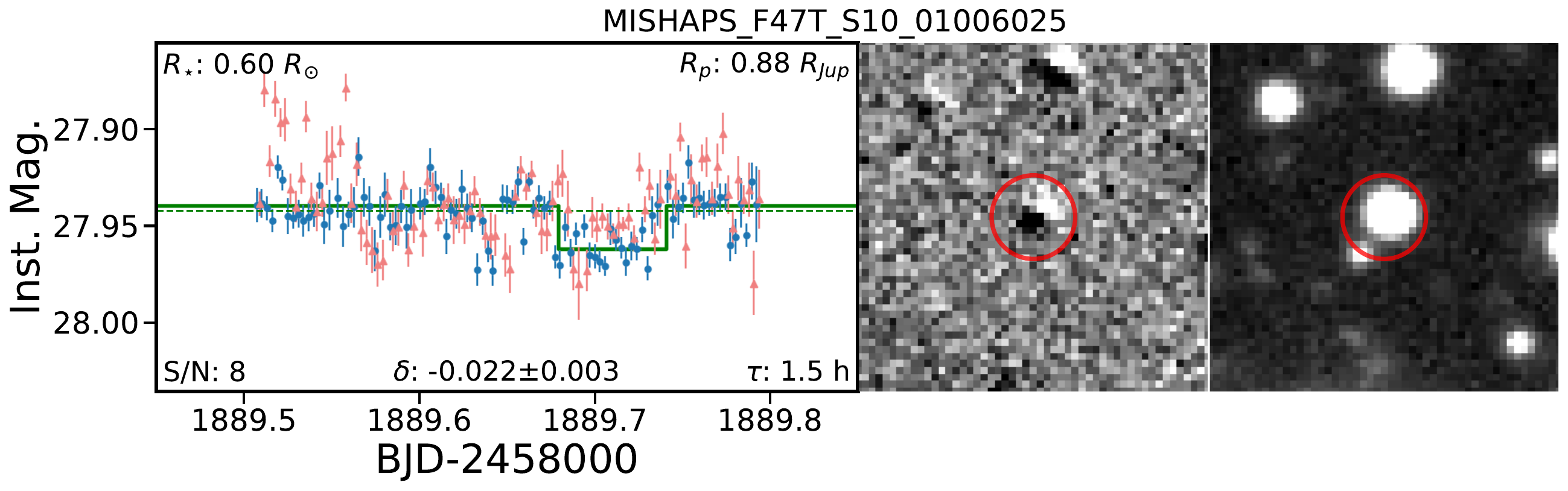}\\
	
	\caption{Detection lightcurves (left column), stacked in-transit difference images (middle column), and $r-$band reference images (right column) for rejected candidates N5\_01007377, N9\_01001945, N9\_01002359, N9\_01005409, N9\_01010459. N9\_01012179, N10\_01021994, N17\_01009501, S2\_01006929, S9\_01001873, S9\_01005029, and S10\_01006025.}
\end{figure*}

\begin{figure*}[htb]
	\centering      
	\includegraphics[width=0.49\textwidth]{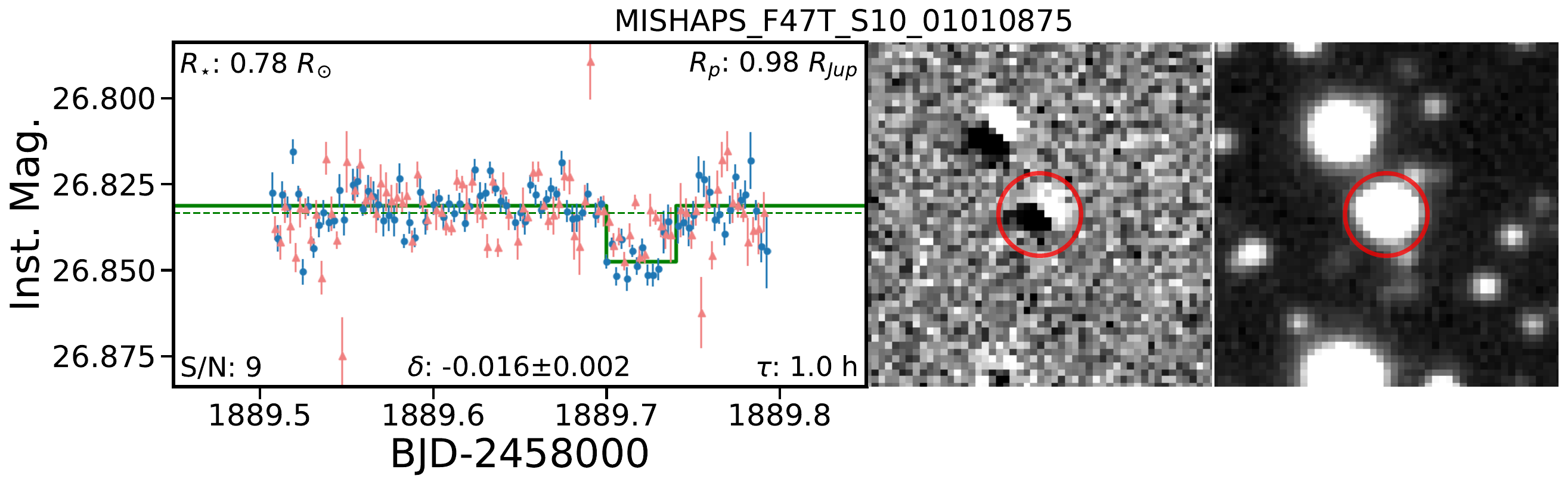}
	\includegraphics[width=0.49\textwidth]{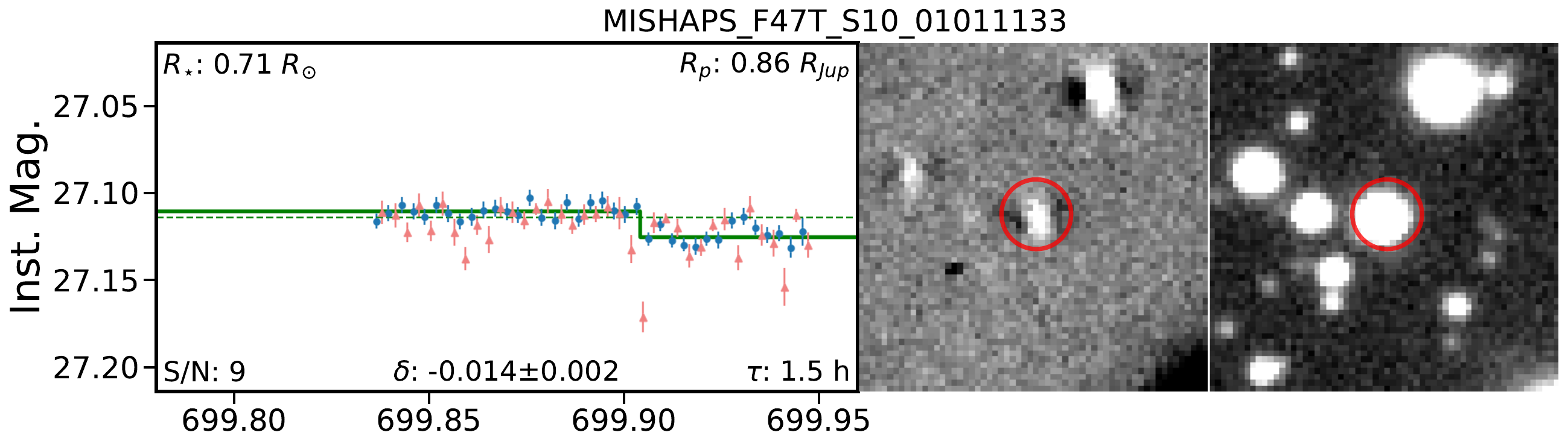}\\
	\includegraphics[width=0.49\textwidth]{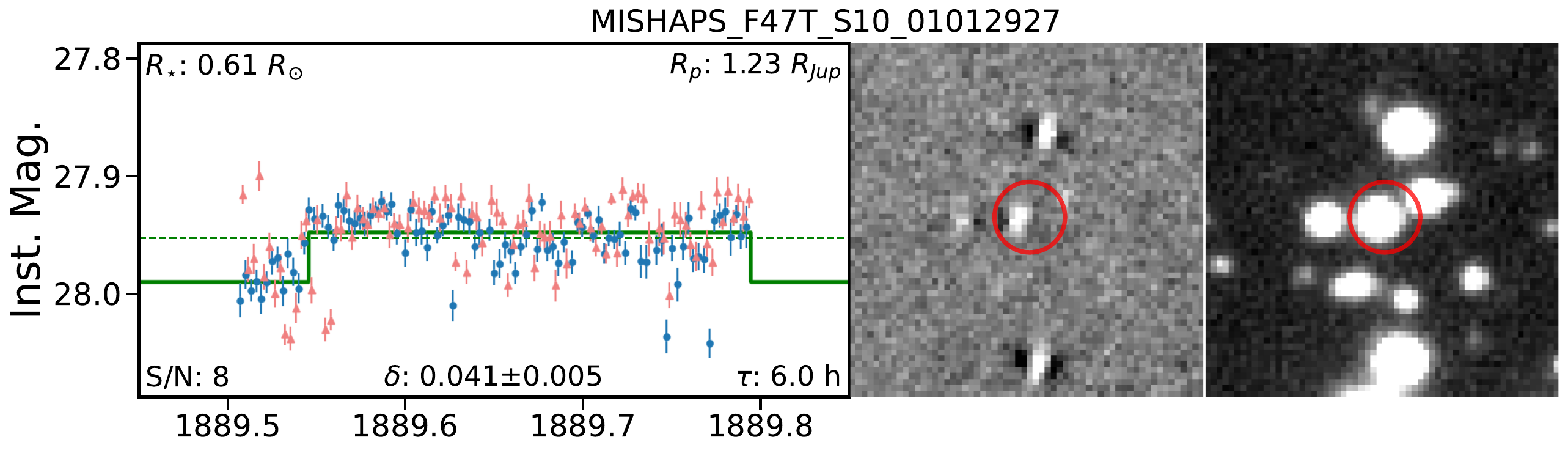}
	\includegraphics[width=0.49\textwidth]{MISHAPS_F47T_S10_01010875.app_plot.pdf}\\
	\includegraphics[width=0.49\textwidth]{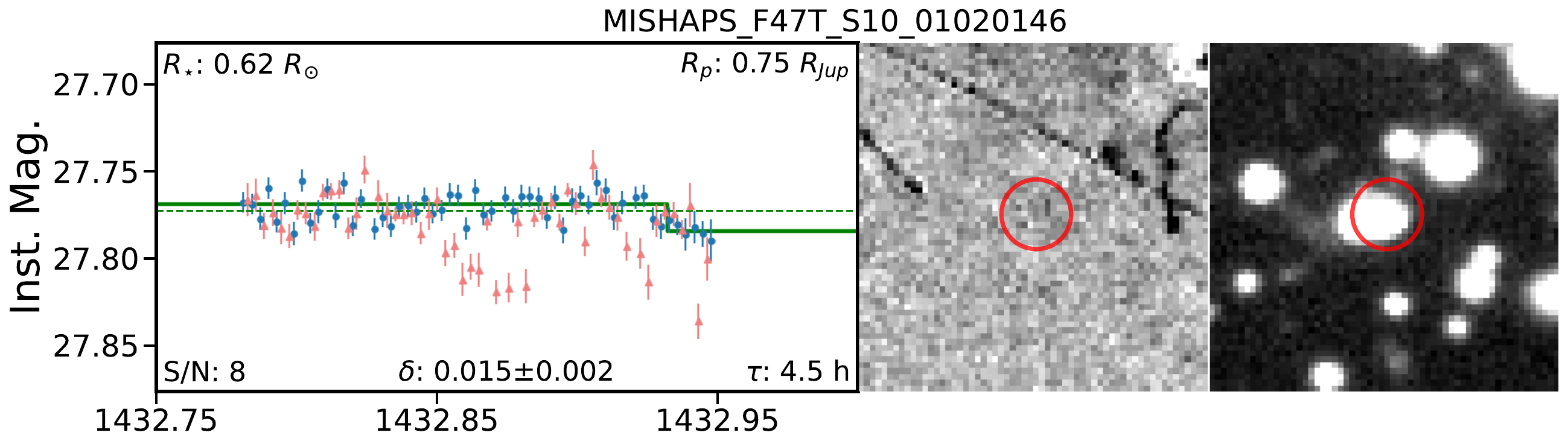}
	\includegraphics[width=0.49\textwidth]{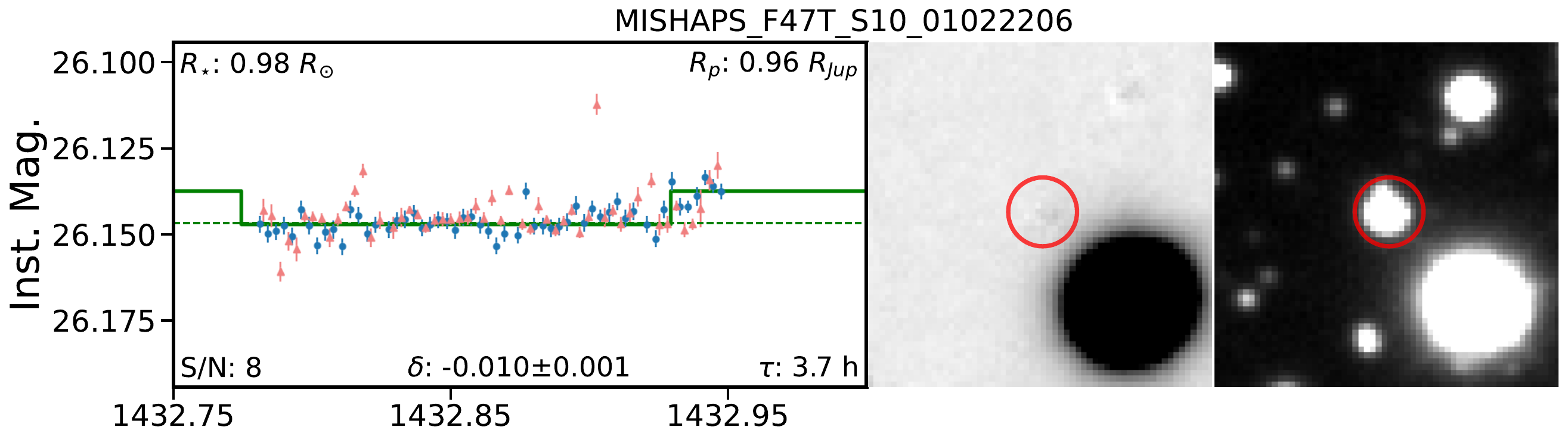}\\
	\includegraphics[width=0.49\textwidth]{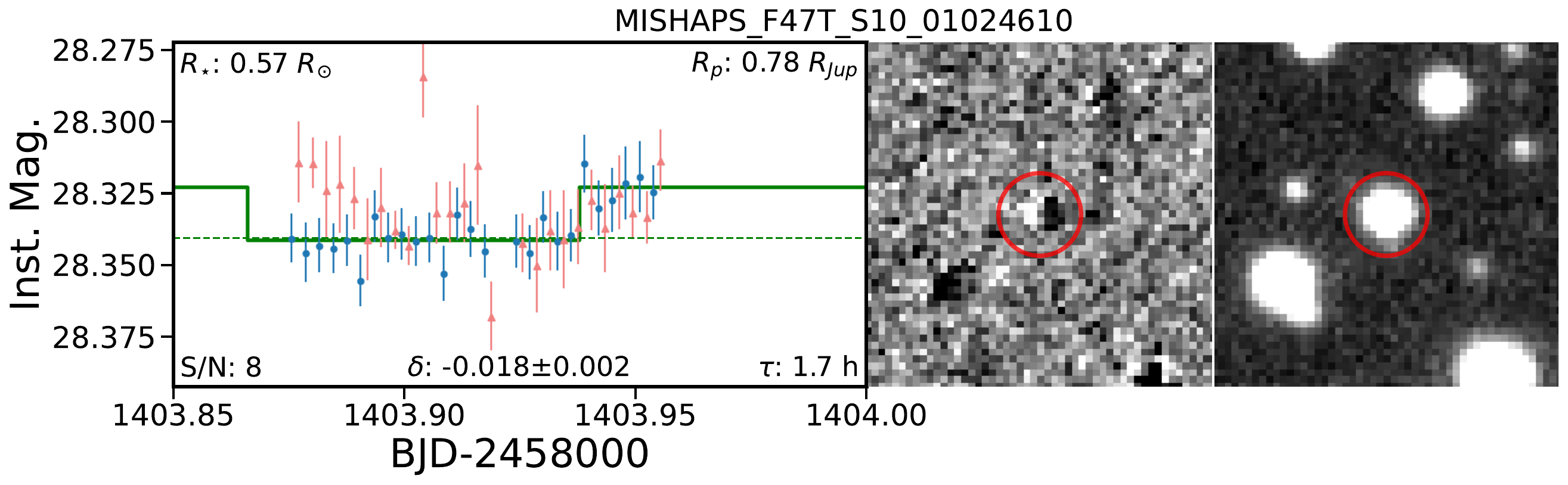}
	\includegraphics[width=0.49\textwidth]{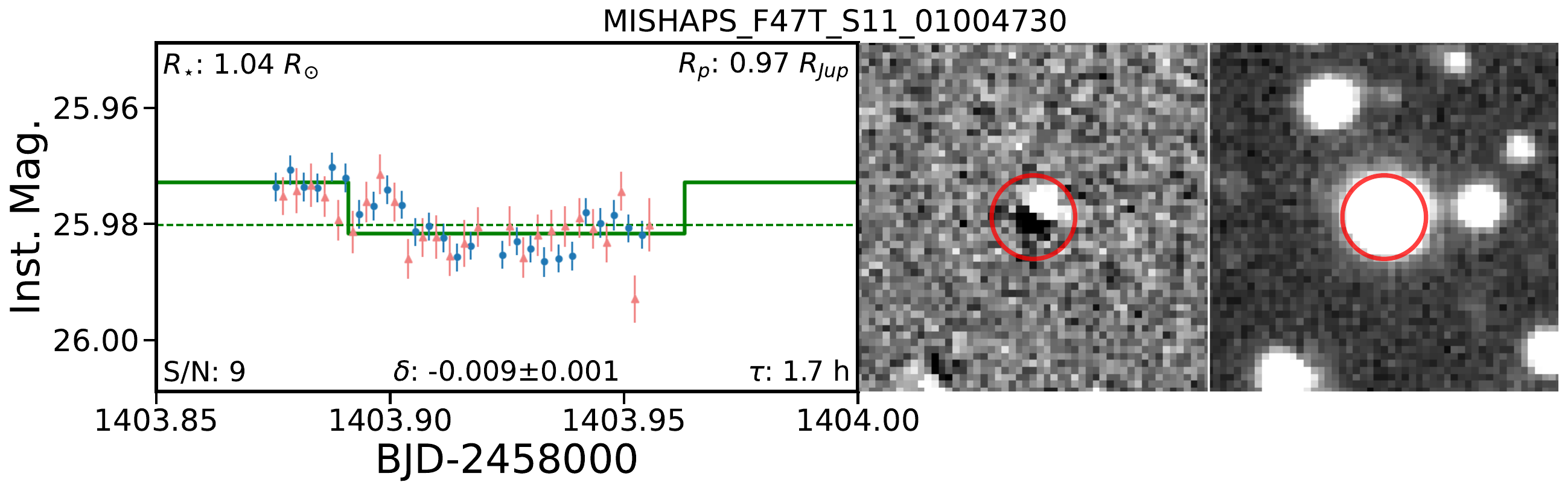}\\
	\includegraphics[width=0.49\textwidth]{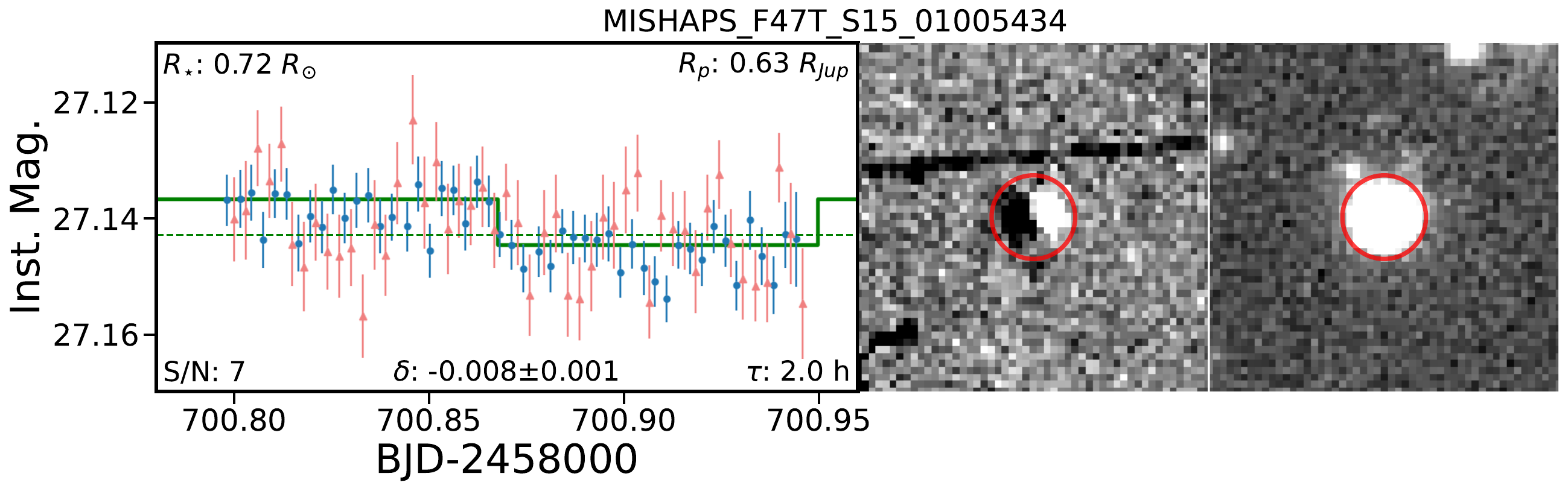}
	\includegraphics[width=0.49\textwidth]{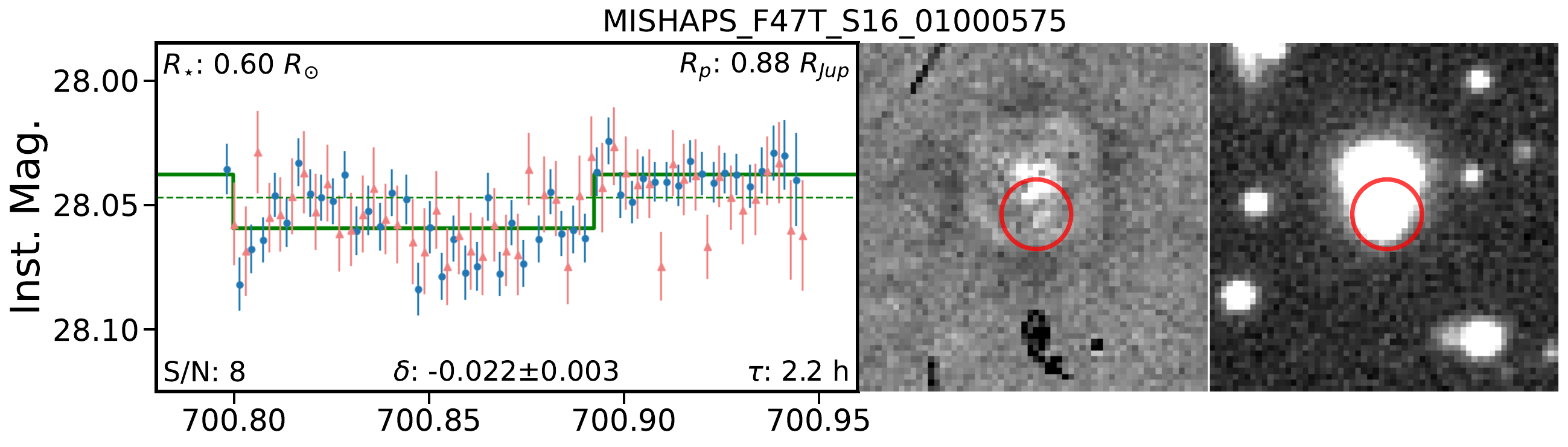}\\
	\includegraphics[width=0.49\textwidth]{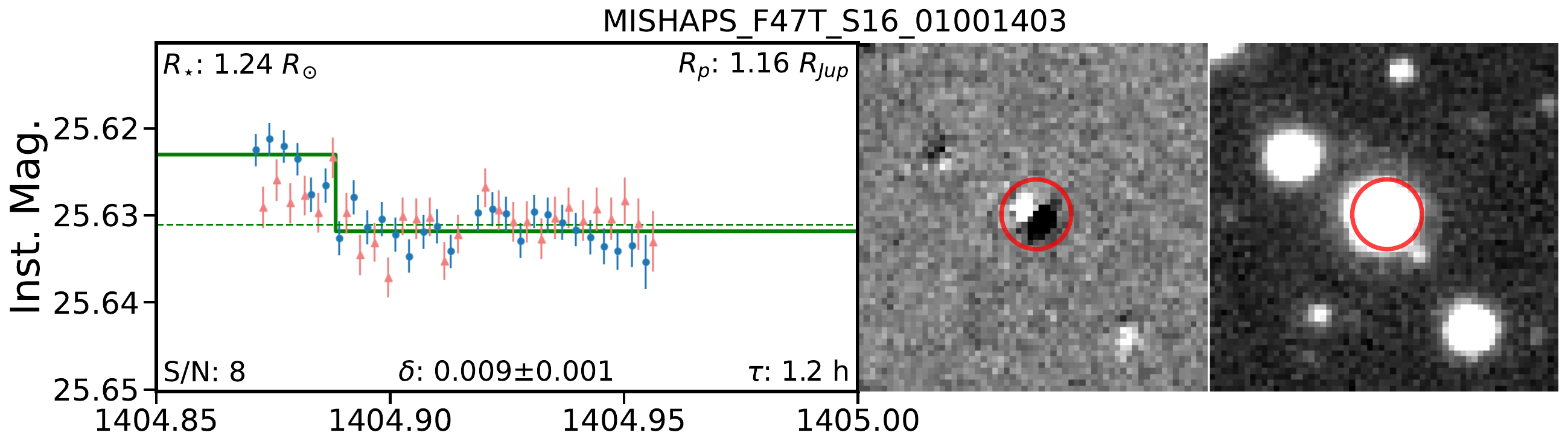}
	\includegraphics[width=0.49\textwidth]{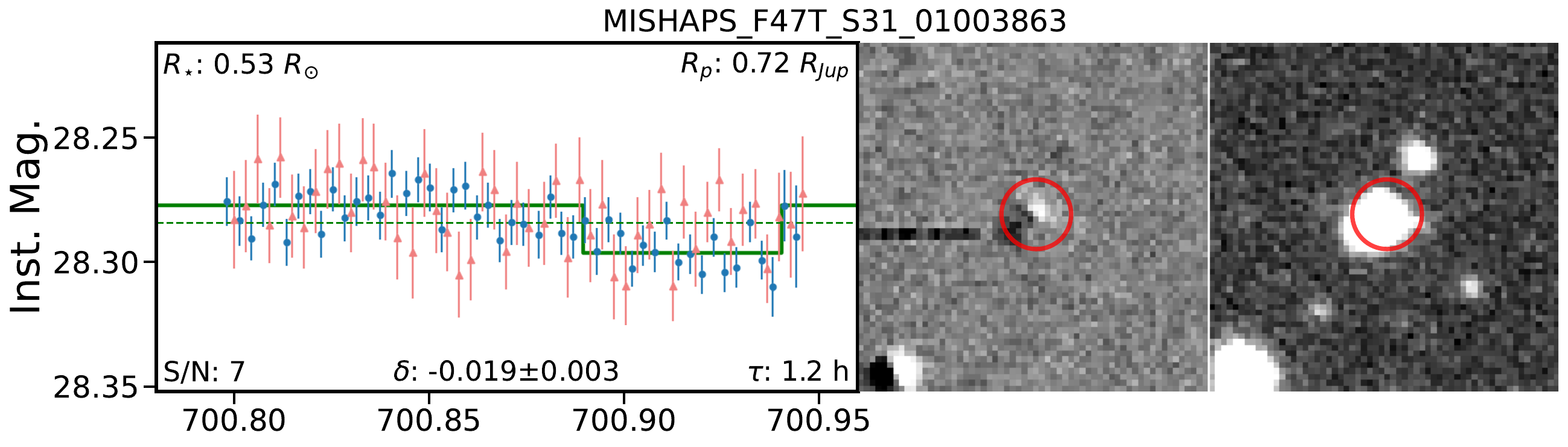}\\
	\caption{Same as previous, for rejected candidates S10\_01010875, S10\_01011133, S10\_01012927, S10\_01010875, S10\_01020146, S11\_01004730, S15\_01005434, S16\_01000575, S16\_01001403, and S31\_01003863.}
\end{figure*}

\begin{figure*}[htb]
	\centering
	\includegraphics[width=0.49\textwidth]{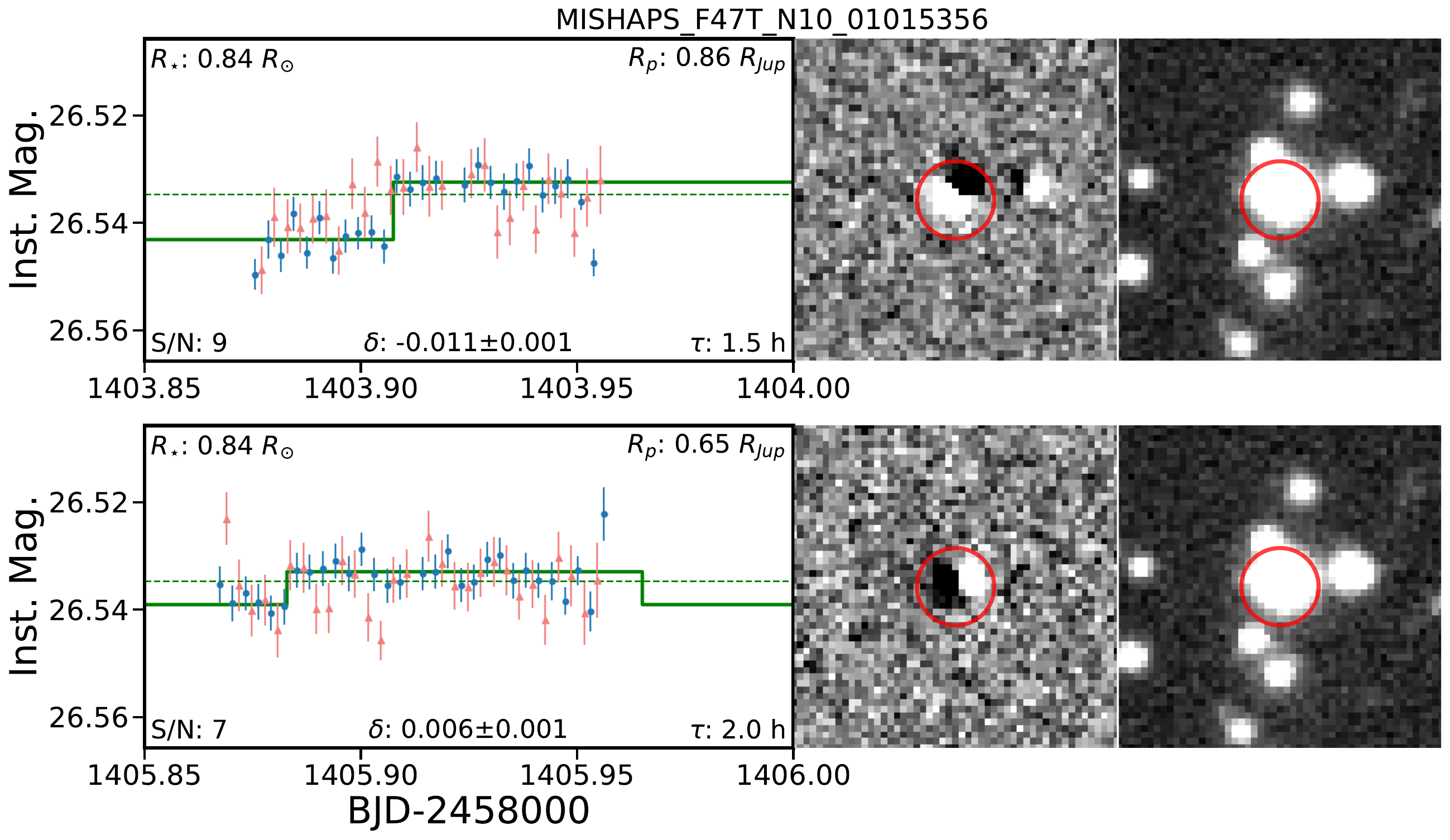}
	\includegraphics[width=0.49\textwidth]{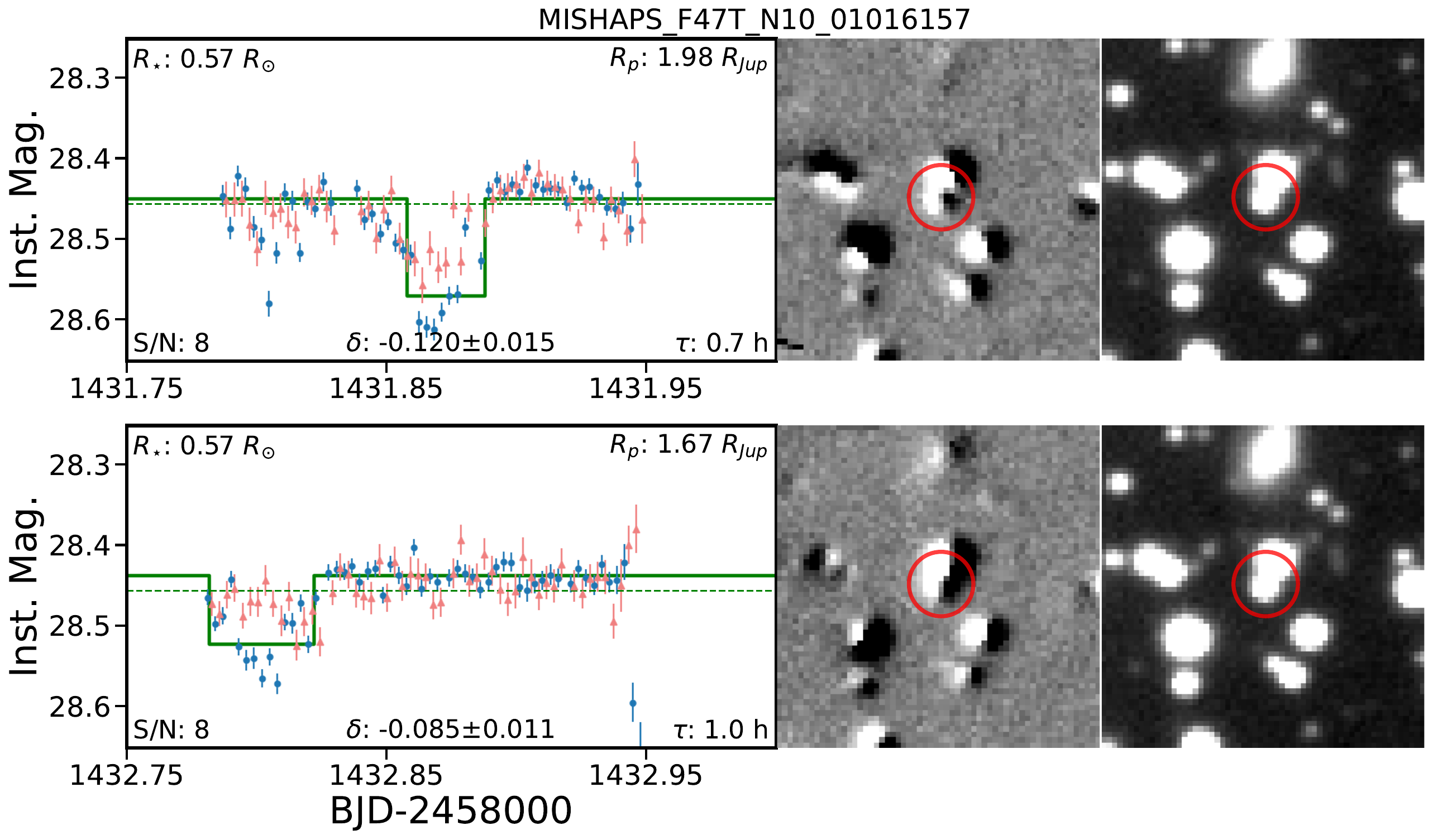}\\
	\includegraphics[width=0.49\textwidth]{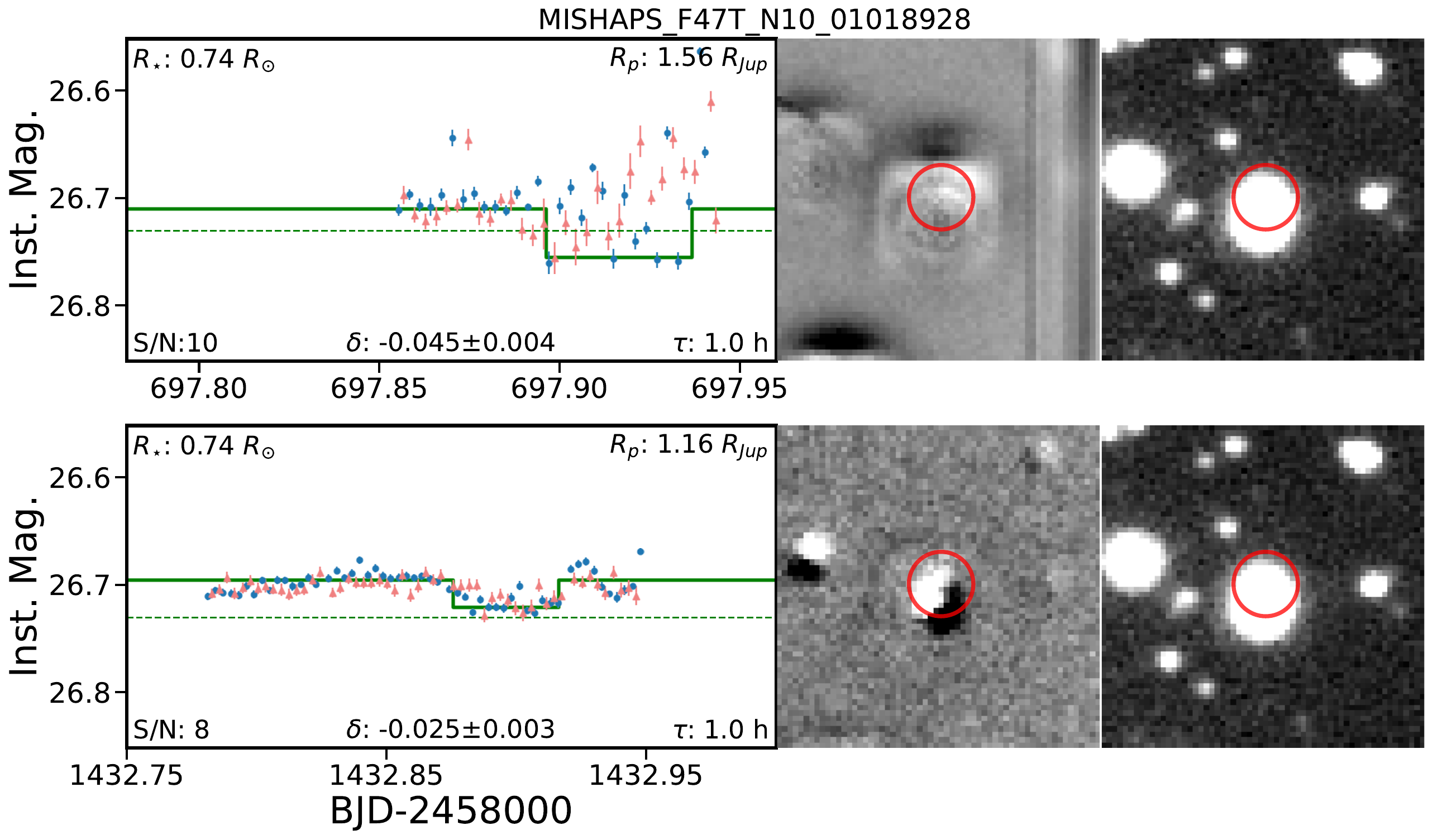}
	\includegraphics[width=0.49\textwidth]{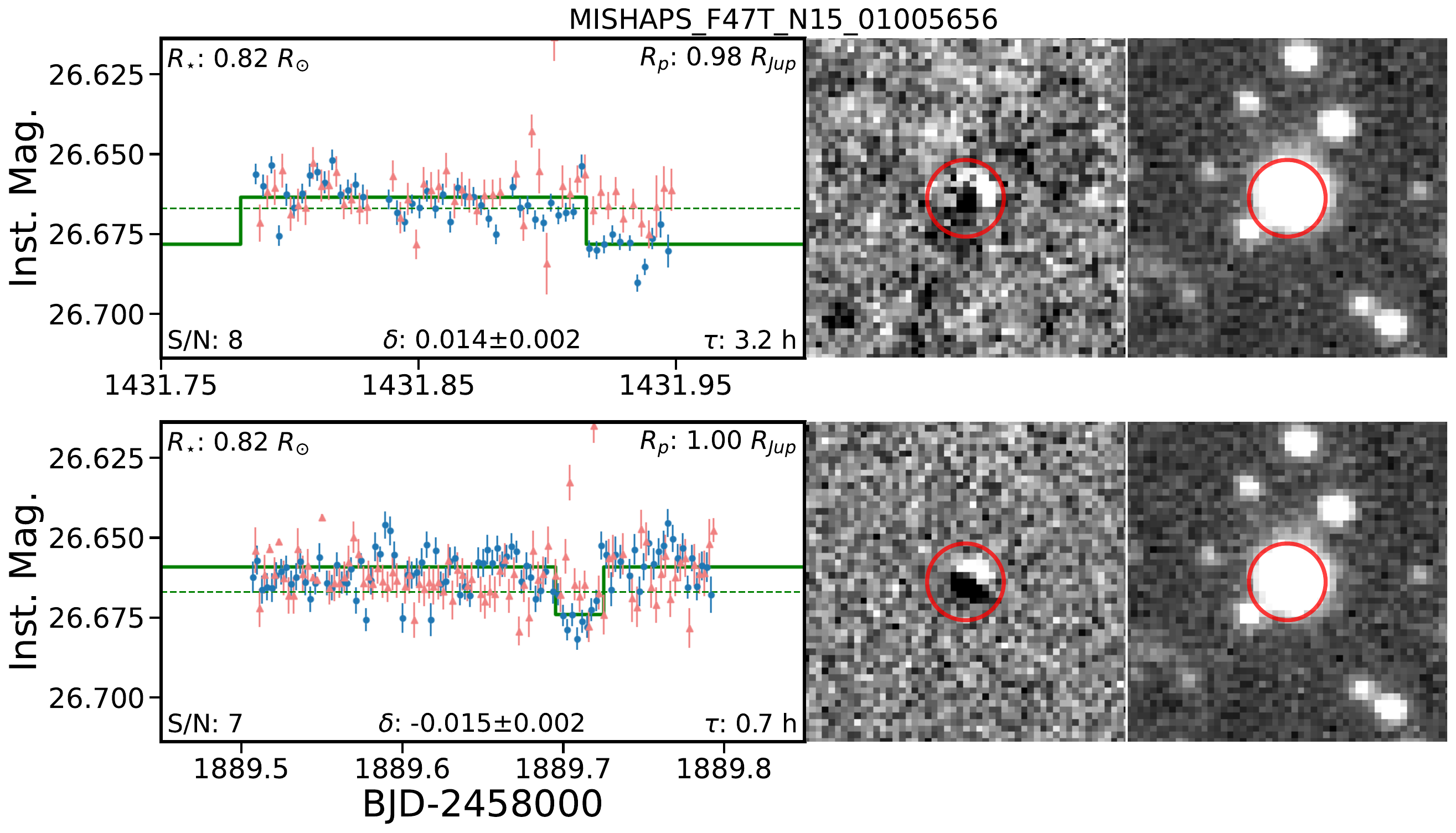}\\
	\includegraphics[width=0.49\textwidth]{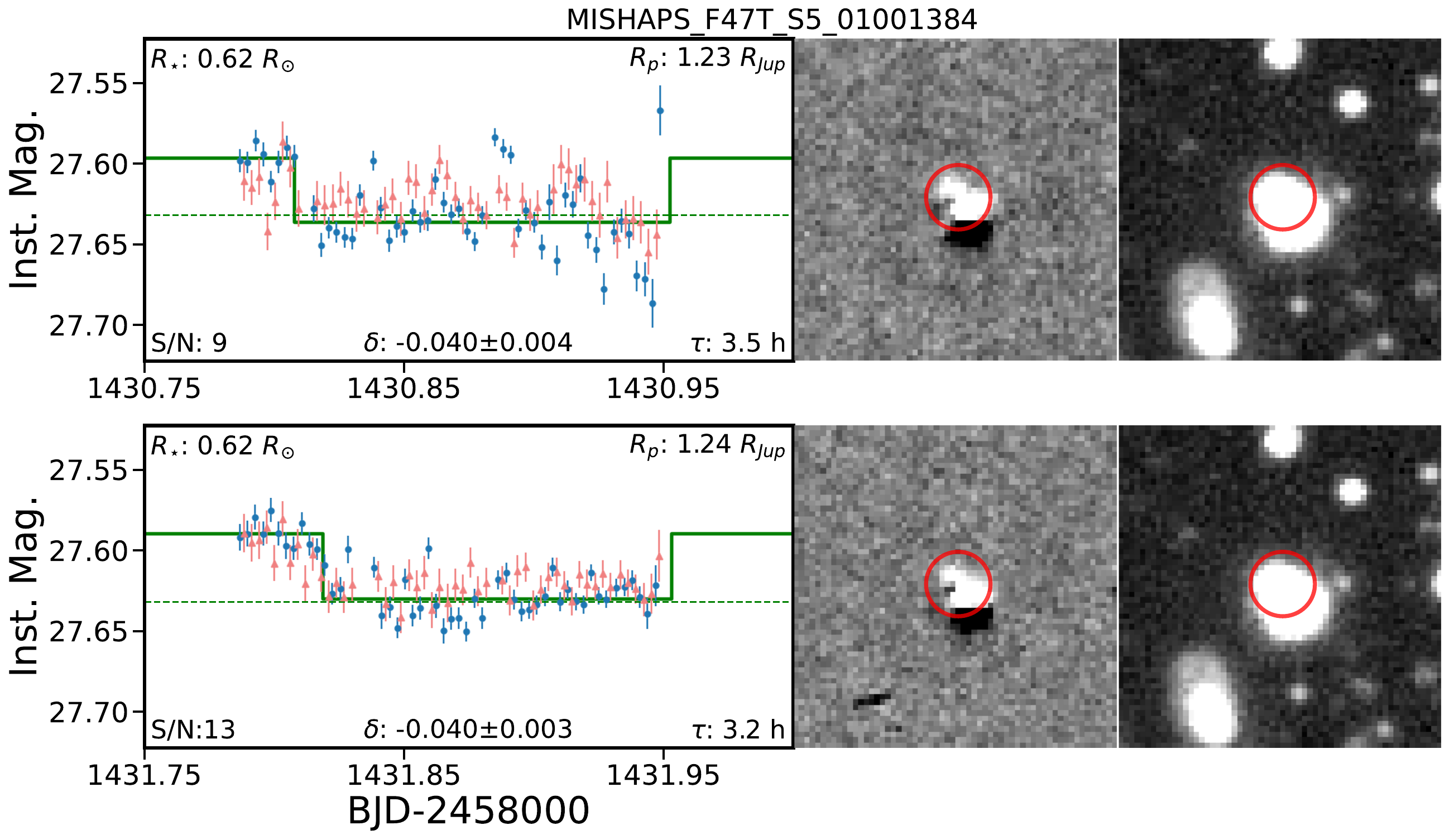}
	\includegraphics[width=0.49\textwidth]{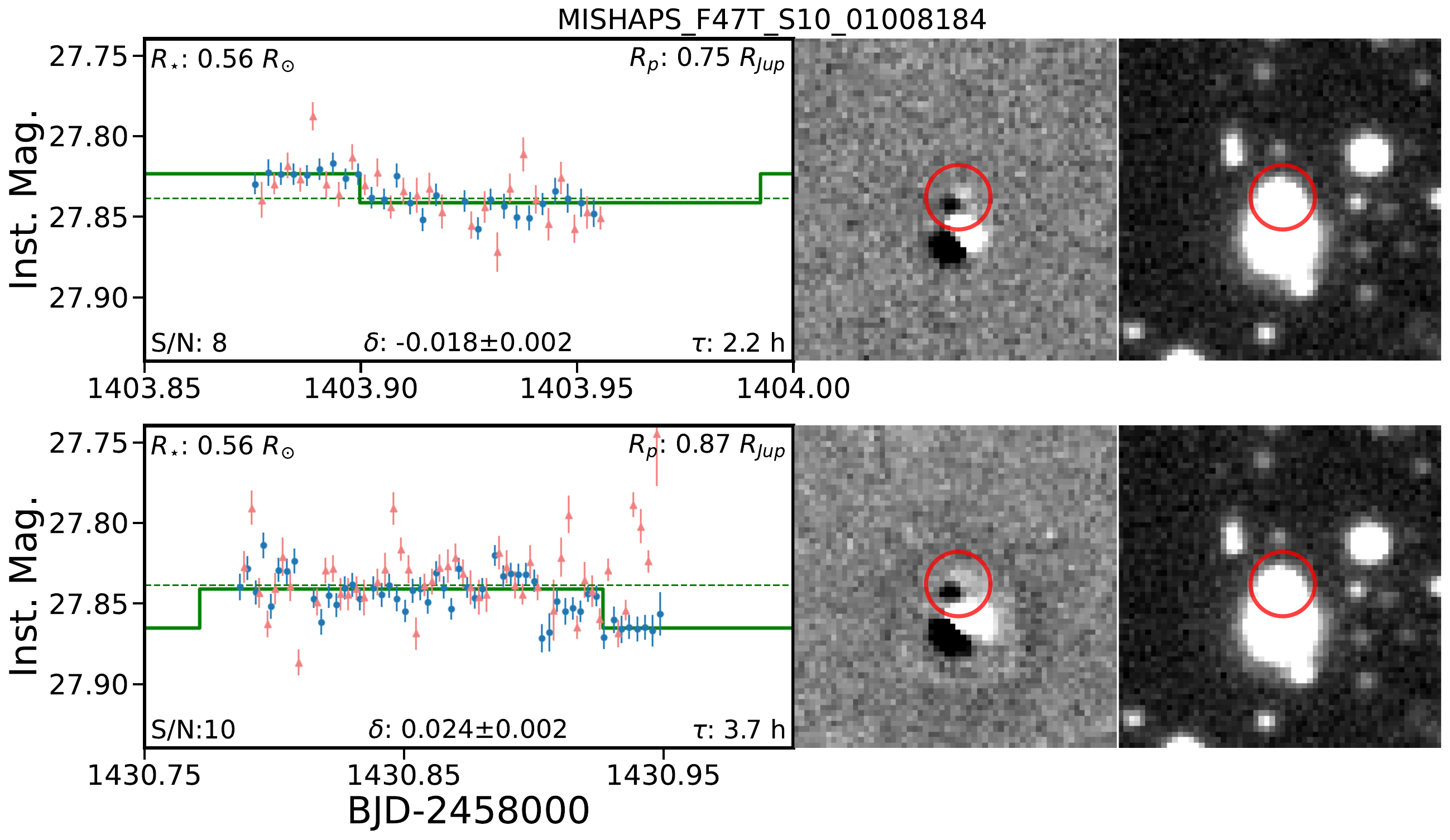}\\
	\includegraphics[width=0.49\textwidth]{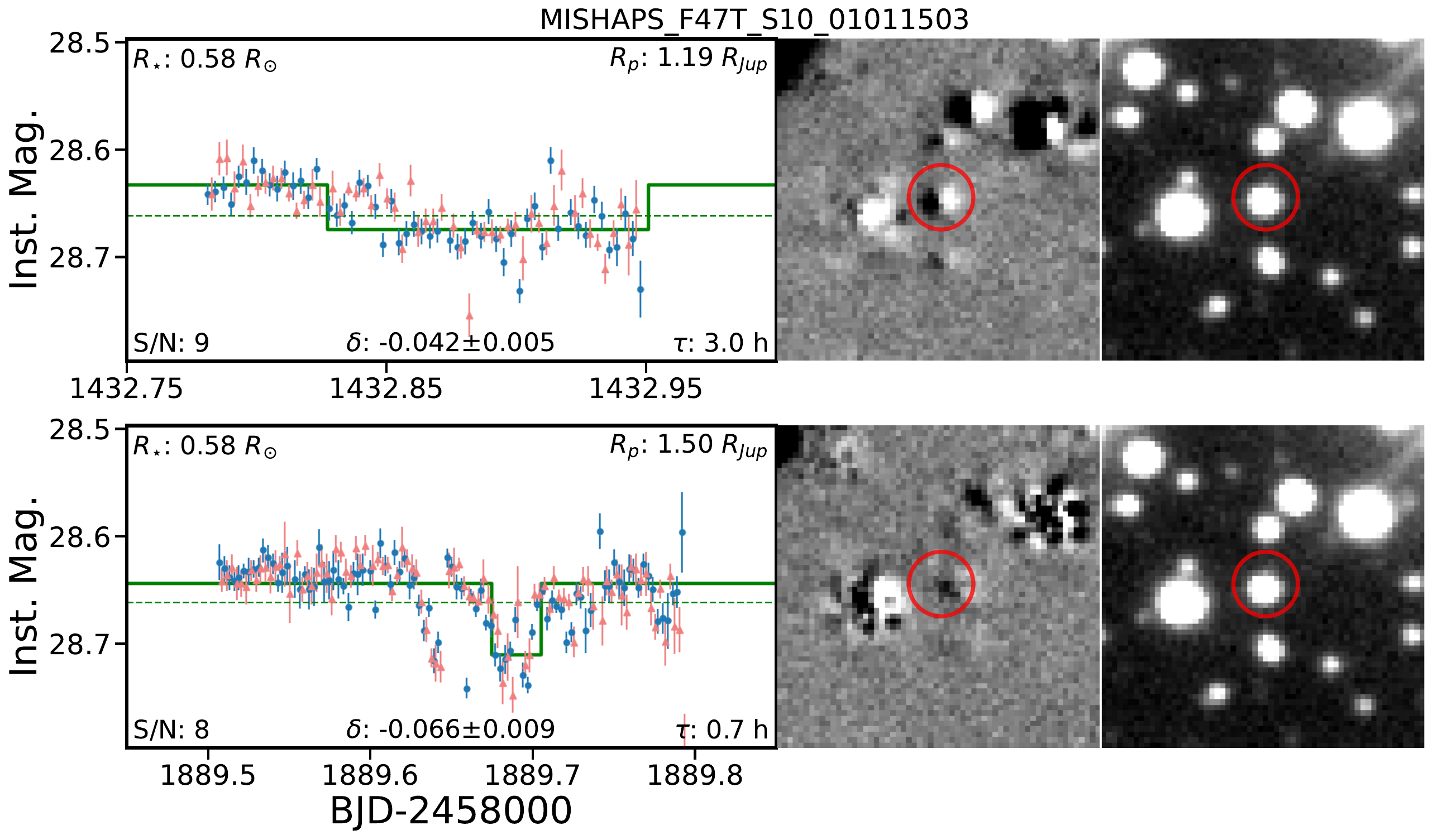}
	\includegraphics[width=0.49\textwidth]{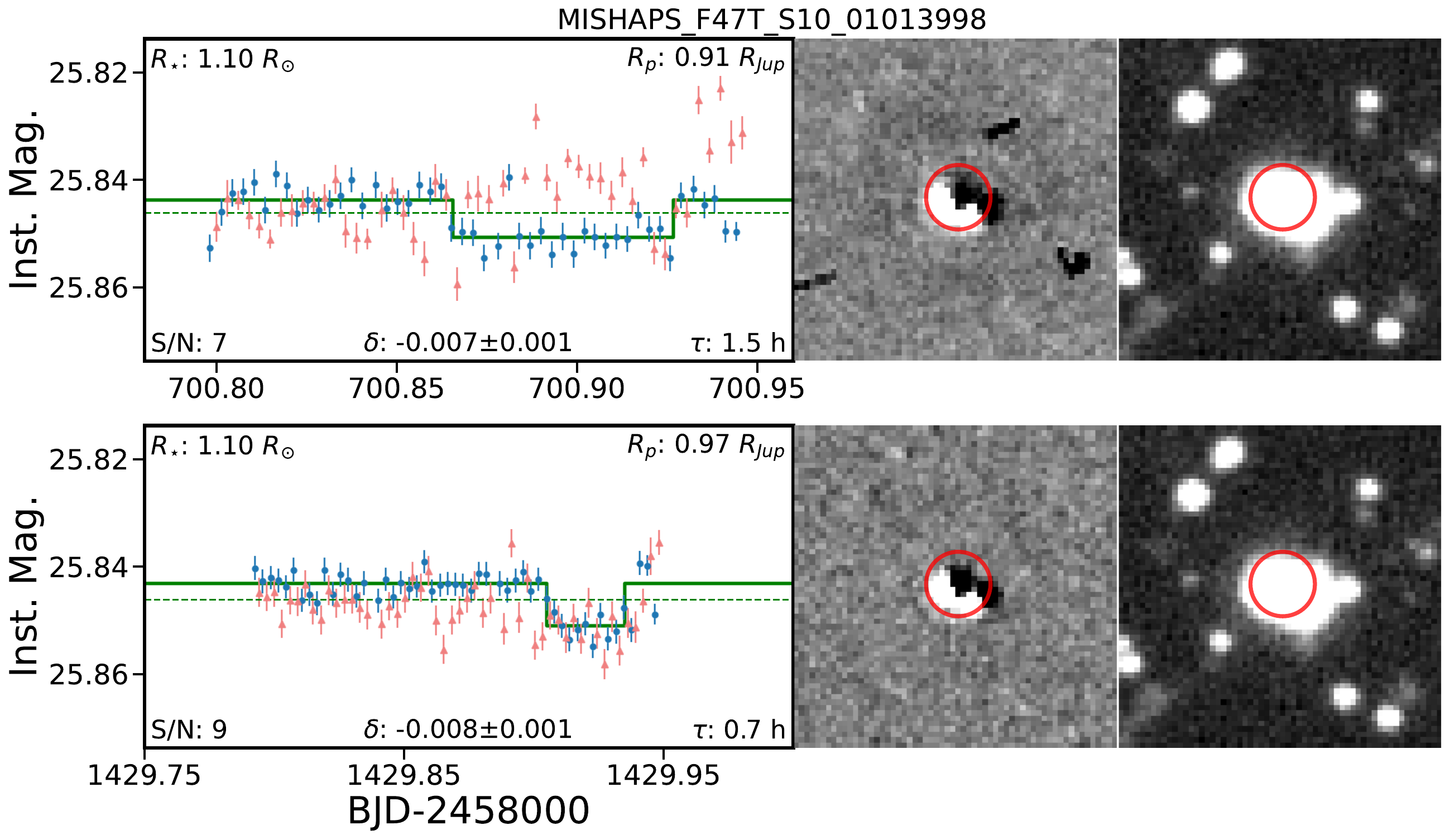}\\
	\caption{Same as previous, for rejected candidates N10\_01015356, N10\_01016157, N10\_01018928, N15\_01005656, S5\_01001384, S10\_01008184, S10\_01011503, and S10\_01013998.}
\end{figure*}

\begin{figure*}[htb]
	\centering
	\includegraphics[width=0.49\textwidth]{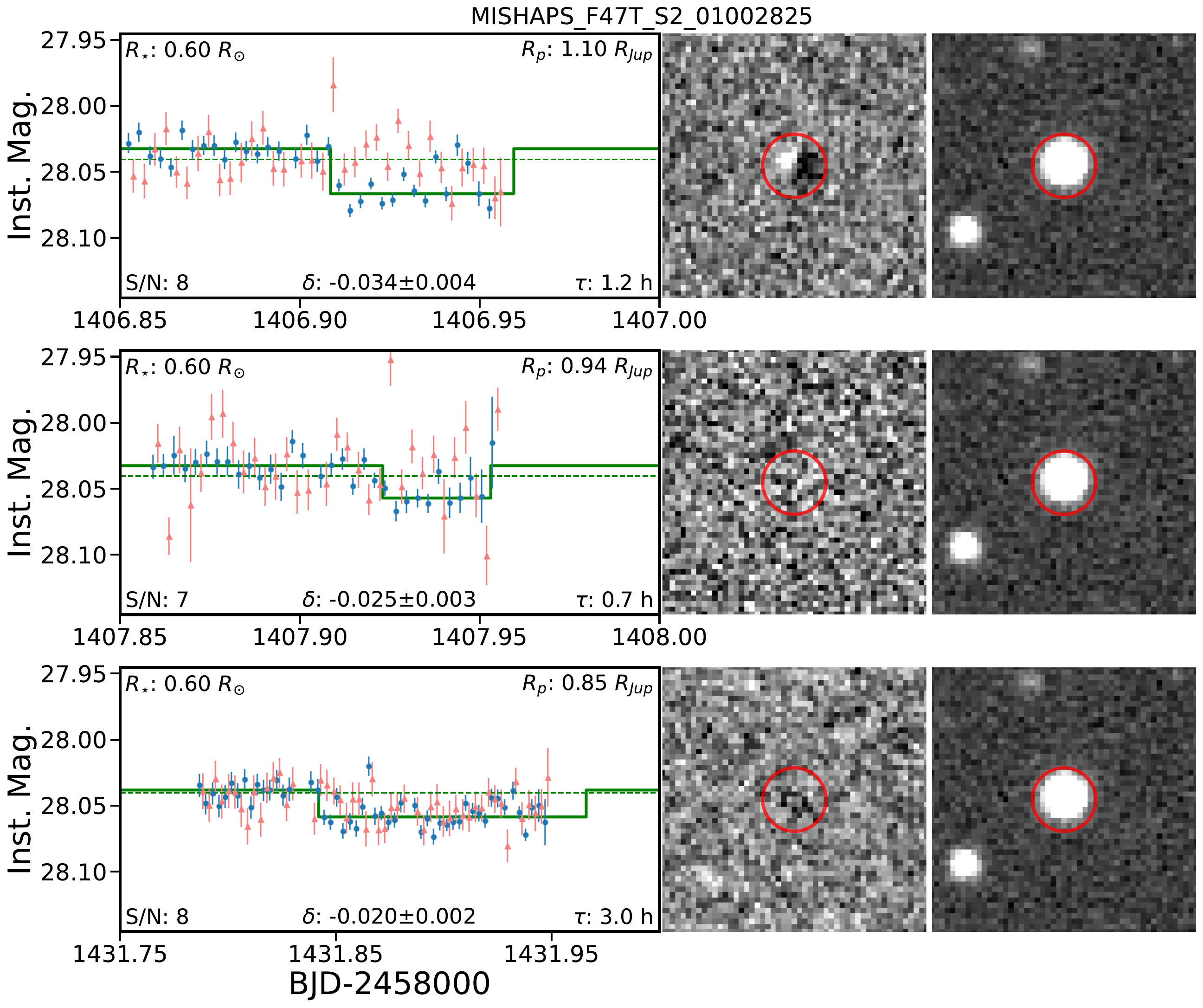}
	\includegraphics[width=0.49\textwidth]{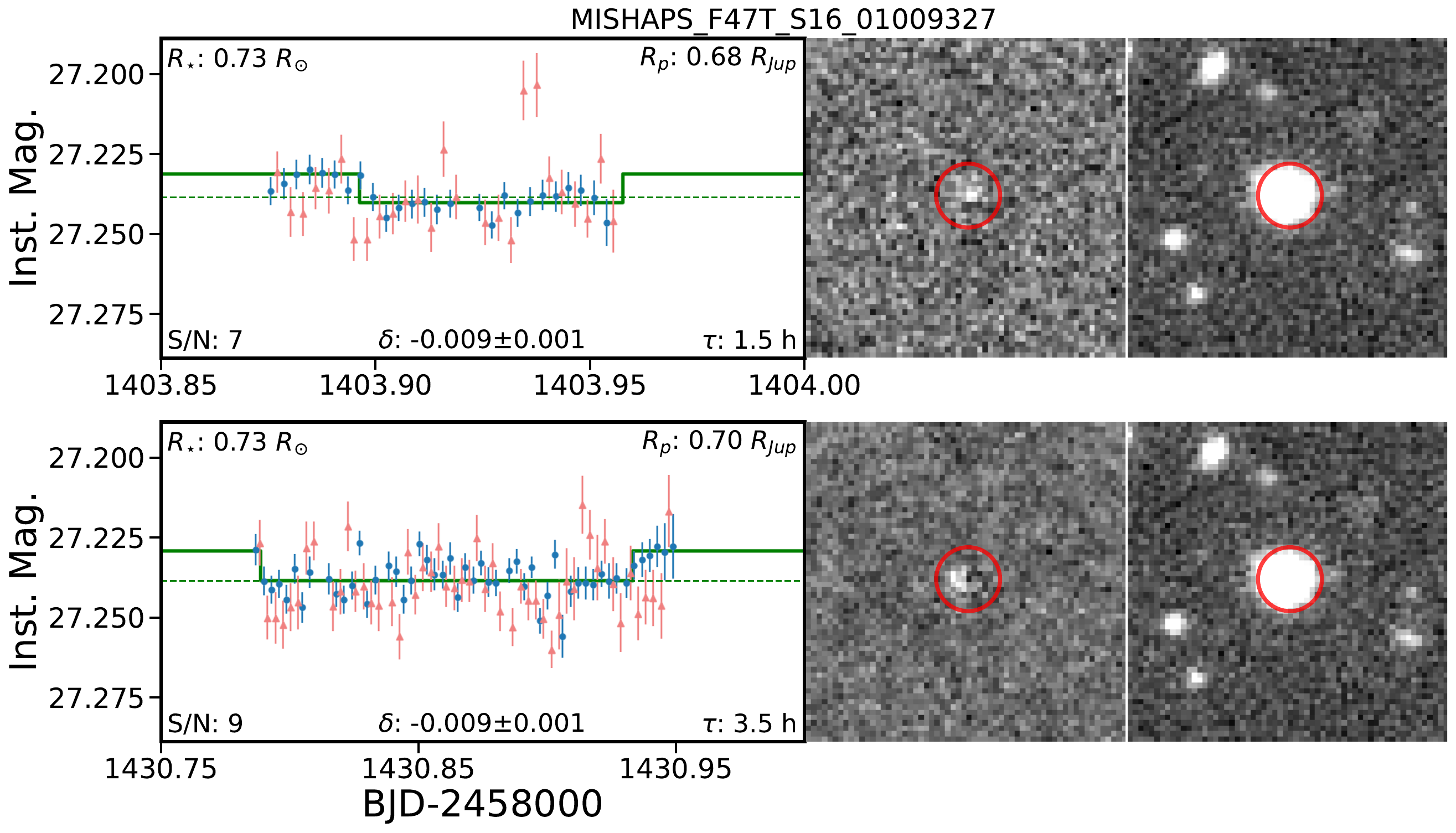}
	\caption{Same as previous, for rejected candidates S2\_01002825 and S16\_01009327.}
\end{figure*}

\begin{figure*}[htb]
	\centering
	\includegraphics[width=0.49\textwidth]{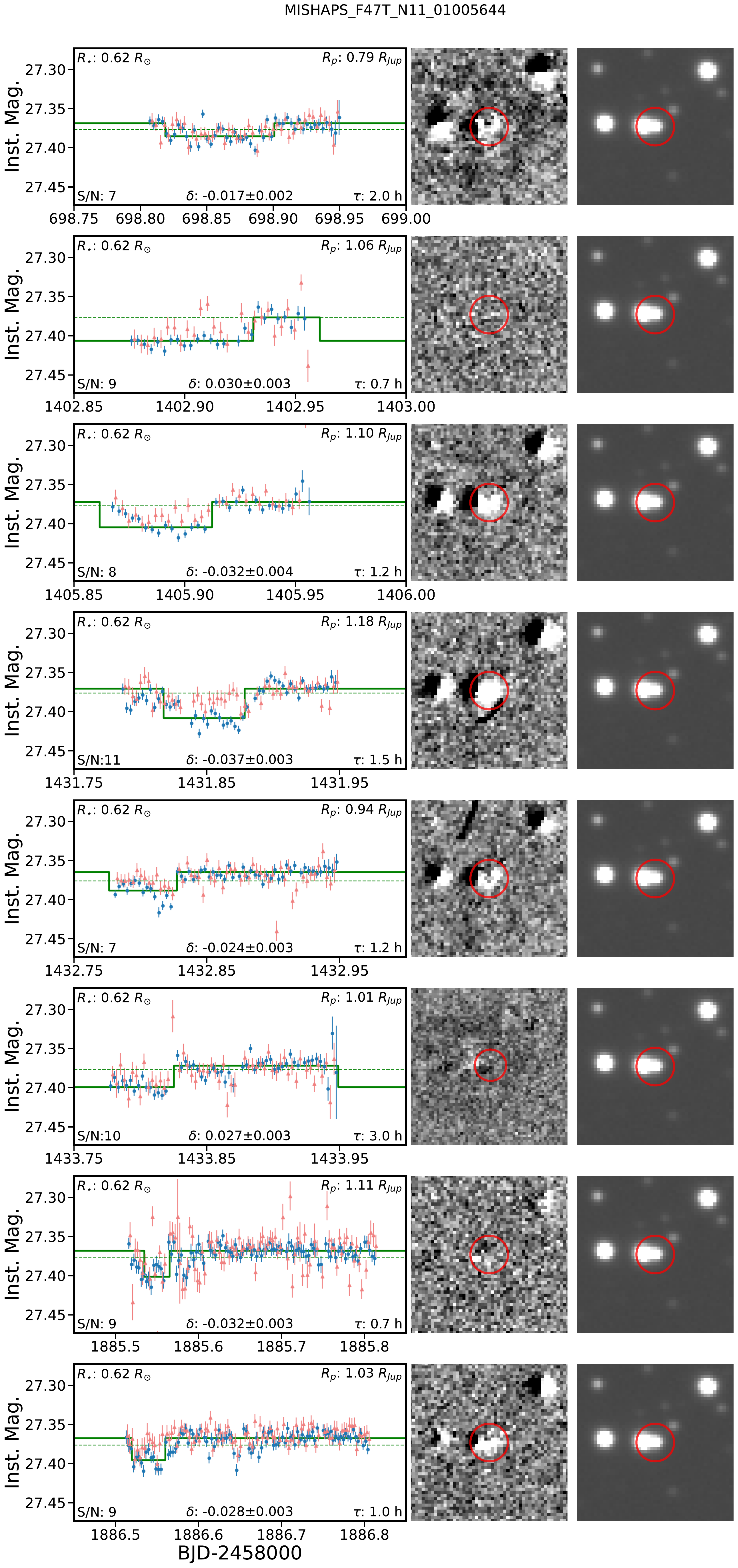}
	\includegraphics[width=0.49\textwidth]{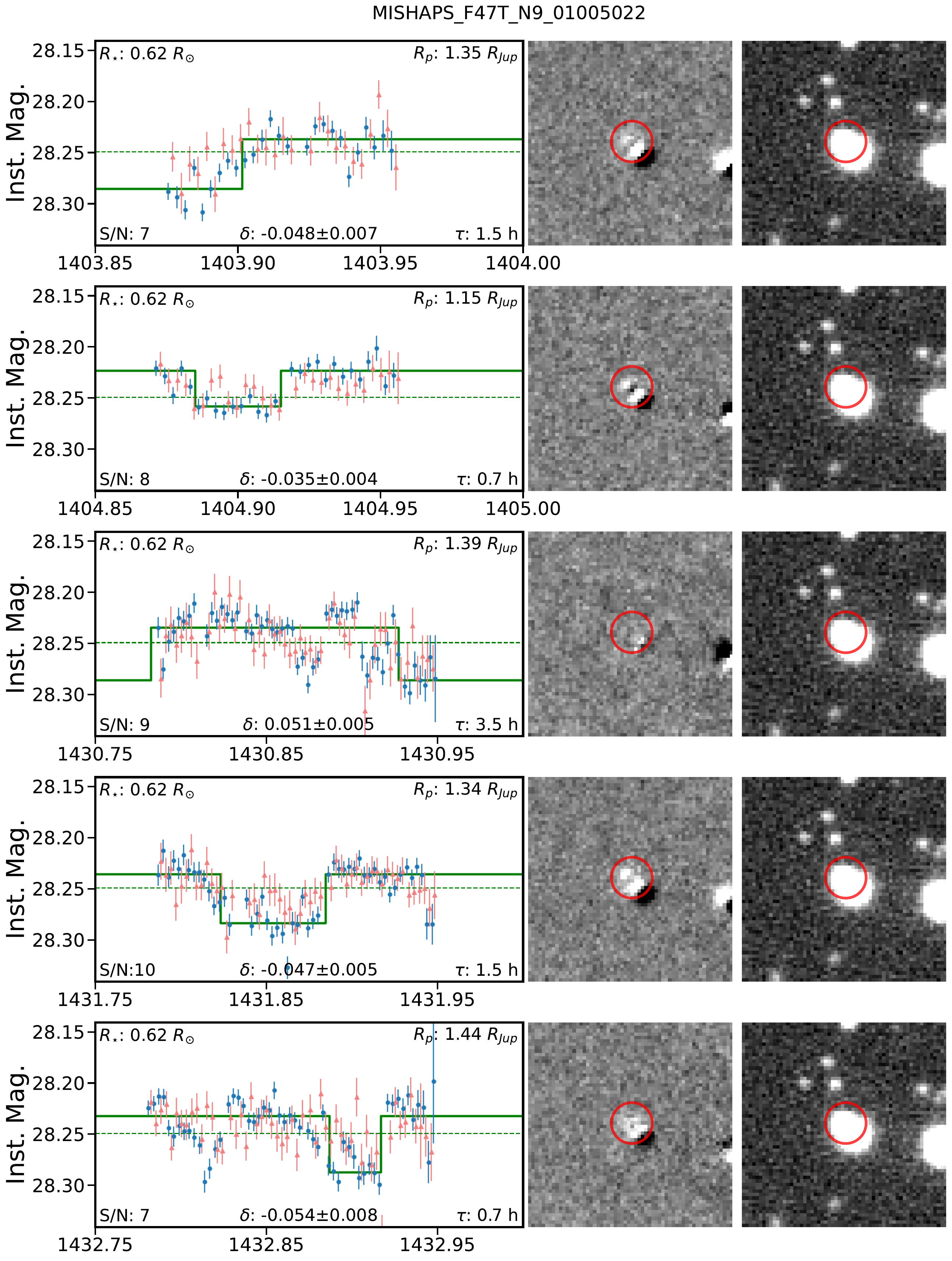}
	\caption{Same as previous, for rejected candidates N11\_01005644 and N9\_01005022.}
\end{figure*}

\end{document}